\documentclass[
    reprint, 
    amsmath, 
    amssymb, 
    aps, 
    prb, 
    floatfix, 
]{revtex4-2}
\pdfoutput=1

\usepackage{soul}
\usepackage[T1]{fontenc} 
\usepackage{lmodern}
\usepackage{amsmath}
\usepackage{array}
\usepackage{dsfont}
\usepackage{mathtools}
\usepackage{graphicx}
\usepackage{dcolumn}
\usepackage{bm}
\usepackage{hyperref}
\usepackage{physics}
\usepackage{empheq}
\usepackage{color}
\usepackage{multirow}
\usepackage{orcidlink}
\usepackage{afterpage}
\usepackage{mathrsfs}
\usepackage{tikz}
\usepackage{ulem}
\usetikzlibrary{shapes}
\usetikzlibrary{tikzmark,calc}

\newcommand{\bs}[1]{\boldsymbol{#1}}

\allowdisplaybreaks

\begin{document}

\title{
Energetics of fractional anomalous Hall crystals in rhombohedral graphene
}

\author{F\'elix Desrochers \orcidlink{0000-0003-1211-901X}}
\email{felix\_desrochers@fas.harvard.edu}
\author{Ashvin Vishwanath \orcidlink{0000-0002-6306-2263}}
\email{avishwanath@g.harvard.edu}
\affiliation{%
Department of Physics, Harvard University, Cambridge, MA 02138, United States
}%

\date{\today}

\begin{abstract}
Fractional anomalous Hall crystals (FAHCs)  replicate the topological order of the fractional quantum Hall effect in the continuum without requiring any external magnetic field. They spontaneously break continuous translation symmetry like a Wigner crystal, but are distinguished by each unit cell holding a fixed fractional number of electrons. Until now, these states have been confined to theoretical speculation or engineered models, leaving open the question of whether they can plausibly emerge in actual physical systems. Here, we establish them as energetically competitive candidate states in a realistic material setting. We study rhombohedral pentalayer graphene (R5G) with variational wavefunctions that are exact zero modes of a recently proposed ideal model of R5G. We evaluate their energies using Monte Carlo, after reinstating realistic dispersion and screened Coulomb interactions. We find FAHCs to be energetically competitive with integer anomalous Hall crystals and Fermi liquids, and their stability follows a simple principle. Each crystal maps onto a parent quantum Hall liquid that fixes its interaction energy, while the kinetic energy favors crystal periods that match the finite-momentum minimum of R5G's Mexican-hat dispersion. A weak periodic potential can then selectively lower and pin the commensurate fractional crystals. This picture predicts how the integer and fractional quantum anomalous Hall stability windows evolve with twist angle and displacement field, which we compare to recent experiments. These results support a continuum-and-interactions-first route to fractional anomalous Hall states in rhombohedral graphene.
\end{abstract}

\maketitle

\section{Introduction}

\begin{figure}
    \centering
    \includegraphics[width=1.00\linewidth]{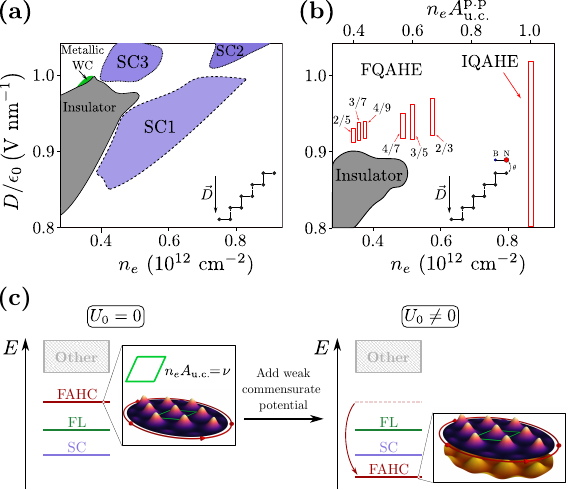}
    \caption{
    {\bf Schematic phase diagram and commensuration mechanism of QAHE stabilization.} (a) Moir\'eless R5G, where superconductivity (SC) and metallic Wigner-crystal behavior have been reported~\cite{han2025signatures,han2026evidence}. (b) R5G aligned with an hBN substrate at a twist angle of $0.77^\circ$ in the strong-displacement-field regime, where integer and fractional quantum anomalous Hall effects (QAHE) have been observed~\cite{lu2024fractional,lu2025extended}. (c) The mechanism proposed in this work. In the continuum ($U_0 = 0$), a fractional anomalous Hall crystal (FAHC) at filling $n_e A_{\rm u.c.}=\nu<1$ can be energetically competitive with nearby candidates such as the Fermi liquid, superconductor, or other ordered states. A weak periodic potential, $U_0\neq0$, acts as a commensuration field: when its unit cell matches that of the FAHC, the crystal gains a pinning energy that can lower it below the competing states, making it the ground state. Red loop: chiral edge mode of the FAHC. Green outline: crystal unit cell.
    }
    \label{fig:experimental_phase_diagram_r5g}
\end{figure}

Rhombohedral multilayer graphene has emerged in recent years as a remarkably flexible platform for strongly correlated topological matter. By tuning the carrier density and external electric field, experiments on graphene stacks with and without a nearly aligned hBN moir\'e substrate have reported integer and fractional quantum anomalous Hall effects (IQAHE and FQAHE)~\cite{lu2024fractional, lu2025extended, xie2025tunable, waters2025chern, aronson2025displacement, Su2025, huo2025does, uzan2025hbn, zhang2025moire, li2025stacking, liu2026odd}, metallic Wigner crystals~\cite{han2026evidence}, and signatures of chiral superconductivity~\cite{han2025signatures, choi2025superconductivity, yang2025magnetic, seo2025family, dutta2026reconfigurable} (Fig.~\ref{fig:experimental_phase_diagram_r5g}(a) and (b)). Within this broad landscape of low-temperature phases, the FQAHE in rhombohedral graphene/hBN~\cite{lu2024fractional, lu2025extended} poses a particularly sharp theoretical puzzle~\cite{parameswaran2024anomalous, devakul2024new, vishwanath2023zero}, as it appears in a regime where the standard moir\'e fractional Chern insulator starting point is not obviously available.

In the usual setting, one begins with an isolated Chern band with favorable quantum geometry, weak residual dispersion, and a clear separation from remote bands~\cite{ju2024fractional, cao2025fractional, bernevig2025fractional}. Rhombohedral graphene/hBN heterostructures differ in several important ways. The FQAHE is observed in a strong displacement field, where the relevant conduction-band states are polarized mainly on the graphene surface away from the hBN substrate. In this regime, the moir\'e potential is not obviously a dominant microscopic scale, and it is not clear that the non-interacting moir\'e problem by itself produces the relevant isolated topological mini-band~\cite{dong2024anomalous, zhou2024fractional, dong2024theory, guo2024theory,  dong2024stability,  soejima2024anomalous,  huang2024self, patri2024extended, arbeitman2024Moire, kwan2025moire, yu2025moire,  Huang2025,  li2025multiband}. Irrespective of the quantitative strength of the moir\'e potential, which remains highly debated~\cite{kwan2025moire, huo2025does, uzan2025hbn}, the Chern band associated with the IQAHE appears to be interaction-induced~\footnote{Current theoretical work suggests that, starting from a metallic state with a very weak moiré potential, interactions lead to the formation of an isolated Chern band~\cite{dong2024theory, dong2024anomalous, zhou2024fractional, kwan2025moire}. Even if the periodic potential is made stronger such that there is an isolated band at charge neutrality, the topology of the isolated conduction Chern band at filling $\nu=1$ of the unit cell still appears to be set by interaction. This arises because application of a periodic potential and interaction in a continuum band with non-trivial quantum geometry typically favor different topologies of the mini-band~\cite{tan2024parent, desrochers2026electronic}. We also refer to this latter case as an interaction-induced Chern band.}. These observations suggest that it may be useful to first understand the continuum interacting problem itself and only then ask how a moderate periodic potential selects among the competing continuum states. This problem is reminiscent of the conventional two-dimensional jellium problem~\cite{wigner1934interaction, pines1952collective, bohm1953collective, bonsall1977some, tanatar1989ground, chitra2001pinned, fogler2000dynamical}, but with nontrivial quantum geometry~\cite{Parameswaran2012, parameswaran2013fractional, Roy2014, yu2025quantum, wang2021exact, ledwidth2023vortexability, ledwith2020fractional, estienne2023ideal} and a tunable non-monotonic dispersion. 

Electronic crystals provide a natural starting point for such a continuum perspective. In an ordinary Wigner crystal, Coulomb interactions break continuous translation symmetry but do not produce a quantized Hall response. In contrast, an integer anomalous Hall crystal (IAHC) combines translation symmetry breaking with Berry curvature, producing a filled Chern band and an integer anomalous Hall response without an external magnetic field~\cite{dong2024anomalous, dong2024theory, zhou2024fractional, dong2024stability, soejima2024anomalous, tesanovic1989hall, halperin1986compatibility, kivelson1986cooperative, kivelson1987cooperative, tan2024parent, valenti2025quantum, miao2026various, bernevig2025berry, zhou2025new, desrochers2026elastic, desrochers2026elastic, hirsbrunner2026topological, zeng2025berry, zheng2024sublattice, dong2025phonons, soejima2025Jellium, soejima2026topological, guo2025correlation}. Fractional anomalous Hall crystals (FAHCs) extend this idea by combining a crystalline charge texture with fractional quantum Hall topological order~\cite{tan2024wavefunction, song2024intertwined, zhou2025new,lu2026generic}. In such a state, the crystal unit cell contains a fractional electron number, and, once the crystal is pinned, the Hall response and fractional charge are inherited from an underlying quantum Hall (QH) liquid. A non-integer electron count per unit cell runs counter to the usual picture of electron crystallization, in which a crystal locks to an integer commensurate filling, so such states may appear to be exotic outliers. Here we argue that fractional crystals are energetically competitive and naturally compatible with the microscopic features of rhombohedral graphene. Furthermore, signatures of metallic or self-doped Wigner crystals have recently been reported in moir\'eless rhombohedral multilayers~\cite{han2026evidence} (see Fig.~\ref{fig:experimental_phase_diagram_r5g}(b)). This points to electronic crystallization as a robust tendency in this material family. The fact that they are self-doped further suggests that their period is influenced by continuum energetics rather than being rigidly tied to the electronic density~\cite{feng2026self, dong2026crystals}.

In this work, we evaluate the energetic competitiveness of FAHCs in rhombohedral graphene and use the insights gained to make phenomenological predictions and identify trends in experimental observables. We consider the ideal holomorphic limit of rhombohedral graphene~\cite{tan2025ideal} as a parent problem that supplies explicit IAHC and FAHC wavefunctions. In this limit, a repulsive contact interaction has a large zero-mode manifold containing the anomalous Hall crystals (AHCs) of interest. We use these explicit wavefunctions as variational states after restoring the physical rhombohedral-graphene dispersion and a dual-gated screened Coulomb interaction. Our calculation proceeds in two stages. We first evaluate the IAHC and FAHC energies directly by Monte Carlo simulation. We then develop a semi-analytic energy evaluator, benchmarked against Monte Carlo, based on an AHC--QH correspondence highlighted in previous work~\cite{tan2024wavefunction}. This shows that the interaction energy is largely inherited from the corresponding quantum Hall liquid, while the kinetic-energy splitting is controlled by how the crystal period distributes weight over the parent-band momenta.

Applying this framework to rhombohedral pentalayer graphene, we identify a simple mechanism that can favor the formation of fractional crystals. For a monotonic dispersion, a FAHC at fixed density is penalized because its smaller real-space unit cell pushes the unfolded parent-band momentum distribution to larger momenta. The Mexican-hat dispersion of rhombohedral graphene changes this logic. A fractional crystal can lower its kinetic energy when its reciprocal-lattice scale is larger and matches the finite-momentum ring where the conduction band is minimized. This momentum-space locking selects a preferred range of lattice constants and naturally produces a sequence of favorable fractional unit-cell fillings. 

This leads to a continuum-and-interactions-first route to fractional anomalous Hall phenomenology. The moir\'e potential need not first create an isolated topological mini-band into which interactions are projected. Instead, the continuum system can contain nearly competitive anomalous Hall crystals, while the moir\'e perturbation subsequently pins and selectively lowers crystals whose period is commensurate with the substrate, as illustrated in Fig.~\ref{fig:experimental_phase_diagram_r5g}(c). This framework provides a conceptually clear alternative picture when an isolated single-particle miniband is not the natural starting point. We anticipate that this perspective will motivate and aid the interpretation of future fully microscopic multiband calculations of graphene/hBN. Moreover, when combined with the momentum-locking feature of the rhombohedral dispersion, this perspective yields concrete predictions for the phase diagram. These predictions are consistent with existing experiments and motivate further systematic mapping of the stability windows of the integer/fractional quantum anomalous Hall effect as functions of the displacement field and the twist angle.

The rest of this work is organized as follows. In Sec.~\ref{sec:ideal_model_zero_modes}, we introduce the ideal rhombohedral parent band and the contact-interaction zero modes used as variational states. In Sec.~\ref{sec:energetics}, we benchmark their energetics using Monte Carlo simulations and formulate semi-analytic energy evaluators. In Sec.~\ref{sec:r5g_phase_diagram}, we apply the framework to rhombohedral pentalayer graphene and discuss the momentum-locking and twist-angle trends. We conclude in Sec.~\ref{sec:discussion} with implications, limitations, and experimental tests.

\section{Ideal rhombohedral graphene and contact-interaction zero modes}\label{sec:ideal_model_zero_modes}

In this work, we use the ideal limit of rhombohedral graphene introduced in Ref.~\cite{tan2025ideal} as a parent problem to supply variational wavefunctions studied below. We thus begin by briefly reviewing the ideal limit, its justification, and the contact-interaction zero modes that contain the IAHC and FAHC wavefunctions. We note that related ideal continuum models have also been considered in Refs.~\cite{desrochers2026electronic, han2025exact, bernevig2025berry}.

\subsection{Ideal parent band}

\begin{figure}
    \centering
    \includegraphics[width=1.00\linewidth]{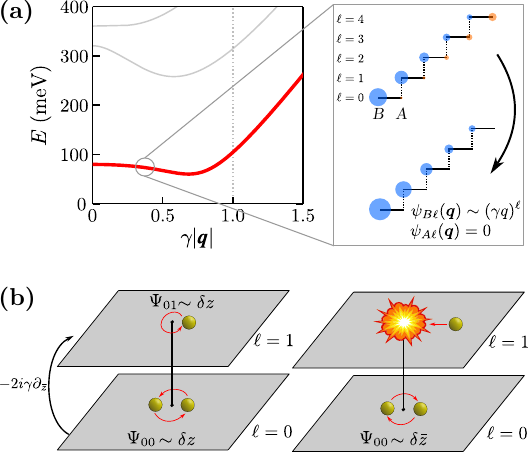}
    \caption{
    {\bf Illustration of the ideal limit of RNG and many-body constraint in the presence of contact interactions.} (a) Dispersion of R5G in the presence of a finite displacement field, where the inset presents the orbital occupation in the $\gamma |\bs{q}|\lesssim 1$ regime (top) and in the ideal limit (bottom). (b) Contact interaction constraint. For a chiral zero between a pair of particles (left), the zero remains for any layer component. For an achiral zero of low order, wavefunctions with higher-layer components can have a non-vanishing amplitude for configurations with electrons at the same location. 
}
\label{fig:ideal_model_andf_contact_interaction}
\end{figure}

Rhombohedral multilayer graphene consists of multiple graphene sheets that are stacked in a regular staircase pattern. The different orbitals in its unit cell are labeled by graphene sublattices $\{B, A\}$ and layers $\ell\in\{0,1,\ldots, N_{\ell}-1\}$ (see Fig.~\ref{fig:ideal_model_andf_contact_interaction}). The minimal tight-binding model for a single valley and single spin species consists of graphene Dirac cones with Fermi velocity $v_D$, coupled through interlayer hopping $t_1$ between orbitals $(A,\ell)$ and $(B,\ell+1)$. At the valley center $\bs{q}=\bs{0}$ this results in completely isolated degenerate edge states that are supported on the $(B,0)$ and $(A, N_{\ell}-1)$ orbitals as well as dimers that form bonding and antibonding states between $(A,\ell)$ and $(B,\ell+1)$ with energy $\pm t_1$. At finite momentum $\bs{q}$ the two surface states start to couple through a leading order coupling $\Delta_{N_\ell}(\bs q)= - (v_D q)^{N_\ell}/t_1^{N_\ell - 1}$, with $q=q_x+iq_y$, which leads to a dispersion $\mathcal{E}(\bs{q})\propto \pm|\bs{q}|^{N_\ell}$ with approximate eigenstates
\begin{align}
    \ket{s_{\pm}^{(0)}(\bs q)} = \frac{1}{\sqrt 2} \left( \ket{\varphi_A(\bs q)} \pm e^{-i \arg[\Delta_{N_\ell}(\bs q)] }  \ket{\varphi_B(\bs q)} \right).
\end{align}
The two chiral basis states are supported only on the $A$ or $B$ sublattice and have layer components
\begin{equation}
\begin{aligned}
    \ket{\varphi_B(\bs q)}
    &=
    N_{\bs q}
    \sum_{\ell=0}^{N_\ell-1}
    (\gamma q)^\ell
    \ket{B,\ell},\\
    \ket{\varphi_A(\bs q)}
    &=
    N_{\bs q}
    \sum_{\ell=0}^{N_\ell-1}
    (\gamma \bar q)^{N_\ell-1-\ell}
    \ket{A,\ell}. 
\end{aligned}
\label{Eqn:Edgestates}
\end{equation}
Here we use the interlayer hopping length $\gamma\equiv v_D/t_1\approx 1.73~\mathrm{nm}$ and $ N_{\bs q}^{-2}=\sum_{\ell=0}^{N_{\ell}-1} |\gamma\bs{q}|^{2\ell}$ is a normalization constant. This approximate form of the eigenstates is only valid for $|\gamma\bs{q}|\lesssim 1$. For larger momentum, the dominant energy scale is the intralayer hopping $v_D q$, which results in an approximately linear dispersion~\footnote{See Supplemental Material for the derivation and validity analysis of the ideal rhombohedral-graphene limit, the construction and band geometry of the anomalous Hall crystal single-particle basis, the AHC--QH correspondence and the bound justifying it, closed-form expressions for the kinetic, interaction, and periodic-potential energies, the Monte Carlo methodology and benchmarks, and the Hartree-Fock Fermi-liquid energy.
}.

In the presence of a finite displacement field, which we model as an interlayer potential $u_D$, the two edge states of Eq.~\ref{Eqn:Edgestates} are now split by a gap of $\Delta_D \equiv (N_\ell-1)u_D$. As shown in Fig.~\ref{fig:ideal_model_andf_contact_interaction}(a), this results in a relatively flat band in the regime $|\gamma\bs{q}|\lesssim  1$. At finite momentum, the two edge states still hybridize, which results in eigenstates for the conduction band of the form 
\begin{align}
    \ket{s^{(e)}(\bs{q})}
    &\approx
    \ket{\varphi_B(\bs q)}
    +
    \frac{\Delta_{N_\ell}(\bs q)}{\Delta_D}
    \ket{\varphi_A(\bs q)}. 
\end{align}
The ideal limit of Refs.~\cite{tan2025ideal, han2025exact} then considers the regime $\Delta_{N_\ell}(\bs q)/\Delta_D \ll 1$ such that we can drop the small admixture between surface states. This results in an ideal~\cite{Parameswaran2012, parameswaran2013fractional, Roy2014, yu2025quantum, wang2021exact, ledwidth2023vortexability, ledwith2020fractional, estienne2023ideal, cano2026ideal} continuum conduction band supported a single sublattice with eigenstates $\ket{\psi(\bs{q})}=\exp(i\bs{q}\cdot\bs{r})\ket{s(\bs{q})}$, where
\begin{align}
    \braket{B,\ell}{s(\bs{q})} = N_{\bs{q}} (\gamma q)^{\ell}, \qquad \braket{A,\ell}{s(\bs{q})}  = 0. \label{eq:ideal_rng_wavefunction}
\end{align}
The form factor of this parent band is
\begin{align}
    F(\bs{q}, \bs{q}') &= \braket{s(\bs{q})}{s(\bs{q}')}= N_{\bs{q}} N_{\bs{q}'} \sum_{\ell=0}^{N_{\ell}-1} (\gamma^{2} \bar{q} q')^\ell . \label{eq:parent_band_form_factor}
\end{align}
The holomorphic structure of Eq.~\eqref{eq:ideal_rng_wavefunction} also fixes the quantum geometry. The Berry curvature of the ideal parent band is $\Omega(\bs q) = -2\partial_q\partial_{\bar q} \log N^{-2}_{\bs{q}}$, and the corresponding Fubini--Study metric satisfies the ideal isotropic relation $g_{\alpha\beta}(\bs q)=\frac{1}{2}|\Omega(\bs q)|\delta_{\alpha\beta}$~\cite{wang2021exact, ledwidth2023vortexability,cano2026ideal}.

Such an ideal limit is only valid in the regime where the displacement field is strong enough to isolate the two surface states, but not so strong as to hybridize them with dimers of energy on the order of $t_1$. That is, we expect it to work best when $t_1|\gamma q|^{N_\ell} \ll (N_\ell-1)u_D \ll t_1$ (see Supplemental Material~\cite{Note2} for more details). 

\subsection{Many-body states with contact interaction}

To consider the interacting many-body problem, it is convenient to first recast the ideal limit in real space. Projection to the ideal continuum band~\eqref{eq:ideal_rng_wavefunction} imposes the single-particle constraint
\begin{align}
    \braket{\bs{r},\ell}{\psi}
    =
    \psi_{\ell}(\bs{r})
    =
    (-2i\gamma \partial_{\bar z})^{\ell}\psi_0(\bs{r}),
    \label{eq:antiholomorphic_condition_sp_wavefunction}
\end{align}
with $\partial_{\bar z}=(\partial_x+i\partial_y)/2$ and $z=x+iy$. For an $N_e$-particle state, this becomes
\begin{align}
    \Psi_{\{\ell\}}(\{\bs r\})
    =
    \left[
    \prod_{j=1}^{N_e}
    (-2i\gamma\partial_{\bar z_j})^{\ell_j}
    \right]
    \Psi_{\{0\}}(\{\bs r\}),
    \label{eq:antiholomorphic_condition_mb_wavefunction}
\end{align}
where $\{\bs r\}\equiv(\bs r_1,\bs r_2,\ldots,\bs r_{N_e})$ and
$\{\ell\}\equiv(\ell_1,\ell_2,\ldots,\ell_{N_e})$. Projection to the ideal parent band imposes a specific relation between every layer component and the parent many-body wavefunction $\Psi_{\{0\}}(\{\bs r\})$, in which all particles occupy the zeroth layer.

To understand what form of $\Psi_{\{0\}}(\{\bs r\})$ is favored by interactions, we consider a repulsive contact interaction
\begin{align}
    H_{\rm int}
    =
    V_0\sum_{i<j}
    \delta^{(2)}\left( \bs{r}_i- \bs{r}_j\right),
    \qquad V_0>0,
\end{align}
where $\delta^{(2)}(\bs r)$ is the two-dimensional Dirac delta function. A zero-energy state of the contact interaction must vanish whenever any two particles coincide for every layer component. Fermi statistics guarantees a zero when two particles coincide in the zeroth-layer component, but derivatives in Eq.~\eqref{eq:antiholomorphic_condition_mb_wavefunction} can remove a generic zero in higher-layer components. A holomorphic, or chiral, zero is special in this context. If $\Psi_{\{0\}} \sim z_i-z_j$ as $\bs r_i\to \bs r_j$, then the anti-holomorphic derivatives $\partial_{\bar z_i}$ and $\partial_{\bar z_j}$ cannot remove this zero. By contrast, as illustrated in Fig.~\ref{fig:ideal_model_andf_contact_interaction}(b), an anti-chiral factor $(\bar z_i-\bar z_j)^p$ can be weakened or eliminated by the same derivatives unless its order is sufficiently high. Hence, strong short-range interactions naturally favor a chirality of pairwise zeros in $\Psi_{\{0\}}$ \cite{tan2025ideal}.

Restricting ourselves to the case where $\Psi_{\{0\}}(\{\bs r\})$ contains only chiral nodes for every particle pair, this identifies a broad class of contact interaction zero modes of the form
\begin{align}
    \Psi_{\{\ell\}}(\{\bs r\})
    = \mathcal{J}(\{z\})\Xi_{\{\ell\}}(\{\bs r\}), 
    \label{eq:chiral_jastrow_zero_modes}
\end{align}
where $\mathcal{J}(\{z\}) = \prod_{i<j}(z_i-z_j)$ and $\Xi_{\{\ell\}}(\{\bs r\})$ is symmetric under particle exchange and nonsingular when particles coincide. Since $\mathcal J$ is holomorphic, we have 
\begin{align}
    \Xi_{\{\ell\}}(\{\bs r\}) = \left[ \prod_{j=1}^{N_e} (-2i\gamma\partial_{\bar z_j})^{\ell_j} \right] \Xi_{\{0\}}(\{\bs r\}).
\end{align}

\subsection{Translation symmetry breaking ansätze}

The discussion above identifies the local structure required of contact-interaction zero modes, but it leaves considerable freedom in their long-distance organization. We now specialize to a physically motivated submanifold of such zero modes: states that spontaneously break continuous translation symmetry and simultaneously carry an anomalous Hall response.

A natural starting point is the observation that spin-polarized lowest-Landau-level (LLL) quantum Hall wavefunctions, $\Phi^{(\nu)}_{\rm QH}$~\cite{laughlin1983anomalous, haldane1983fractional,jain1989composite, prange1990quantum, jain2007composite}, have precisely the required local structure since they have chiral zeros and are antisymmetric under exchange. For example at filling $\nu=1/p$ with $p$ odd, $\Phi^{(\nu)}_{\rm QH}$ could be
\begin{align}
    \Phi^{(1/p)}_{\rm QH} = \prod_{i<j}(z_i-z_j)^p \exp(-\sum_{i}\frac{ |\bs{r}_i|^2 }{4 l_B^2}), \label{eq:laughlin_state_wf}
\end{align}
which describes a fully filled LLL at $p=1$ or Laughlin states for $p>1$~\cite{laughlin1983anomalous}. 

To obtain a zero-field crystal rather than an ordinary quantum Hall liquid in an external magnetic field, we multiply the quantum Hall state by a one-body Abrikosov-lattice factor in the opposite emergent magnetic field. We choose one emergent flux quantum per crystal unit cell,
\begin{align}
    A_{\rm u.c.}=2\pi l_B^2 = |\bs{a}_1 \wedge \bs{a}_2|,
    \label{eq:one_flux_per_unit_cell}
\end{align}
so that the number of electrons per crystal unit cell is
\begin{align}
    \nu
    =
    n_e A_{\rm u.c.},
    \label{eq:crystal_filling_relation}
\end{align}
where we denote the lattice basis vectors $\bs{a}_1, \bs{a}_2$ and use $\wedge \bs{v} = (v_y, -v_x)$. Therefore, $\nu$ is both the filling of the quantum Hall factor and the electronic filling of the self-generated crystal unit cell.

More explicitly, let $\chi_{0}(\bs r)$ denote an Abrikosov-lattice orbital in the magnetic field opposite to that of the quantum Hall factor. The most general choice within this magnetic-translation sector can be expanded in complex-conjugated Landau-level orbitals as~\cite{tan2024parent}
\begin{align}
    \chi_0(\bs r)
    =
    \sum_{n\geq 0}
    c_n\,
    \overline{\varphi_{\bs 0}^{n{\rm LL}}(\bs r)} ,
    \label{eq:general_abrokosov_factor_main}
\end{align}
where the overline denotes complex conjugation, corresponding to the opposite emergent magnetic field, and $\bra{\bs{r}} c_{\bs{k}}^{n\rm{LL}\dagger}\ket{0} = \varphi_{\bs k}^{n{\rm LL}}(\bs r)$ is a magnetic Bloch state with momentum $\bs{k}$ in the $n^{\rm th}$ LL~\cite{Note2}. In what follows, we use the simplest representative of this family $\chi_0(\bs r) = \overline{\varphi_{\bs 0}^{\rm LLL}(\bs r)}$. This choice fixes the detailed shape of the crystalline texture and leaves no additional variational parameter in the wavefunction. Allowing higher conjugated-Landau-level components in Eq.~\eqref{eq:general_abrokosov_factor_main} would enlarge the variational space by optimizing the Abrikosov texture, but this refinement is not included below. Higher layer components are given by
\begin{align}
    \chi_{\ell}(\boldsymbol{r}) &= \left(-2 i \gamma\left(\partial_{\bar{z}}-z / 4 l_B^2\right) \right)^{\ell} \chi_{0}(\boldsymbol{r})\nonumber \\
    &=  \sqrt{ \ell ! } \left( -\frac{\sqrt{2} \gamma}{l_B} \right)^\ell \bar{\varphi}_{\bs{0}}^{\ell \mathrm{LL}}(\bs{r}).
    \label{eq:chi_layer_generation_simple}
\end{align}

The AHC wavefunctions of interest are then products of the LLL fermionic wavefunction and of a bosonic condensate in the Abrikosov lattice subjected to an opposite field~\cite{tan2025ideal}
\begin{align}
    \Psi_{\nu,\{\ell\}}^{\rm AHC}(\{\bs r\})
    =
    \Phi^{(\nu)}_{\rm QH}(\{\bs r\})
    \prod_{i=1}^{N_e}
    \chi_{\ell_i}(\bs r_i).
    \label{eq:ahc_ansatz_main_text}
\end{align}
The quantum Hall part provides the chiral zero, and the Abrikosov factor cancels the magnetic translation phases and produces a periodic charge texture. For $\nu=1$, the quantum Hall factor is the fully filled LLL. It is therefore an IAHC. For fractional incompressible quantum Hall states, such as Laughlin or Jain states at $\nu = 1/3,\, 2/5,\, 3/5,\, 2/3,\ldots$, Eq.~\eqref{eq:ahc_ansatz_main_text} describes FAHCs with fractional electron number per crystal unit cell. In the ideal flat-band limit with contact interaction, all these states are degenerate with many other zero modes. The remainder of the paper asks how physical dispersion and longer-ranged screened interactions lift this degeneracy.

This construction also has a useful parton interpretation. One may write the physical electron as $c_\ell(\bs r) = f(\bs r)b_\ell(\bs r)$, where the $f$ fermion forms the quantum Hall state $\Phi^{(\nu)}_{\rm QH}$, while the $b_\ell$ boson condenses into the Abrikosov lattice in the opposite emergent magnetic field. The two emergent magnetic fields cancel in the gauge-neutral electron, so the physical state is realized in the absence of an external magnetic field. The boson condensate breaks continuous translation symmetry, while the quantum Hall parton carries the topological response. Pinning the crystal then transfers the Hall response of the quantum Hall fermionic parton to the physical electronic state~\cite{ioffe1989gapless}.

This construction also naturally realizes a chiral superconductor with chiral central charge $c_-=-1/2$ relative to the band Berry curvature, as well as a correlated Fermi liquid (FL). In both cases, the $f$ fermions occupy a $\nu=1$ integer Chern band, while the bosons at $\nu=-1$ either form a bosonic Pfaffian state~\cite{moore1991nonabelions}, giving the chiral superconductor, or a composite Fermi liquid~\cite{halperin1993theory, read1998lowest, pasquier1998dipole}, giving the correlated FL. Unlike the wavefunctions derived from the Abrikosov vortex lattice state of the bosons, the energetics of these states are more challenging to evaluate and are left for future work.

\subsection{Emergent ideality}

\begin{figure}
    \centering
    \includegraphics[width=1.00\linewidth]{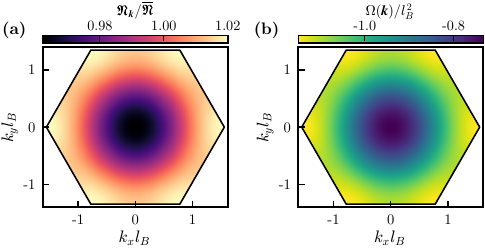}
    \caption{
    {\bf Momentum-space geometry of the emergent ideal  AHC mini-band in the first Brillouin zone.} (a) Normalization factor $\mathfrak{N}_{\boldsymbol{k}}$ relative to its Brillouin zone average $\overline{\mathfrak{N}}$. (b) Berry curvature distribution $\Omega(\boldsymbol{k})/l_B^2$. Both panels use $N_\ell=5$ and $\gamma/l_B=0.3$.}
    \label{fig:berry_curvature_distribution}
\end{figure}

The AHC ansätze in Eq.~\eqref{eq:ahc_ansatz_main_text} arise from partially ($\nu<1$) or fully filling ($\nu=1$) an emergent Chern band generated by the spontaneously formed periodic texture. This can be seen by first noting that any LLL QH wavefunction admits a general expansion in terms of Slater determinants of magnetic Bloch states
\begin{align}
    \Phi_{\mathrm{QH}}(\{\bs{r}\}) = \sum_{\{\bs{k}\}} C_{\{\bs{k}\}} \det_{j,\bs{k}}[ \varphi_{\bs{k}}^{\mathrm{LLL}}(\bs{r}_j) ].\label{eq:expansion_qh_first_quantized}
\end{align}
Our AHC wavefunction of interest~\eqref{eq:ahc_ansatz_main_text} can similarly be written as
\begin{align}
    \Psi_{\{\ell\}}(\{\bs{r}\}) 
    &= \sum_{\{ \bs{k} \}} \frac{C_{\{\bs{k}\}}}{\mathfrak{N}_{\{\bs{k}\}}} \det_{j,\bs{k}}[ \mathfrak{N}_{\bs{k}} \varphi_{\bs{k}}^{\mathrm{LLL}}(\bs{r}_j) \chi_{\ell_j}(\bs{r}_j) ] \nonumber \\
    &= \sum_{\{ \bs{k} \}} C_{\{\bs{k}\}} \left( \det_{j,\bs{k}}[ \varphi_{\bs{k}}^{\mathrm{LLL}}(\bs{r}_j)] \prod_j \chi_{\ell_j}(\bs{r}_j) \right) \nonumber \\
    &= \Phi_{\mathrm{QH}}(\{\bs{r}\}) \prod_{j=1}^{N_e}\chi_{\ell_j}(\bs{r}_j), \label{eq:expansion_ahc_first_quantized}
\end{align}
where $\mathfrak{N}_{\{\bs{k}\}}= \prod_{\bs{k} \in \{\bs{k}\}} \mathfrak{N}_{\bs{k}}$ with the normalization constant
\begin{align}
    \mathfrak{N}_{\bs{k}}^{-2} = \int_{\rm u.c.} d^{2}r \sum_{\ell} \left| \varphi^{\mathrm{LLL}}_{\bs{k}}(\bs{r}) \chi_{\ell}(\bs{r}) \right|^2.
\end{align}
and $\{\bs{k}\}\equiv\{ \bs{k}_1, \bs{k}_2, \ldots, \bs{k}_{N_e} \}$. From the first line in \eqref{eq:expansion_ahc_first_quantized}, it directly follows that the IAHC and FAHC of interest are written in a normalized single-particle basis of the form
\begin{align}
    \braket{\bs{r},\ell}{\psi_{\bs{k}}} = \psi_{\bs{k},\ell}(\bs{r}) = \mathfrak{N}_{\bs{k}} \varphi_{\bs{k}}^{\mathrm{LLL}}(\bs{r}) \chi_{\ell}(\bs{r}), \quad \bs{k}\in {\rm BZ}\label{eq:ahc_single_particle_states}
\end{align}
which represent states forming the self-generated crystal band obtained by folding the continuum parent band in the associated emergent first Brillouin zone (BZ). From the single-particle states~\eqref{eq:ahc_single_particle_states}, we directly see that this interaction-induced mini-band represents a $|C|=1$ band with ideal quantum geometry~\cite{wang2021exact, ledwidth2023vortexability, Note2}. For such an ideal band, the Berry curvature distribution is controlled by the momentum dependence of the normalization $\mathfrak{N}_{\bs{k}}$ and is given by 
\begin{align}
     \Omega(\bs{k}) = - l_B^2 + \nabla_{\bs k}^2 \log( \mathfrak{N}_{\bs{k}} ).
\end{align}
Figure~\ref{fig:berry_curvature_distribution} presents the normalization $\mathfrak{N}_{\bs{k}}$ and associated Berry curvature for representative parameters. Both quantities are mostly uniform and have a relatively small momentum dependence. We emphasize that an ideal parent band does not, on its own, imply that the emergent crystal mini-band also has ideal quantum geometry~\cite{tan2024parent, desrochers2026electronic}. Here, the emergent ideality of the mini-band instead originates from the requirement of chiral zeros between every pair of particles, which ties the single-particle states to holomorphic lowest-Landau-level orbitals.

The form factor for the mini-band factorizes as~\cite{Note2}
\begin{align}
    \mathcal F^{\rm AHC}(\bs k+\bs q,\bs k)
    =
    \frac{\mathcal R_{\bs q}(\bs k)}
    {\sqrt{
    \mathcal R_{\bs 0}(\bs k)
    \mathcal R_{\bs 0}(\lceil\bs k+\bs q\rceil)}}
    \mathcal F^{\rm LLL}(\bs k+\bs q,\bs k),
    \label{eq:main_ahc_form_factor_factorization}
\end{align}
where $\lceil\bs k+\bs q\rceil$ denotes the momentum folded back into the emergent mini-BZ, and $\mathcal F^{\rm LLL}$ is the LLL form factor entering the Girvin-MacDonald-Platzmann algebra~\cite{girvin1986magneto}. We also introduced the periodic coefficients
\begin{align}
    \mathcal{R}_{\bs{q}}(\bs{k})
    &= \frac{1}{A_{\rm u.c.}}\sum_{\bs{g}} \mathcal{P}_{\bs{q}}(\bs{k}+\bs{g})
\end{align}
with
\begin{align}
    \mathcal P_{\bs q}(\bs p)
    =&
    F(\bs p +\bs q, \bs p)
    N_{\bs p}^{-1}N_{\bs p+\bs q}^{-1}\exp\left(-\frac{l_B^2}{2}
    \left(
        |\bs p|^2 + p\bar q 
    \right)
    \right)
    \label{eq:main_R_definition}
\end{align}
where $F(\bs{p},\bs{p}')$ is the parent band form factor~\eqref{eq:parent_band_form_factor} and $\bs{g}=n_1 \bs{b}_1 + n_2 \bs{b}_2$ are reciprocal lattice vectors (i.e., $\bs{a}_i \cdot \bs{b}_j = 2\pi \delta_{ij}$). From these definitions, we note that $\mathcal{R}_{\bs{0}}(\bs{k}) = \mathfrak{N}_{\bs{k}}^{-2}$.

$\mathcal R_{\bs q}(\bs k)$ is a periodic function of $\bs k$ that can thus be rewritten as a sum over lattice harmonics using Poisson resummation. The resulting nonzero lattice harmonics are exponentially suppressed by the real-space lattice scale (see Supplemental Material for details~\cite{Note2}). Keeping only the lowest-order harmonic, which we refer to as the {\it zeroth-harmonic approximation}, gives
\begin{align}
    \mathcal R_{\bs q}(\bs k)\approx \mathcal R,
    \quad
    \mathcal F^{\rm AHC}(\bs k+\bs q,\bs k)
    \approx 
    \mathcal F^{\rm LLL}(\bs k+\bs q,\bs k),
    \label{eq:main_zeroth_harmonic_W}
\end{align}
so the AHC form factor reduces to the LLL form factor. All corrections and Berry curvature inhomogeneity in the mini-BZ are captured by the leading nonzero lattice harmonics of $\mathcal R_{\bs q}(\bs{k})$. In this zeroth-harmonic approximation, all information about the underlying lattice is lost (i.e., observables are independent of the underlying electronic crystal lattice shape). A similar zeroth-harmonic approximation was considered previously in Refs.~\cite{tan2024parent, tan2024wavefunction}. In what follows, we always consider the {\it full} momentum dependence of the form factor~\eqref{eq:main_ahc_form_factor_factorization} and only invoke the zeroth-harmonic approximation to illustrate how certain expressions simplify.

\section{Energetics}
\label{sec:energetics}

\begin{figure}
    \centering
    \includegraphics[width=1.00\linewidth]{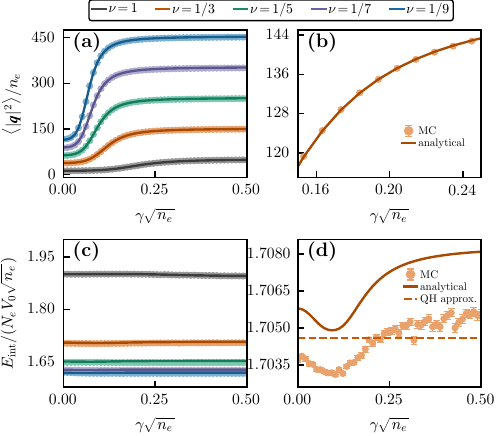}
    \caption{
    {\bf Monte Carlo benchmarks for the energetics of AHC variational states.} Panels (a,b) show the kinetic-energy moment $\langle |\bs{q}|^2\rangle/n_e$. Panels (c,d) show the interaction energy per particle, in units of $V_0\sqrt{n_e}$, for a dual-gated screened interaction with gate distance $d_s=1/\sqrt{n_e}$. Results are for a triangular crystal in the $N_\ell=4$ ideal parent band, including the IAHC at $\nu=1$ and Laughlin FAHCs at $\nu=1/3,1/5,1/7,1/9$. Markers are direct MC estimates, solid curves are the semi-analytic AHC--QH estimate, and dashed curves in the interaction-energy panels are the zeroth-harmonic QH approximation. Panels (b,d) show zoomed-in views of the $\nu=1/3$ data from panels (a,c) respectively.
    }
    \label{fig:mc_evaluation_energy}
\end{figure}

To study the energetics of IAHCs and FAHCs, we now restore two perturbations that lift the ideal contact-interaction degeneracy: a physical single-particle dispersion and a longer-ranged density-density interaction. For a variational state $\ket{\Psi}$ in the ideal parent band we evaluate
\begin{align}
    \frac{E[\Psi]}{N_e}
    =
    \frac{1}{N_e}
    \left\langle
    \sum_i \mathcal E(-i\nabla_i)
    +
    \frac{1}{2}\sum_{i\neq j}V(|\bs r_i-\bs r_j|)
    \right\rangle_{\Psi},
    \label{eq:main_variational_energy_functional}
\end{align}
where $\mathcal E(\bs q)$ is a single-particle dispersion. In this section, we first benchmark the energetics using direct Monte Carlo simulations and then derive approximate semi-analytical formulas that will be used for the R5G parameter scans.

\subsection{Monte Carlo evaluation}
\label{subsec:mc_energetics}

We directly sample the AHC wavefunctions~\eqref{eq:ahc_ansatz_main_text} with MC to evaluate both kinetic and interaction energies~\cite{metropolis1953equation, morf1986monte, levesque1984crystallization, morf1986microscopic, wang2019lattice, ceperley1986quantum, becca2017quantum}. The Monte Carlo calculation is performed in disk geometry, with the layer indices analytically traced over, using up to $N_e=500$ electrons. Details of the sampling procedure and additional benchmarks are given in the Supplemental Material~\cite{Note2}. Figure~\ref{fig:mc_evaluation_energy} shows representative results for triangular crystals and Laughlin fillings.

The kinetic-energy benchmark in Fig.~\ref{fig:mc_evaluation_energy}(a,b) uses $\mathcal E(\bs q)=|\bs q|^2$. The dominant trend is an approximate linear growth of $\langle |\bs{q}|^2\rangle$ with $1/\nu$. This has a simple geometric origin. At fixed electron density, the crystal unit-cell area is $A_{\rm u.c.}=\nu/n_e$. Lowering $\nu$ decreases the real-space unit cell and increases the reciprocal-lattice scale by a factor of order $1/\sqrt{\nu}$. The unfolded parent-band momentum distribution then samples momenta on the order of $\sqrt{n_e/\nu}$. The kinetic energy also exhibits a pronounced dependence on $\gamma$. For all fillings, $\langle |\bs{q}|^2\rangle$ initially grows approximately as $\gamma^2$ before crossing over to a broad plateau at larger $\gamma\sqrt{n_e}$.

The interaction-energy benchmark is shown in Fig.~\ref{fig:mc_evaluation_energy}(c,d) for a dual-gated screened interaction $V(\bs{q}) = 2 \pi V_0 \tanh(|\bs{q}|d_s)/|\bs{q}|$ with $V_0 = e^2 / \epsilon$. The interaction energy depends only weakly on $\gamma$. This near-independence is already apparent in Fig.~\ref{fig:mc_evaluation_energy}(c), while the magnified view in Fig.~\ref{fig:mc_evaluation_energy}(d) reveals a small but systematic residual variation. The main qualitative behavior is a decrease in the interaction energy as the filling $\nu$ decreases. As seen from the Jastrow factor in \eqref{eq:laughlin_state_wf}, this follows from the fact that smaller $\nu$ Laughlin states suppress short-distance pair probability more strongly and consequently are more efficient at keeping electrons apart. 

\begin{figure}
    \centering
    \includegraphics[width=1.00\linewidth]{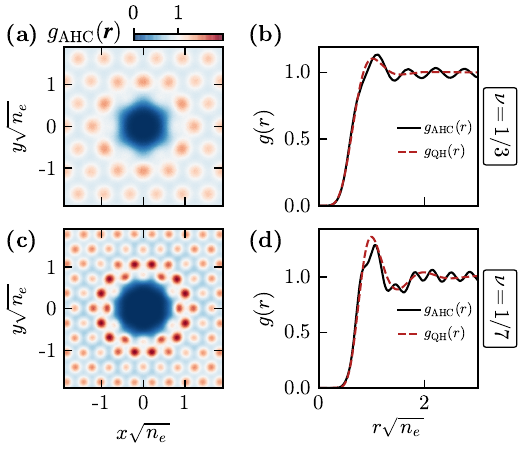}
    \caption{
    {\bf Pair correlations of the AHC variational state for $\boldsymbol{N_\ell=2}$ and $\bs{\gamma\sqrt{n_e}=0.5}$, evaluated by Monte Carlo.} Panels (a,c) show the two-dimensional pair-correlation function $g_{\mathrm{AHC}}(\bs r)$ for $\nu=1/3$ and $\nu=1/7$, respectively, with distances measured in units of $1/\sqrt{n_e}$. Panels (b,d) show the corresponding radial averages $g_{\mathrm{AHC}}(r)$, compared with the Laughlin quantum Hall result $g_{\mathrm{QH}}(r)$ at the same filling. 
    }
    \label{fig:pair_correlation_energy}
\end{figure}

A striking feature of the MC data is that the interaction energy is approximately the same as the bare Laughlin state (see dashed line in Fig.~\ref{fig:mc_evaluation_energy}(c,d)). The real-space pair correlation function~\cite{girvin1984anomalous, girvin1986magneto, tan2024wavefunction} 
\begin{align}
    g^{\rm AHC}(\bs r)
=
\frac{1}{n_e N_e}
\left\langle
\sum_{i\neq j}
\delta^{(2)}\!\left(\bs r-\bs r_i+\bs r_j\right)
\right\rangle_{\rm AHC}.
\label{eq:real_space_pair_corr_main}
\end{align}
in Fig.~\ref{fig:pair_correlation_energy} illustrate this point. The two-dimensional $g_{\rm AHC}(\bs r)$ in Fig.~\ref{fig:pair_correlation_energy}(a) and (c) has Bragg-like oscillations associated with broken translation symmetry. However, its radial short-distance structure in Fig.~\ref{fig:pair_correlation_energy}(b) and (d) closely follows the corresponding Laughlin $g_{\rm QH}(r)$. Since the screened repulsion is most sensitive to this short-distance structure, the AHC interaction energy is naturally close to the quantum Hall interaction energy, with only small corrections from lattice harmonics.

\subsection{Analytics}
\label{subsec:analytic_energetics}

The Monte Carlo results suggest a simple correspondence between the interaction energy of the AHCs and their corresponding QH states. They also indicate that the kinetic energy is controlled primarily by how the AHC single-particle basis maps the emergent crystal Brillouin zone onto momentum occupation of the original parent band. In this section, we provide a formal justification for these observations and derive simple approximations to the energies of our IAHC and FAHC ansätze.

\subsubsection{AHC--QH correspondence}

From Eqs.~\eqref{eq:expansion_qh_first_quantized} and~\eqref{eq:expansion_ahc_first_quantized}, we see that the QH and AHC wavefunctions are closely related. In fact, if we neglect the momentum dependence of the normalization factor (i.e., $\mathfrak{N}_{\bs{k}} \approx \mathfrak{N}$), the second-quantized wavefunctions have the same coefficients $C_{\{\bs{k}\}}$ but different single-particle bases. Under such an approximation, there exists an isometry between the AHC and LLL Fock space,
\begin{align}
    c_{\bs k}^{\rm LLL\dagger}
    \longleftrightarrow
    \psi_{\bs k}^{\dagger},
    \label{eq:main_fock_space_isometry}
\end{align}
where $\psi_{\bs k}^\dagger$ is a creation operator for the single-particle state $\ket{\psi_{\bs{k}}}$ in~\eqref{eq:ahc_single_particle_states}. This equivalence allows us to relate AHC expectation values to QH ones, such as
\begin{equation}
    \left\langle
    c_{\bs k_1}^{\rm LLL\dagger}c_{\bs k_2}^{\rm LLL}
    \right\rangle_{\rm QH}
    =
    \left\langle
    \psi_{\bs k_1}^\dagger\psi_{\bs k_2}
    \right\rangle_{\rm AHC}.
    \label{eq:ahc_qh_mapping_one_body_corrected}
\end{equation}
The expectation value on the left is evaluated with respect to~\eqref{eq:expansion_qh_first_quantized} and the one on the right with~\eqref{eq:expansion_ahc_first_quantized}. Under such assumptions, we can thus relate AHC observables to known ones for QH states (e.g., they may either be restricted by continuous magnetic translations or have been previously tabulated in the QH literature). We provide more details on the justification for this AHC-QH correspondence in the Supplemental Material~\cite{Note2}, where we present explicit bounds on the momentum dependence of $\mathfrak{N}_{\bs{k}}$. In any case, the validity of the AHC--QH correspondence will also be justified \textit{a posteriori} through comparison with our unbiased MC results.

We emphasize that, althought we will always assume the AHC--QH correspondence to evaluate matrix elements in what follows, we keep the underlying lattice dependence of all other quantities. For example, in our dual scheme, the average density is approximated by
\begin{align}
    \expval{\hat{\tilde{\rho}}^{\rm AHC}_{\bs{q}}} &= \sum_{\bs{k}\in \mathrm{BZ}} \mathcal{F}^{\rm AHC}(\bs{k}+\bs{q}, \bs{k})
    \expval{\psi^\dagger_{\lceil \bs{k} + \bs{q} \rceil} \psi_{\bs{k}}}_{\rm AHC} \nonumber \\
    &\approx \sum_{\bs{k}\in \mathrm{BZ}} \mathcal{F}^{\rm AHC}(\bs{k}+\bs{q}, \bs{k})
    \left\langle
    c_{\lceil\bs{k} + \bs{q} \rceil}^{\rm LLL\dagger}c_{\bs k}^{\rm LLL}
    \right\rangle_{\rm QH},
\end{align}
where the full lattice dependence of the form factor~\eqref{eq:main_ahc_form_factor_factorization} is preserved.

\subsubsection{Kinetic energy}

It is now straightforward to show that the kinetic energy is given by
\begin{align}
    \frac{E_{\rm kin}}{N_e}
    &= \frac{1}{N_{e}} \sum_{\bs{k} } \matrixel{\psi_{\bs{k}}}{ \mathcal{E}(-i \nabla) }{\psi_{\bs{k}}} \expval{ \psi^\dagger_{\bs{k}} \psi_{\bs{k}}} \nonumber \\
    &\approx \frac{1}{A_{\rm u.c.} N_{\mathrm{u.c.}}} \sum_{\bs{k}} \frac{ \sum_{\bs g}\mathcal P_{\bs 0}(\bs k+\bs g)\, \mathcal E(\bs k+\bs g)}{\mathcal R_{\bs 0}(\bs k)}, \label{eq:kinetic_energy_ahc}
\end{align}
where we have used the AHC--QH correspondence~\eqref{eq:ahc_qh_mapping_one_body_corrected} in the second line. The relevant QH expectation value is then evaluated using the fact that continuous magnetic translation symmetry makes it always possible to pick a representative ground state with a uniform momentum space occupation (i.e., $\langle c_{\bs k}^{\rm LLL\dagger}c_{\bs k}^{\rm LLL} \rangle_{\rm QH} = \nu$)~\cite{Note2}. The analytical formula~\eqref{eq:kinetic_energy_ahc} is directly compared to MC results in Fig.~\ref{fig:mc_evaluation_energy}(a)-(b) and shows excellent agreement. 

To gain further analytical insight, one can take the zeroth-order approximation~\eqref{eq:main_zeroth_harmonic_W} of the denominator. In this case, the kinetic energy simplifies to 
\begin{align}
    \expval{|\bs{q}|^{2\alpha}}
    \approx
    \left(\frac{4 \pi n_e}{\nu}\right)^{\alpha}
    \frac{
    \sum_{\ell=0}^{N_\ell-1}
    \left(\frac{4 \pi n_e \gamma^2}{\nu}\right)^{\ell}
    (\ell+\alpha)!}
    {
    \sum_{\ell=0}^{N_\ell-1}
    \left(\frac{4 \pi n_e \gamma^2}{\nu}\right)^{\ell}
    \ell!} \label{eq:kinetic_energy_zeroth_order_approx}.
\end{align}
This approximate form makes the various trends observed in the MC data of Fig.~\ref{fig:mc_evaluation_energy} explicit. We directly see that the dominant scaling with filling is $\expval{|\bs{q}|^{2\alpha}}\sim (n_e/\nu)^\alpha$. For the $\gamma$ dependence, in the small $\gamma\sqrt{n_e}$ regime, we can expand both sums at leading order in $\gamma$, which leads to a dominant quadratic behavior. In contrast, for large $\gamma\sqrt{n_e}$, we can keep only the highest power, $\gamma^{2(N_\ell - 1)}$, in the numerator and denominator, which cancels out, yielding an approximately $\gamma$-independent kinetic energy. We give additional details in the Supplemental Material~\cite{Note2}, where we show that even the zeroth-order approximation~\eqref{eq:kinetic_energy_zeroth_order_approx} agrees with the MC data for $\expval{|\bs{q}|^2}$ and $\expval{|\bs{q}|^4}$ to within 6\% across all parameters studied ($N_\ell=2,3,4$, $\nu=1/3$, 1/5, 1/7, 1/9, and $\gamma\sqrt{n_e}\in[0, 0.5]$).

\subsubsection{Interaction energy}

The interaction energy can be expressed in terms of the AHC pair correlation in momentum space~\cite{girvin1984anomalous, tan2024wavefunction},
\begin{align}
    \frac{E_{\rm int}}{N_e}
    =
    \frac{n_e}{2}
    \int\frac{d^2q}{(2\pi)^2}
    V(\bs q)g^{\rm AHC}(\bs q).
    \label{eq:main_interaction_pair_correlation}
\end{align}
The AHC--QH correspondence then gives a compact approximate expression for $g^{\rm AHC}$. The density correlations separate into a coherent Bragg contribution from the crystalline density modulation and a connected contribution inherited from the QH liquid,
\begin{equation}
\begin{aligned}
    g^{\rm AHC}(\bs q)
    \approx&
    \sum_{\bs g}
    (2\pi)^2\delta^{(2)}(\bs q-\bs g)
    e^{-l_B^2|\bs g|^2/2}
    |\rho(\bs g)|^2 \\
    &+
    \sum_{\bs R}
    |a_{\bs R}(\bs q)|^2
    g^{\rm QH}_{\rm conn}(\bs q + \bs G_{\bs R}) .
    \label{eq:main_gahc_pair_correlation}
\end{aligned}
\end{equation}
Here $\bs g$ runs over reciprocal lattice vectors, $\bs R$ over direct lattice vectors, and $\bs G_{\bs R}$ is the reciprocal vector associated with $\bs R$ defined by $l_B^2\bs G_{\bs R}\wedge\bs k=\bs k\cdot\bs R$. The connected QH pair correlation is $g^{\rm QH}_{\mathrm{conn}}(\bs q)=\int d^2 r \exp(i \bs{q} \cdot \bs{r})[g_{\rm QH}(\bs{r})-1]$. Pair correlations for quantum Hall states have been extensively documented in the literature~\cite{girvin1984anomalous, girvin1986magneto, prange1990quantum, tan2024wavefunction} and can thus be easily evaluated. In the rest of this work, we use the expansion and results from Ref.~\cite{fulsebakke2023parametrization} to evaluate $g_{\rm QH}(\bs{r})$. The coefficients $\rho(\bs G)$ and $a_{\bs R}(\bs Q)$ are completely determined by the lattice harmonics of $\mathcal R_{\bs q}(\bs k)/\sqrt{\mathcal R_{\bs 0}(\bs k)\mathcal R_{\bs 0}(\lceil\bs k+\bs q\rceil)}$, and are given explicitly in the Supplemental Material~\cite{Note2}. 

Figure~\ref{fig:mc_evaluation_energy}(c) shows excellent agreement between the MC results and analytical approximation~\eqref{eq:main_interaction_pair_correlation}. In particular, Figure~\ref{fig:mc_evaluation_energy}(d) further shows that the approximation reproduces the weak $\gamma$-dependence observed in the MC data. The additional analysis in the Supplementary Material~\cite{Note2} shows agreement within 0.4\% between the MC results and our semi-analytical interaction energy estimator for $N_\ell=2,3,4$, $\nu=1/3$, 1/5, 1/7, 1/9, and $\gamma\sqrt{n_e}\in[0, 0.5]$. This overall agreement between the MC data and our analytical approximation gives further evidence for the validity of the AHC--QH correspondence.

In the zeroth-harmonic approximation, all nontrivial lattice corrections are dropped, and Eq.~\eqref{eq:main_gahc_pair_correlation} reduces to
\begin{align}
    g^{\rm AHC}(\bs q)\approx g^{\rm QH}(\bs q),
    \quad
    \frac{E_{\rm int}}{N_e}
    \approx
    \frac{n_e}{2}
    \int\frac{d^2q}{(2\pi)^2}
    V(\bs q)g^{\rm QH}(\bs q),
    \label{eq:main_interaction_qh_approx}
\end{align}
which reproduces the result of Ref.~\cite{tan2024wavefunction}. This explains why the QH approximation in Fig.~\ref{fig:mc_evaluation_energy}(c,d) already captures the dominant interaction energy, but misses any $\gamma$-dependence.

\begin{figure*}[t]
    \centering
    \includegraphics[width=1.0\textwidth]{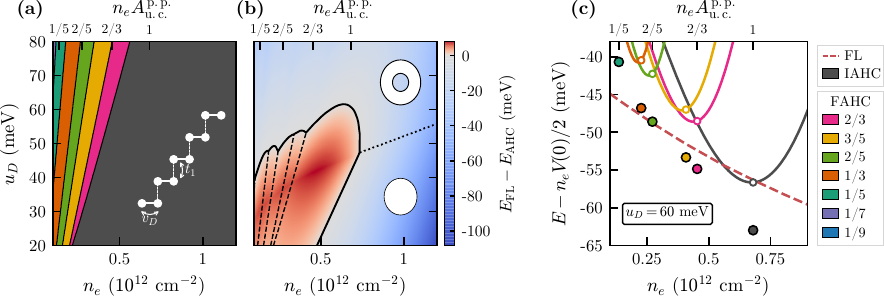}\vspace{-3mm}
    \caption{
    {\bf Ground state phase diagram of candidate anomalous Hall crystal states and their competition with Fermi liquids.} Calculations use the band geometry of the ideal rhombohedral pentalayer graphene model with $\gamma=1.73~\mathrm{nm}$, the dispersion of the physical model with intralayer Dirac velocity $v_D= 0.66~{\rm eV\,nm}$ and interlayer hopping $t_1=0.38~\mathrm{eV}$ (see inset of panel (a)), and a dual-gated screened interaction with $\epsilon=4$ and gate distance $d_s=30~\mathrm{nm}$. (a) Lowest-energy AHC candidate as a function of electron density $n_e$ and interlayer potential $u_D$, restricted to the variational set $\nu=1,2/3,3/5,2/5,1/3,1/5,1/7,1/9$. (b) Energy difference between the optimized spin- and valley-polarized Hartree-Fock Fermi liquid and the lowest AHC candidate, $E_{\rm FL}-E_{\rm AHC}$. The solid black line marks the FL--AHC crossing, dashed black lines mark transitions between distinct AHCs, and the dotted black line marks the disk-to-annulus transition of the optimized Fermi liquid. The upper axis is $n_e A^{\rm p.p.}_{\rm u.c.}$, where $1/A^{\rm p.p.}_{\rm u.c.}=0.68\times10^{12}~\mathrm{cm}^{-2}$ is the density scale of an illustrative triangular periodic potential. (c) Fixed-$u_D$ cut at $u_D=60~\mathrm{meV}$, with the common Hartree contribution $n_e V(0)/2$ subtracted. Colored circular markers indicate the energy of a candidate AHC at the density at which it is commensurate with a first-harmonic periodic potential of strength $|U_0|=15~\mathrm{meV}$. White markers are the energy of the state in the absence of a periodic potential.
    }
    \label{fig:phase_diagram_r5g_ideal_numerics}
\end{figure*}

\subsubsection{Periodic potential}

Finally, we record how an external periodic potential couples to the AHC states. We keep only the smallest reciprocal lattice vectors and consider a real triangular potential
\begin{align}
    U(\bs r)
    =
    \sum_{j=1}^{3}
    \left[
    U_{\bs G_j}e^{i\bs G_j\cdot\bs r}
    +
    U_{\bs G_j}^{*}e^{-i\bs G_j\cdot\bs r}
    \right],
    U_{\bs G_j}=|U_0|e^{i\alpha_j},
\end{align}
where $\{\pm\bs G_j\}$ are the six smallest nonzero reciprocal lattice vectors. The phases $\alpha_j$ specify the relative registry between the imposed potential and the AHC. In the estimates below, we choose this registry to maximize the first-order energy lowering, so only the amplitude $|U_0|$ is kept as an input parameter. The periodic potential is taken to act uniformly on all layers. The minimized first-order energy shift per particle is then
\begin{align}
    \frac{E_{\rm p.p.}^{\rm min}}{N_e}
    =
    -\frac{2|U_0|}{N_e}
    \sum_{j=1}^{3}
    \left|
    \expval{\hat{\tilde\rho}^{\rm AHC}_{\bs G_j}}
    \right|,
\end{align}
with $\hat{\tilde\rho}^{\rm AHC}_{\bs q}$ the density projected to the AHC mini-band.
The first-order energy shift is nonzero only when the AHC density has a Bragg harmonic at the external potential's wavevector. Thus, the state must break continuous translations, and its reciprocal lattice must be commensurate with that of the imposed potential. Using the AHC--QH correspondence, the commensurate density harmonics are
\begin{align}
    \left|
    \expval{\hat{\tilde\rho}^{\rm AHC}_{\bs G_i}}
    \right|
    &=
    \left|
    \sum_{\bs k\in{\rm BZ}}
    \mathcal F^{\rm AHC}(\bs k+\bs G_i,\bs k)
    \expval{\psi_{\bs k}^{\dagger}\psi_{\bs k}}
    \right|
    \nonumber\\
    &\approx
    \nu e^{-l_B^2|\bs G_i|^2/4}
    \left|
    \sum_{\bs k\in{\rm BZ}}
    e^{i l_B^2\bs G_i\wedge\bs k}
    \frac{\mathcal R_{\bs G_i}(\bs k)}
         {\mathcal R_{\bs 0}(\bs k)}
    \right|.
\end{align}

\section{Application to rhombohedral pentalayer graphene}
\label{sec:r5g_phase_diagram}

We now apply the semi-analytic energy functional developed above to R5G. The calculation uses a hybrid scheme: the wavefunctions and form factors are those of the ideal holomorphic parent band with $N_\ell=5$ and $\gamma=v_D/t_1=1.73~{\rm nm}$, while the kinetic energy is evaluated using the physical R5G conduction-band dispersion $\mathcal E_{\rm R5G}(\bs q,u_D)$ of the minimal continuum model. The displacement field enters only through the dispersion, not the band geometry. For interactions, we use a dual-gated screened potential with $\epsilon=4$ and gate distance $d_s=30~{\rm nm}$. We compare AHC variational states with $\nu\in\{1,2/3,3/5,2/5,1/3,1/5,1/7,1/9\}$ against an optimized spin- and valley-polarized Hartree-Fock Fermi liquid, allowing both disk-like and annular Fermi surfaces as described in the Supplemental Material~\cite{Note2}.

\subsection{Phase diagram}

For each density and displacement field, we first minimize only within the AHC variational family. The resulting AHC-only phase diagram is shown in Fig.~\ref{fig:phase_diagram_r5g_ideal_numerics}(a). The preferred filling increases with density. Small-$\nu$ FAHCs are favored at low density, while larger fractional fillings and eventually the IAHC are favored at higher density. Figure~\ref{fig:phase_diagram_r5g_ideal_numerics}(b) then compares the best AHC state with the optimized FL by plotting their energy difference. In the low-density, low-displacement-field region, the AHC variational states can lie below the FL. We do not consider this region particularly relevant to an experimentally observable FAHC. At such low densities, topologically trivial Wigner crystals, which are not included in our variational comparison, are likely strong competitors. In this low-density region where an insulator is seen in experiment (see Fig.~\ref{fig:experimental_phase_diagram_r5g}(a) and (b)), disorder may also dominate the insulating response~\cite{babbar2026wigner}. Rather, the more relevant observation for our purposes is that outside this extremely low-density regime, the AHCs often remain only slightly above the FL, by an energy scale that a commensurate periodic potential can plausibly overcome.

To illustrate this effect, the upper axis in Fig.~\ref{fig:phase_diagram_r5g_ideal_numerics}(b,c) uses the dimensionless density $n_e A^{\rm p.p.}_{\rm u.c.}$ of a triangular periodic potential with an inverse unit cell area of $1/A^{\rm p.p.}_{\rm u.c.}=0.68\times10^{12}~\mathrm{cm}^{-2}$. This would correspond to the moiré unit cell area of a graphene/hBN heterostructure with a small twist angle of approximately $0.47^{\circ}$. The first-order pinning energy derived in Sec.~\ref{subsec:analytic_energetics} is then nonzero only at the corresponding commensurate densities such as $\nu = n_e A_{\rm{u.c.}}^{\rm p.p.}$. The fixed-$u_D$ cut in Fig.~\ref{fig:phase_diagram_r5g_ideal_numerics}(c) shows this explicitly. Before adding the periodic potential, several FAHCs are competitors but higher in energy than the FL. When the periodic potential is included at commensurate densities, those particular crystals are selectively lowered in energy. For the illustrative first-harmonic strength $|U_0|=15~\mathrm{meV}$ used in the figure, this lowering can make the commensurate FAHC the lowest state among the candidates shown. We emphasize that this is not intended as a microscopic model of the hBN potential. Rather, it demonstrates the mechanism by which a moir\'e perturbation can pin and select a continuum FAHC that is already nearly competitive.

\subsection{Momentum-space locking of the crystal Brillouin zone}
\label{subsec:r5g_bz_locking}

\begin{figure}
    \centering
    \includegraphics[width=1.0\linewidth]{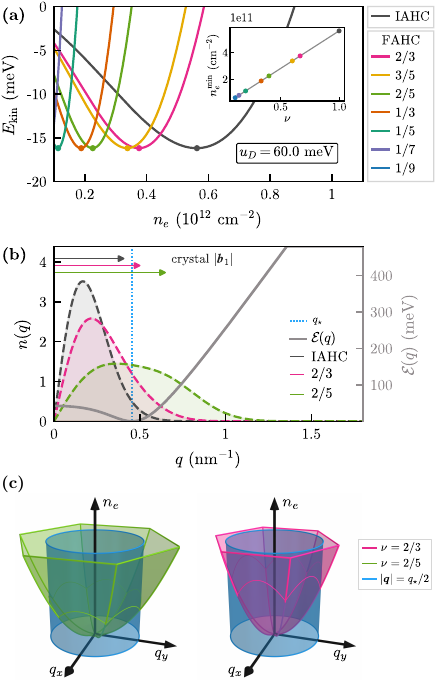}
    \caption{
    {\bf Kinetic-energy trends and momentum-space-locking mechanism for AHCs in R5G.} (a) Kinetic energy of each candidate AHC crystal (at $\nu=1,\,2/3,\,3/5,\,2/5,\,1/3,\,1/5,\,1/7,\,1/9.$) as a function of density at fixed interlayer potential $u_D=60~\mathrm{meV}$, computed using the physical R5G dispersion with $v_D= 0.66~{\rm eV\,nm}$ and $t_1=0.38~\mathrm{eV}$. Each curve has a minimum at some density $n_e^{\rm min}$; the inset shows that this minimizing density scales linearly with unit-cell filling $\nu$. (b) Radial parent-band momentum distribution $n(q) \equiv q\int_0^{2\pi} d\theta\, n(q, \theta)$ for selected AHC states at $n_e=0.38\times10^{12}~\mathrm{cm}^{-2}$, overlaid with the conduction-band dispersion $\mathcal E(q)$ on the right axis. Horizontal arrows mark the reciprocal-lattice scale $|\bs b_1|$ for each crystal, and the vertical dotted line marks the minimum of $\mathcal E(q)$, i.e., $q_\star$. States whose $|\bm b_1|$ lands near $q_\star$ have their occupied momenta take advantage of the band minimum, thereby lowering their kinetic energy; this is the numerical signature of the momentum-locking mechanism. For the fillings shown, the $\nu=2/3$ FAHC has the lowest kinetic energy at this density. (c) Schematic of the momentum-space-locking mechanism. For a triangular AHC, the reciprocal-lattice scale satisfies $|\bs b_1|=(8\pi^2 n_e/\sqrt{3}\nu)^{1/2}$, so fixed-$\nu$ crystals trace cones in the $(k_x,k_y,n_e)$ representation whose radius grows as $\sqrt{n_e/\nu}$. The blue cylinder marks the momentum scale selected by the Mexican-hat minimum of the R5G dispersion, shown here as the zone-edge scale $|\bs{q}|\simeq q_\star/2$. The kinetic energy is minimized when the AHC 1BZ scale locks to this radius. Lower filling AHCs therefore reach the locking condition at lower density, thereby producing the linear trend $n_e^{\rm min}\propto\nu$.
    }
    \label{fig:kinetic_energy_analysis}
\end{figure}

The fixed-$u_D$ cut in Fig.~\ref{fig:phase_diagram_r5g_ideal_numerics}(c) further reveals a simple regularity. Each AHC candidate has a preferred density at which its energy is minimized, and these preferred densities scale approximately linearly with the unit-cell filling $\nu$. This pattern arises from the kinetic energy. As shown in Fig.~\ref{fig:kinetic_energy_analysis}(a), each crystal has a preferred density $n_e^{\rm min}$ where its kinetic energy is minimized. This optimal density $n_e^{\rm min}$ scales approximately linearly with $\nu$. Figure~\ref{fig:kinetic_energy_analysis}(b) explains the origin of this behavior. The AHC folds the continuum band into the Brillouin zone of the self-generated crystal, producing an unfolded parent-band occupation $n(\bs{q})$. Reducing $\nu$ at fixed density shrinks the real-space unit cell and shifts $n(\bs{q})$ to larger momenta. The optimal crystal is therefore the one whose occupation best matches the finite-momentum ring where $\mathcal E_{\rm R5G}(\bs{q};u_D)$ is smallest.

A crude estimate is obtained by equating the shortest reciprocal-lattice vector to the radius of the Mexican-hat minimum. This yields the heuristic estimate
\begin{align}
    |\bs b_1|\sim q_{\star}(u_D)
    \quad\Longrightarrow\quad
    n_e^{\rm min}(\nu,u_D)
    \sim
    \frac{\sqrt{3}}{8\pi^2}\,\nu\,q_{\star}^2(u_D),
    \label{eq:r5g_min_density_scaling}
\end{align}
where $q_{\star}(u_D)$ denotes the momentum scale of the dispersion minimum. This ``momentum-space-locking'' mechanism reverses the usual intuition from monotonic dispersions, where the larger reciprocal-lattice scale of a fractional crystal would be energetically prohibitive~\cite{tan2024wavefunction}. The minimum in the dispersion of R5G provides a natural length scale that favors a specific range of periodicity and can make electronic crystals with fewer than one electron per unit cell energetically competitive for realistic interaction strengths. This mechanism is illustrated schematically in Fig.~\ref{fig:kinetic_energy_analysis}(c).

\subsection{Twist angle stability}
\label{subsec:twist_angle_stability}

\begin{figure}
    \centering
    \includegraphics[width=1.00\linewidth]{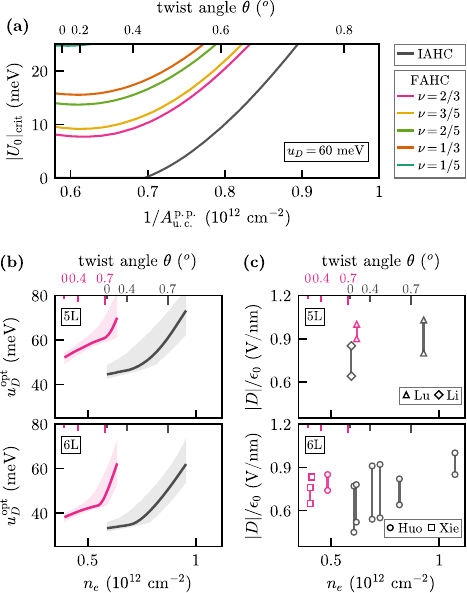}
    \caption{{\bf Twist angle dependence.} (a) Critical first-harmonic periodic-potential strength $|U_0|_{\rm crit}$ required for each commensurate candidate crystal to become lower in energy than the optimized Fermi liquid, plotted as a function of the moir\'e density $1/A_{\rm u.c.}^{\rm p.p.}$ at fixed internal interlayer potential $u_D=60~{\rm meV}$. The upper axis gives the corresponding graphene/hBN twist angle $\theta$. (b) Optimal internal interlayer potential $u_D^{\rm opt}$ for the IAHC and $\nu=2/3$ FAHC in five- and six-layer rhombohedral graphene for twist angles between $0^\circ$ and $0.8^\circ$. At each twist angle, the candidate is evaluated at its commensurate density $n_e=\nu/A_{\rm u.c.}^{\rm p.p.}(\theta)$, and $u_D^{\rm opt}$ is defined as the value of $u_D$ that minimizes the continuum energy difference $E_{\rm AHC}-E_{\rm FL}$. Shaded regions indicate values of $u_D$ for which this energy difference stays within $1~{\rm meV}$ of its optimum. (c) Experimental displacement-field windows where IQAHE and $\nu=2/3$ FQAHE signatures are reported in Lu et al.~\cite{lu2024fractional}, Li et al.~\cite{li2025stacking}, Huo et al.~\cite{huo2025does}, and Xi et al.~\cite{xie2025tunable}. Vertical bars mark the reported displacement-field ranges; marker shapes label the experimental references.
}\label{fig:twist_angle_stability}
\end{figure}

The same scaling explains why the fractional sequence selected by the continuum kinetic energy naturally aligns with commensuration of an underlying moir\'e-scale periodic potential. For graphene/hBN, twist angles between $0^\circ$ and $1^\circ$ give reciprocal-lattice scales of order $0.50$--$0.72~{\rm nm}^{-1}$. This is naturally on the same scale as the R5G dispersion minimum, which lies at a momentum on the order of $1/\gamma\simeq0.58~{\rm nm}^{-1}$ in the strong-displacement-field regime. Hence, the imposed periodicity has a reciprocal scale comparable to the intrinsic momentum scale of the R5G dispersion. The commensurate densities satisfy $n_e=\nu/A^{\rm p.p.}_{\rm u.c.}$, while the kinetic minima satisfy $n_e^{\rm min}\propto\nu$. The two sequences therefore track one another, and the periodic potential can lower the energy of the crystal in a regime where it is already the most energetically competitive. This produces a coherent physical picture where both effects cooperate.

This also implies that AHCs should become harder to stabilize as the graphene/hBN twist angle becomes too large. Figure~\ref{fig:twist_angle_stability}(a) quantifies the periodic-potential strength needed to stabilize each crystal at the commensurate density $n_e=\nu /A_{\rm u.c.}^{\rm p.p.}$  The rapid growth of $|U_0|_{\rm crit}$ at larger twist angle reflects the increasing mismatch between the imposed reciprocal-lattice scale and the intrinsic momentum scale favored by the continuum dispersion. This is consistent with the fact that reported IQAHE/FQAHE signatures in R5G and R6G occur primarily in devices with twist angles below $1^\circ$~\cite{lu2024fractional,li2025stacking,huo2025does,xie2025tunable}.

The locking criterion also predicts the qualitative behavior of the displacement-field window where IQAHE and FQAHE are stabilized as functions of twist angles. Once $\theta$ fixes the moir\'e reciprocal scale, the optimal internal potential is determined by $q_\star(u_D)\sim|\bs b|$. Since $q_\star(u_D)$ grows with $u_D$ in the relevant regime~\cite{Note2}, smaller twist angles should shift the windows to smaller displacement fields. Figure~\ref{fig:twist_angle_stability}(b) shows this dependence in the variational calculation, and Fig.~\ref{fig:twist_angle_stability}(c) shows the same qualitative trend in experiments on five- and six-layer devices

\section{Discussion}
\label{sec:discussion}

We have developed a continuum variational framework for integer and fractional anomalous Hall crystals in rhombohedral graphene. The starting point is the ideal holomorphic limit, in which a projected repulsive contact interaction has a large zero-mode manifold containing both IAHCs and FAHCs. By evaluating the energies of these wavefunctions after restoring physical dispersion and longer-ranged screened interactions, we obtain a controlled way to ask how the ideal zero-mode degeneracy is lifted. The main result is that the interaction energy is largely inherited from the corresponding quantum Hall liquid, while the kinetic-energy splitting between different anomalous Hall crystals is governed by the unfolded parent-band momentum distribution set by the crystal period.

Through momentum-space locking, this separation explains how FAHCs become competitive. For a monotonic dispersion, the enlarged reciprocal lattice of a fractional crystal is costly in kinetic energy~\cite {tan2024parent}. The finite-momentum dispersion minimum of rhombohedral graphene instead rewards it, leaving the FAHCs near-degenerate with the polarized Fermi liquid over a range of experimentally relevant densities and displacement fields. A commensurate moir\'e potential can then act as a selection field, pinning and lowering the members of the AHC family whose periods match the substrate. This points to a ``moir\'e-enabled'' interpretation of fractional anomalous Hall phenomenology~\cite{dong2024stability}, in which the fractional order is prepared by continuum interactions and merely selected by the moir\'e.

This ``continuum-first'' perspective fits naturally with the growing experimental evidence for nearby electronic-crystal physics in rhombohedral multilayers, including self-doped Wigner-crystal regimes in moir\'eless devices and extended QAH states in graphene/hBN~\cite{han2026evidence, lu2025extended}. It may also help organize the nearly fraction-independent apparent activation scales extracted from the initial R5G/hBN data~\cite{xie2024integer}. In a pinned FAHC, the measured $R_{xx}$ activation scale need not equal the intrinsic quasiparticle gap of the underlying QH liquid. It can instead be limited by pinning, domain walls, dislocations, or nearby competing phases. { This FAHC framework may also provide a simple route to the approximate particle-hole symmetry of the fractional gaps recently observed in R5G/hBN~\cite{butler20261}: when the emergent crystal mini-band is close to the zeroth-harmonic limit, its projected form factors and interaction energy are inherited from the corresponding particle-hole symmetric LLL problem. Quantifying how lattice-harmonic corrections, residual dispersion, and band mixing perturb this emergent particle-hole symmetry is an interesting direction for future work.} A further implication of the moiré-enabled FAHC scenario is that both the IQAHE and FQAHE phases should have similar collective modes associated with pinned excitations of the underlying Abrikosov lattice~\cite{tan2025ideal}. FAHC should include additional Girvin-MacDonald-Platzmann collective modes that are absent from IAHC~\cite{girvin1986magneto}. It will be interesting in future work to study the mixing between those collective modes and crystalline excitations. This suggests several experimental tests of the moir\'e-enabled FAHC scenario, including searches for pinned-crystal collective modes and nonlinear depinning or current-induced melting.

It is worth noting several assumptions underlying the present analysis. Our R5G calculation uses a hybrid scheme in which the trial wavefunctions and form factors are taken from the ideal holomorphic parent band, while the energy is evaluated using a physical, yet simplified, R5G dispersion. This isolates the effect of the nonmonotonic dispersion, but it does not yet include the full physical band geometry, trigonal warping, and remote hoppings. Moreover, the periodic potential used above is phenomenological and should be viewed as a minimal commensuration field rather than as a quantitative model of the graphene/hBN interface. Another important limitation is that the IAHC and FAHC wavefunctions used here are fixed by the contact-interaction zero-mode construction and contain no variational parameters. Allowing the Abrikosov texture or quantum Hall correlations to relax could modify the energetic competition, as suggested by previous work~\cite{tan2024wavefunction}. In particular, as we allow for more general AHC ansätze, we expect a shallower kinetic energy dependence on the electronic density than in Fig.~\ref{fig:kinetic_energy_analysis}(a). Our restricted variational comparison also cannot rule out other nearby phases. Future microscopic calculations and more expressive variational approaches, including neural quantum states~\cite{CarleoTroyer2017NQS, valenti2025quantum, Geier2025SelfAttentionSolids, LuoDaiFu2024MoireNQS, TengDaiFu2025FQHAttention, Qian2025LandauLevelMixingNQS, Nazaryan2025NeedleHaystack, AbouelkomsanGeierFu2026TopologicalOrderNQS, AbouelkomsanFu2026CrystallizationFQH, Zhu2026DisorderAwareFQHCrystallization}, will therefore be important to determine the true ground state phase diagram and to assess the competitiveness of FAHC states. It will also be interesting to extend these variational wavefunctions to include spin. We have specialized here to fully spin-polarized states, but the weak spin-orbit coupling in graphene leaves open the possibility of spin-textured or unpolarized competitors.

The framework developed here should also be useful beyond rhombohedral pentalayer graphene, as it applies to general holomorphic continuum parent bands. Within such settings, the semi-analytic energy estimates we have derived, along with the various features that make FAHCs energetically competitive, should be broadly relevant. In particular, our results suggest that nonmonotonic dispersion and commensurability locking are general mechanisms for stabilizing fractional anomalous Hall crystals. Constructing minimal continuum or material models in which these ingredients make an FAHC the true ground state would provide a striking example of fractional quantum Hall topological order coexisting with spontaneous breaking of continuous translation symmetry at zero external magnetic field.

\begin{acknowledgments}
We thank Trithep Devakul, Zhihuan Dong, Zhaoyu Han and Patrick Ledwith for stimulating and helpful discussions. A.V. and F.D. acknowledge funding from NSF DMR-2220703 and the Simons Collaboration on Ultra-Quantum Matter, which is a grant from the Simons Foundation (651440, A.V.).
\end{acknowledgments}

%

\end{document}


\setcounter{secnumdepth}{3}

\setcounter{section}{0}
\renewcommand{\thesection}{S.\Roman{section}}
\renewcommand{\theequation}{S\arabic{equation}}
\setcounter{table}{0}
\renewcommand{\thetable}{S\arabic{table}}

\setcounter{figure}{0}
\renewcommand{\thefigure}{S\arabic{figure}}
\setcounter{equation}{0}
\renewcommand{\theequation}{S\arabic{equation}}
\setcounter{table}{0}
\renewcommand{\thetable}{S\arabic{table}}
\setcounter{figure}{0}
\renewcommand{\thefigure}{S\arabic{figure}}

\title{
Supplemental Material for: \linebreak 
``Energetics of fractional anomalous Hall crystals in rhombohedral graphene''
}

\author{F\'elix Desrochers \orcidlink{0000-0003-1211-901X}}
\email{felix\_desrochers@fas.harvard.edu}
\author{Ashvin Vishwanath \orcidlink{0000-0002-6306-2263}}
\email{avishwanath@g.harvard.edu}
\affiliation{%
Department of Physics, Harvard University, Cambridge, MA 02138, United States
}%

\date{\today}

\maketitle
\vspace{-1.2cm}
\begingroup
\small
\tableofcontents
\endgroup

\thispagestyle{empty}
\section*{\label{Sec: Conventions} Conventions and notations}

To simplify reading, we summarize here the conventions and main symbols used throughout the Supplemental Material and the Main Text.

\begin{listliketab} 
\storestyleof{itemize} 
    \begin{tabular}{ll}
        $\bs{a}_1,\bs{a}_2$: &Real space basis vectors\\
        $\bs{b}_1,\bs{b}_2$: &Reciprocal space basis vectors\\
        $\bs{R}=n_1 \bs{a}_1 + n_2 \bs{a}_2$: &Real space lattice site position\\
        $\bs{g}=n_1 \bs{b}_1 + n_2 \bs{b}_2$: &Momentum space lattice site position\\
        $\wedge \bs{v} = (v_y,-v_x)$: &2d ``cross-product''\\
        $v = v_x + iv_y$: &Complex representation of a two-dimensional vector $\bs{v}$\\
        $A_{\mathrm{u.c.}}=|\bs{a}_1\wedge\bs{a}_2|$: &Unit cell area\\
        $\Omega_{\mathrm{BZ}} = |\bs{b}_1\wedge\bs{b}_2| = (2\pi)^2/A_{\text{u.c.}}$: &First Brillouin zone area\\
        $l_B=\sqrt{A_{\mathrm{u.c.}}/(2\pi)}$: &Emergent magnetic length\\
        $\gamma = v_D/t_1$: &Interlayer hopping length\\
        $\ell$: &Layer (or internal d.o.f.) index\\
        $N_{\mathrm{u.c.}}$: &Number of unit cells\\
        $N_{\ell}$: &Number of layers \\
        $N_e$: &Number of electrons\\
        $\nu=N_e/N_{\mathrm{u.c.}}$: &Unit cell filling\\
        $n_e=N_e/A = N_e/(N_{\mathrm{u.c.}} A_{\mathrm{u.c.}})$: &Electronic density\\
        $\lceil \bs{q}\rceil$: &Momentum $\bs{q}$ reduced modulo reciprocal lattice vectors back into $\mathrm{BZ}$\\
        $c_{\bs{q}}^\dagger$: &Parent band single-particle creation operator (with $\bs{q}$ unbounded)\\
        $\psi_{\bs{k}}^\dagger$: &Crystal state single-particle creation operator (with $\bs{k}\in\mathrm{BZ}$)
    \end{tabular} 
\end{listliketab}

\vfill \newpage
\section{Validity of the ideal limit of rhombohedral multilayer graphene} \label{si_sec:validity_ideal_limit_rng}

In this Section, we explore in greater detail the justification for the ideal or chiral single-particle approximation used in the main text and assess its validity by comparing selected band properties with those of more realistic models for rhombohedral N-layer graphene (RNG). 

\subsection{General tight-binding model}

We first discuss the general tight-binding model of RNG. To introduce the model, it is convenient to start from the minimal tight-binding model for a single graphene sheet, which consists of $p_z$ orbitals sitting on the two carbon atom sites of the unit cell with nearest-neighbor hopping $t_0$,
\begin{align}
    h_{\mathrm{Graphene}}(\bs{q}) = \mqty(
    0 & -t_0 f_{\bs{q}} \\
    -t_0 \bar{f}_{\bs{q}} & 0
),
\end{align}
where 
\begin{align}
    f_{\bs{q}} &= \sum_{i=0}^2 e^{i \bs{q} \cdot \bs{\delta}_i}
\end{align}
and 
\begin{align}
    \bs{\delta}_0 =  \frac{a_{\mathrm{G}}}{\sqrt{3}} \left(0, 1 \right)^T, \quad 
    \bs{\delta}_1 = C_3 \bs{\delta}_0 , \quad  
    \bs{\delta}_2 = C_3^2 \bs{\delta}_0
\end{align}    
are the nearest-neighbor vectors, where $C_3$ denotes a $120^{\circ}$ rotation and $a_{\mathrm{G}} \approx 2.46 \AA$ is the graphene lattice constant. Such a model has Dirac points at the $\bs{K}$ and $\bs{K}'$ corners of the Brillouin zone, which we refer to as the $\bs{K}$ and $\bs{K}'$ valleys. Writing $\eta=+1$ for $\bs K$ and $\eta=-1$ for $\bs K'$, and measuring $\bs q$ from the corresponding valley center, the effective monolayer continuum Hamiltonian can be written as
\begin{align}
    h_\eta(\bs q)
    =
    v_D\left(\eta q_x\sigma_x+q_y\sigma_y\right),
\end{align}
where
\begin{align}
    v_D\equiv \hbar v_F
    =
    \frac{\sqrt{3}}{2}t_0a_G
    \approx 0.66~{\rm eV\,nm}.
\end{align}
Here $v_F\approx 10^6~{\rm m/s}$ is the graphene Dirac velocity. In the rest of this Section, we work in the $\bs{K}'$ valley, for which
\begin{align}
    h_{\bs{K}'}(\bs{q})=
    \mqty(
        0 & -v_D q \\
        -v_D \bar  q & 0
    ),
\end{align}
and the $\bs{K}$-valley model is obtained by time reversal.

RNG consists of regularly stacked graphene sheets shifted by $\bs{\delta}_1$. The effective model is obtained by coupling multiple copies of the effective model for a single graphene sheet described above. It thus has sublattice $\alpha,\beta\in\{A,B\}$, layer $\ell,\ell'\in\{0,1,\ldots,N_{\ell}-1\}$, spin $\sigma\in\{\uparrow, \downarrow\}$, and valley $\eta\in\{\bs{K},\bs{K}'\}$ degrees of freedom. Expanding around the $\bs{K}'$ valley, the effective model for a single spin species in the $(\alpha \ell)\in\{ (A,0), (B,0), (A,1), (B,1), \ldots, (A,N_{\ell}-1), (B,N_{\ell}-1) \}$ basis (note that we are here using a different basis convention than in the main text) is
\begin{align}
    h_{\mathrm{RG}}^{(N_{\ell})}(\bs{q}) &= \mqty(
    h^{(0)}_0 & h^{(1)} & h^{(2)} &  &  &  \\
    h^{(1)\dagger} & h^{(0)}_1 & h^{(1)} & h^{(2)} &  &  \\
    h^{(2)\dagger} & h^{(1)\dagger} & h^{(0)}_2 & \ddots & \ddots &  \\
    & h^{(2)\dagger} & \ddots & \ddots & h^{(1)} & h^{(2)}  \\
    & & \ddots & h^{(1)\dagger} & h^{(0)}_{N_{\ell}-2} & h^{(1)}  \\
    &   & & h^{(2)\dagger} & h^{(1)\dagger} & h^{(0)}_{N_{\ell}-1} \\
    ),
\end{align}
with momentum measured from the valley center. The intralayer term splits into a kinetic part $h^{(0)}(\bs{q})$, an inversion-symmetric potential $h^{\text{ISP}}_{\ell}$, and a displacement-field potential $h^{\text{D}}_{j}$,
\begin{align}
    h_{\ell}^{(0)}(\bs{q}) =  h^{(0)}(\bs{q}) + h^{\text{ISP}}_{\ell} + h^{\text{D}}_{\ell}. 
\end{align}
The layer-dependent inversion-symmetric potential is
\begin{align}
    h_{0}^{\text{ISP}} = \mqty(0 & 0 \\ 0 & u_{\mathrm{ISP},1}), \quad
    h_{\ell}^{\text{ISP}} = \mqty(u_{\mathrm{ISP},2} & 0 \\ 0 & u_{\mathrm{ISP},2}) \quad (1\le \ell\le N_{\ell}-2), \quad
    h_{N_{\ell}-1}^{\text{ISP}} = \mqty(u_{\mathrm{ISP},1} & 0 \\ 0 & 0).
\end{align}
The effect of the displacement field is modeled as a constant potential difference $u_D$ between adjacent layers,
\begin{align}
    h^{\rm D}_\ell = u_D\left(\ell-\frac{N_\ell-1}{2}\right)\mathds{1}_{2\times2}.
\end{align}
The intralayer kinetic and interlayer coupling terms are
\begin{align}
    h^{(0)}_{\ell}(\bs{q}) &= \mqty(
        u_{A \ell} & -v_D q \\
        -v_D \bar{q} & u_{B \ell}
    ), \\
    h^{(1)}(\bs{q}) &= \mqty(
        v_4 q &  v_3 \bar{q} \\
        t_1 & v_4 q
    ), \\
    h^{(2)} &= \mqty(
        0 & \frac{t_2}{2} \\
        0 & 0
    ).
\end{align}
Here $u_{A \ell}$ and $u_{B \ell}$ include the diagonal contributions from $h_\ell^{\mathrm{ISP}}$ and $h_\ell^{\mathrm{D}}$. In what follows, unless explicitly stated otherwise, we keep only the dominant couplings $v_D$, $t_{1}\approx 380~\mathrm{meV}$, and $u_D$.

\subsection{Eigenfunctions without displacement field}

In the limit where we keep only $v_D$ and $t_{1}$, the model has an effective $\mathrm{SO}(2)$ symmetry such that the spectrum only depends on $|\bs{q}|$. It further has a chiral symmetry
\begin{align}
    \Sigma h_{\mathrm{RG}}^{(N_{\ell})}(\bs{q}) \Sigma^\dagger = - h_{\mathrm{RG}}^{(N_{\ell})}(\bs{q}),
\end{align}
where $\Sigma=\mathds{1}_{N_\ell\times N_\ell}\otimes\sigma_3$ with $\mathds{1}_{N_\ell\times N_\ell}$ acting on the layer subspace. It implies that every eigenstate $\ket{\psi_{\bs{q}}}$ with energy $E(\bs{q})$ has a partner $\Sigma\ket{\psi_{\bs{q}}}$ with energy $-E(\bs{q})$.

At $\bs{q}=0$, the intralayer hopping vanishes, $v_D q =0$, such that the edge modes $A_0$ and $B_{N_{\ell}-1}$ are completely isolated and form zero-energy states. The remaining orbitals form vertical dimers: $B_\ell$ couples to $A_{\ell+1}$ for $\ell=0,\ldots,N_{\ell}-2$. These dimers form bonding and antibonding orbitals at energies $\pm t_{1}$, while $A_0$ and $B_{N_\ell-1}$ remain as the two isolated edge states. At finite $\bs{q}$, expanding the low-energy characteristic polynomial in powers of $\bs{q}$ gives the approximate dispersion
\begin{align}
    E_{\pm}(\bs{q}) = \pm \frac{v_D^{N_\ell} |\bs{q}|^{N_{\ell}}}{t_{1}^{N_{\ell} - 1}} + \cdots .
\end{align}
A convenient basis for this low-energy two-dimensional subspace is given by the two surface-localized spinors
\begin{align}
    \ket{\varphi_B(\bs q)}
    &=
    N_{\bs q}
    \sum_{\ell=0}^{N_\ell-1}
    (\gamma \bar{q})^\ell
    \ket{B,\ell} + \mathcal{O}\left((\gamma \bar{q})^{N_\ell}\right) ,
    \nonumber\\
    \ket{\varphi_A(\bs q)}
    &=
    N_{\bs q}
    \sum_{\ell=0}^{N_\ell-1}
    (\gamma q)^{N_\ell-1-\ell}
    \ket{A,\ell} + \mathcal{O}\left((\gamma q)^{N_\ell}\right),
\end{align}
where
\begin{align}
    \gamma \equiv \frac{v_D}{t_{1}} \approx 1.73~\mathrm{nm}.
\end{align}
The states $\ket{\varphi_A(\bs{q})}$ and $\ket{\varphi_B(\bs{q})}$ should not be confused with the actual finite-$\bs q$ eigenstates. They are a convenient basis for the low-energy subspace. Indeed, the residual boundary terms left by the recursion generate an effective coupling between the two surfaces. Projecting the minimal Hamiltonian onto the basis $\{\ket{\varphi_A(\bs{q})}, \ket{\varphi_B(\bs{q})} \}$ gives, to leading order in $|\gamma q|$,
\begin{align}
    H_{\rm surf}^{(0)}(\bs q)
    =
    \begin{pmatrix}
        0 & \Delta_{N_\ell}(\bs q) \\
        \Delta_{N_\ell}^{*}(\bs q) & 0
    \end{pmatrix},
    \qquad
    \Delta_{N_\ell}(\bs q)
    =
    -t_1(\gamma q)^{N_\ell}
    +
    \mathcal O\!\left(t_1|\gamma q|^{N_\ell+2}\right).
    \label{eq:surf_eff_zero_displacement}
\end{align}
The overall sign of $\Delta_{N_\ell}$ depends on the relative phase convention chosen for $\ket{\varphi_A(\bs{q})}$ and $\ket{\varphi_B(\bs{q})}$, but its magnitude is
fixed to
\begin{align}
    |\Delta_{N_\ell}(\bs q)|
    \simeq
    \frac{v_D^{N_\ell}|\bs q|^{N_\ell}}{t_1^{N_\ell-1}}
    =
    t_1|\gamma q|^{N_\ell}.
\end{align}
Writing
\begin{align}
    \Delta_{N_\ell}(\bs q)
    =
    |\Delta_{N_\ell}(\bs q)|e^{i\alpha_{\bs q}},
\end{align}
the conduction and valence eigenstates for $u_D=0$ and $\bs q\neq0$ are
\begin{subequations}
\label{eq:zero_displacement_surface_eigenstates}
\begin{align}
    \ket{s_+^{(0)}(\bs q)}
    &=
    \frac{1}{\sqrt 2}
    \left(
        \ket{\varphi_A(\bs q)}
        +
        e^{-i\alpha_{\bs q}}
        \ket{\varphi_B(\bs q)}
    \right),
    \\
    \ket{s_-^{(0)}(\bs q)}
    &=
    \frac{1}{\sqrt 2}
    \left(
        \ket{\varphi_A(\bs q)}
        -
        e^{-i\alpha_{\bs q}}
        \ket{\varphi_B(\bs q)}
    \right).
\end{align}
\end{subequations}
For the convention in Eq.~\eqref{eq:surf_eff_zero_displacement}, and writing $q=|\bs q|e^{i\theta_{\bs q}}$, one has $e^{-i\alpha_{\bs q}}=-e^{-iN_\ell\theta_{\bs q}}$. Thus, for any nonzero $\bs q$, both the conduction and valence eigenstates have equal weight in the two surface sectors. Since the above expansion is valid in the $|\gamma q|\ll1$ regime, $\ket{\varphi_A(\bs{q})}$ and $\ket{\varphi_B(\bs{q})}$ describe modes whose layer weights decay geometrically, as $|\gamma q|^{2\ell}$ or $|\gamma q|^{2(N_\ell-1-\ell)}$, away from the corresponding surface.

In contrast, in the regime $|\gamma \bs q|\gg 1$, the dominant term is the intralayer Dirac Hamiltonian. To see this explicitly, write $q=|\bs q|e^{i\theta_{\bs q}}$. On each layer, the monolayer Dirac Hamiltonian has eigenstates
\begin{align}
    \ket{u_s(\bs q)}
    =
    \frac{1}{\sqrt 2}
    \begin{pmatrix}
        1 \\
        s e^{-i\theta_{\bs q}}
    \end{pmatrix},
    \qquad
    \epsilon_s(\bs q)= - s v_D|\bs q|,
    \qquad s=\pm .
\end{align}
Thus, before including interlayer hopping, each helicity sector $s=\pm$ contains $N_\ell$ nearly degenerate states, one on each layer. Since $v_D|\bs q|\gg t_1$, the conduction and valence helicity sectors are split by an energy scale of $2v_D|\bs q|$, and the interlayer hopping can be treated using degenerate perturbation theory within a fixed helicity sector. The vertical hopping connects $B_\ell$ to $A_{\ell+1}$. Projecting this hopping onto the monolayer eigenstates gives
\begin{align}
    \mel{\ell,s}{H_\perp}{\ell+1,s}
    = t_1
    \braket{u_s}{B}\braket{A}{u_s}
    = \frac{s t_1}{2} e^{i\theta_{\bs q}}.
\end{align}
The factor of $1/2$ comes from the equal $A$- and $B$-sublattice weights of the monolayer Dirac eigenstate. The phase $e^{i\theta_{\bs q}}$ is identical on every interlayer link and can be removed by a layer-dependent gauge transformation. Therefore, in a fixed helicity sector, the projected Hamiltonian is an $N_\ell$-site open chain,
\begin{align}
    H^{(s)}_{\mathrm{eff}}
    =
    -s v_D|\bs q|\,\mathds 1_{N_\ell\times N_{\ell}}
    +
    \frac{s t_1}{2}
    \sum_{\ell=0}^{N_\ell-2}
    \left(
        \ket{\ell}\bra{\ell+1}
        +
        \ket{\ell+1}\bra{\ell}
    \right).
\end{align}
The open-chain eigenmodes are~\cite{economou2006green, noschese2013tridiagonal}
\begin{align}
    \phi_m(\ell)
    =
    \sqrt{\frac{2}{N_\ell+1}}
    \sin(\kappa_m(\ell+1)),
    \qquad
    \kappa_m=\frac{m\pi}{N_\ell+1},
    \qquad
    m=1,\ldots,N_\ell .
\end{align}
Since the adjacency matrix of the open chain has eigenvalues $2\cos\kappa_m$, the large-momentum energies are
\begin{align}
    E_{m,s}(\bs q)
    =
    -s v_D|\bs q|
    +
     s t_1
    \cos\left(\frac{m\pi}{N_\ell+1}\right)
    +
    \mathcal O\left(\frac{t_1^2}{v_D|\bs q|}\right).
\end{align}
The omitted corrections come from virtual mixing between opposite helicity
sectors and are suppressed by $t_1/(v_D|\bs q|)=1/(|\gamma\bs q|)$.

\subsection{Ideal limit approximation}

\begin{figure}
    \centering
    \includegraphics[width=1.00\linewidth]{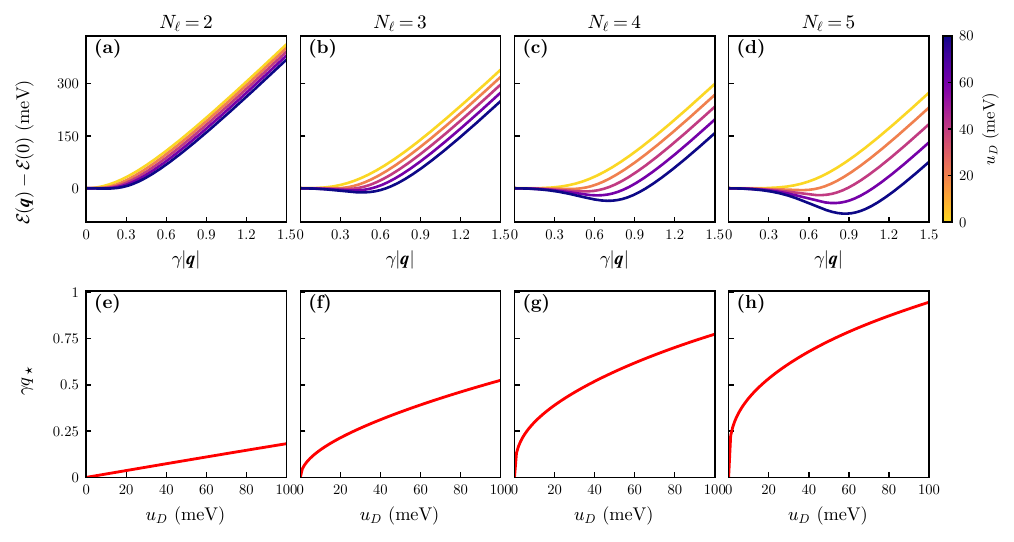}\vspace{-3mm}
    \caption{
    First conduction-band dispersion and band-minimum location of the minimal rhombohedral graphene model for $N_\ell=2,3,4,5$. Top row: first conduction-band dispersion $\mathcal{E}(\bs{q})-\mathcal{E}(\bs{0})$ as a function of normalized momentum $\gamma|\boldsymbol{q}|$. Curves correspond to displacement fields $u_D$ between $0$ and $80\,\mathrm{meV}$, with color indicating $u_D$. Bottom row: normalized momentum $\gamma q_\star$ of the conduction-band minimum, where $q_\star=\mathrm{argmin}_{\bs{q}} \mathcal{E}(\bs{q})$, as a function of $u_D$ between $0$ and $100\,\mathrm{meV}$. The model parameters are $t_0=3.1\,\mathrm{eV}$, $t_1=0.38\,\mathrm{eV}$, and $\gamma=1.73\,\mathrm{nm}$.
    }\label{si_fig:RNG_disp}
\end{figure}

From the above tight-binding model, we now follow arguments laid out in Ref.~\cite{han2025exact} and also used in Ref.~\cite{desrochers2026electronic} to justify the ideal limit considered in the main text and in Ref.~\cite{tan2025ideal}. 

In the presence of a finite displacement field, the degeneracy between surface states at $|\bs q|=0$ is lifted. For definiteness, we take $u_D>0$, so that the $A_0$-surface state is shifted downward and the $B_{N_\ell-1}$-surface state is shifted upward. We denote the resulting surface-state splitting by
\begin{align}
    \Delta_D
    \equiv
    E_B-E_A
    =
    (N_\ell-1)u_D .
    \label{eq:surface_detuning}
\end{align}
We present in Fig.~\ref{si_fig:RNG_disp} a detailed momentum scan of the first conduction band dispersion as a function of displacement field and the number of layers. The band is initially very flat at small momentum, then acquires a linear dispersion at larger momentum when $v_D |\bs{q}|\gg t_1,u_d$. The dispersion further exhibits a local minimum at a momentum of order $\gamma^{-1}$, which becomes more pronounced as the displacement field increases. 

We now discuss in more detail the general form of the eigenstates at small momentum transfer when the dispersion is still relatively flat. Ignoring for the moment the high-order process that couples the two opposite surfaces, the local displacement-dressed surface states have components
\begin{subequations}
\label{eq:local_surface_states_disp}
\begin{align}
    \braket{A,\ell}{\tilde\varphi^{(h)}(\bs q)}
    &\approx
    \tilde{\mathcal{N}}_{\bs q}^{(h)}
    (\gamma\bar q)^\ell,
    &
    \braket{B,\ell}{\tilde\varphi^{(h)}(\bs q)}
    &\approx
    \tilde{\mathcal{N}}_{\bs q}^{(h)}
    (\ell+1)\frac{u_D}{t_1}
    (\gamma\bar q)^{\ell+1},
    \\
    \braket{B,\ell}{\tilde\varphi^{(e)}(\bs q)}
    &\approx
    \tilde{\mathcal{N}}_{\bs q}^{(e)}
    (\gamma q)^{N_\ell-1-\ell},
    &
    \braket{A,\ell}{\tilde\varphi^{(e)}(\bs q)}
    &\approx
    \tilde{\mathcal{N}}_{\bs q}^{(e)}
    (N_\ell-\ell)\frac{u_D}{t_1}
    (\gamma q)^{N_\ell-\ell}.
\end{align}
\end{subequations}
Here $\ket{\tilde\varphi^{(h)}_{\bs q}}$ is localized near the $A_0$
surface and $\ket{\tilde\varphi^{(e)}_{\bs q}}$ is localized near the
$B_{N_\ell-1}$ surface. At finite momentum, however, the two surfaces remain coupled by the same chiral process responsible for the zero-displacement dispersion. In the basis $\{\ket{\tilde\varphi^{(h)}_{\bs q}},\ket{\tilde\varphi^{(e)}_{\bs q}}\}$, the leading surface Hamiltonian is
\begin{align}
    H_{\rm surf}(\bs q)
    =
    \bar E(\bs q)\mathds 1
    +
    \begin{pmatrix}
        -\Delta_D/2 & \Delta_{N_\ell}(\bs q) \\
        \Delta_{N_\ell}^{*}(\bs q) & +\Delta_D/2
    \end{pmatrix}
    +\cdots ,
    \label{eq:surf_eff_disp}
\end{align}
with
\begin{align}
    \Delta_{N_\ell}(\bs q)
    =
    -t_1(\gamma q)^{N_\ell}
    +
    \mathcal O\!\left(
        t_1|\gamma q|^{N_\ell+2},
        u_D|\gamma q|^{N_\ell}
    \right).
    \label{eq:delta_N_def}
\end{align}
Thus, a finite displacement field does not eliminate inter-surface hybridization. It only makes it perturbative when
\begin{align}
    t_1|\gamma q|^{N_\ell}
    \ll
    \Delta_D.
    \label{eq:surface_isolation_condition}
\end{align}

In this isolated-surface regime, ordinary non-degenerate perturbation theory gives
\begin{align}
    \ket{\varphi^{(e)}_{\bs q}}
    &\approx
    \ket{\tilde\varphi^{(e)}_{\bs q}}
    +
    \frac{\Delta_{N_\ell}(\bs q)}{\Delta_D}
    \ket{\tilde\varphi^{(h)}_{\bs q}},
    &
    \ket{\varphi^{(h)}_{\bs q}}
    &\approx
    \ket{\tilde\varphi^{(h)}_{\bs q}}
    - \frac{\Delta_{N_\ell}^{*}(\bs q)}{\Delta_D}
    \ket{\tilde\varphi^{(e)}_{\bs q}} .
    \label{eq:perturbative_surface_eigenstates}
\end{align}

\begin{figure}
    \centering
    \includegraphics[width=1.00\linewidth]{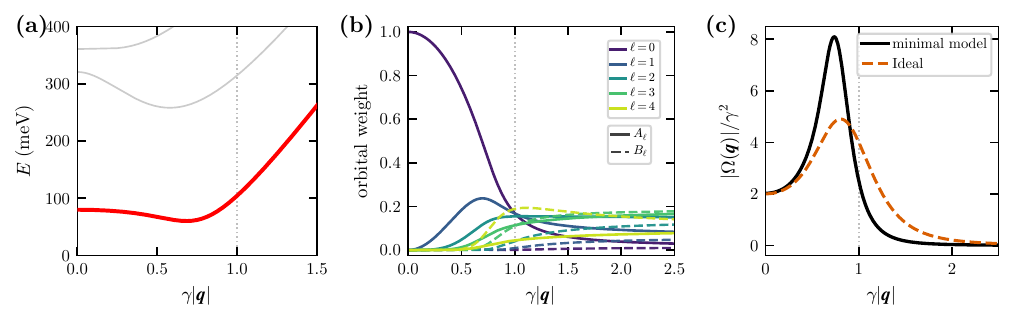}\vspace{-3mm}
    \caption{Basic features of the continuum bands in rhombohedral graphene. All panels are for the single-valley minimal model with only the intralayer Dirac velocity $v_D$, nearest-layer hopping $t_1$, and displacement field $u_D$, for $N_\ell=5$ and $u_D=40\,{\rm meV}$. The horizontal axis is the dimensionless momentum $\gamma |\bs{q}|$, with $\gamma=v_D/t_1$. (a) Energy dispersion of the $2N_\ell$ continuum bands. The first conduction band is highlighted in red. (b) Orbital weights of the first conduction band. Solid lines show $|\psi_{A_\ell}(\bs{q})|^2$ and dashed lines show $|\psi_{B_\ell}(\bs{q})|^2$, with color denoting the layer index $\ell$. (c) Berry curvature of the tracked conduction band compared with the ideal parent band result. The vertical dotted line marks $\gamma|\bs{q}|=1$, which roughly separates the small-momentum regime where the ideal single-sublattice parent band description is expected to be most accurate from the large-momentum regime where the monolayer Dirac kinetic energy dominates.}\label{si_fig:RNG_disp_summary}
\end{figure}
In the regime $u_D/t_1\ll 1$ and $|\Delta_{N_\ell}|/\Delta_D\ll 1$,  the wavefunctions of interest are supported almost entirely on a single sublattice. It then appears natural to define an idealized parent band approximation in which the small opposite-sublattice weight is neglected
\begin{subequations}    
\begin{align}
    \braket{A,\ell}{s^{(h)}(\bs{q})} &=\mathcal{N}_{\bs{q}}^{(h)} (\gamma \bar q)^{ \ell}, 
    &
    \braket{B,\ell}{s^{(h)}(\bs{q})} &= 0,\\
    \braket{B,\ell}{s^{(e)}(\bs{q})} &=  \mathcal{N}_{\bs{q}}^{(e)} (\gamma  q)^{N_\ell-1-\ell},
    &
    \braket{A,\ell}{s^{(e)}(\bs{q})} &= 0.
\end{align}
\end{subequations}
This approximation, where the conduction (valence) band is holomorphic (antiholomorphic) and is supported on a single sublattice, is justified in the window
\begin{align}
    t_1|\gamma q|^{N_\ell}
    \ll
    (N_\ell-1)u_D
    \ll
    t_1 .
    \label{eq:ideal_parent_window}
\end{align}

The same single-sublattice wavefunction can be obtained by deriving an effective ``parent Hamiltonian''. We describe the procedure for the valence band here, which is easy to generalize to the conduction band. This can be achieved by considering the Schur complement of the $A$-sublattice hopping matrix as~\cite{han2025exact}
\begin{subequations}
\begin{align}
    h^{(N_\ell)}_{\mathrm{Schur},A}(\bs{q}) &= h^{(N_\ell)}_{AA}(\bs{q}) -  h^{(N_\ell)}_{AB}(\bs{q})  (h^{(N_\ell)}_{BB}(\bs{q}))^{-1} h^{(N_\ell)}_{BA}(\bs{q})
\end{align}
where
\begin{align}
    h^{(N_\ell)}_{AA}(\bs{q}) &= h^{(N_\ell)}_{BB}(\bs{q}) = u_D \mqty( 1 & & & \\ & 2 & & \\ & & \ddots & \\  & & & N_\ell ) \\
    h^{(N_\ell)}_{AB}(\bs{q}) &= \left( h^{(N_\ell)}_{BA}(\bs{q}) \right)^\dagger = \mqty(-v_D q & & & \\ t_1 & -v_D q & & \\ & \ddots & \ddots & \\ & & t_1 & -v_D q)
\end{align}
\end{subequations}
are the different sublattice blocks, and we have shifted the displacement field by a constant factor of $(N_\ell+1)/2$ for convenience. This results in the effective hopping matrix
\begin{align}
    h^{(N_\ell)}_{\mathrm{Schur},A}(\bs{q}) &= u_D \mqty(1 & & & \\ & 2 & & \\ & & \ddots & \\  & & & N_\ell ) \nonumber \\
    &\qquad\quad  - \frac{t_1^2}{u_D} 
    \mqty( 
    (\gamma |\bs{q}|)^2 & - \gamma q & & & \\
    - \gamma \bar q & (\gamma |\bs{q}|)^2/2 + 1 & - \gamma q/2 & & \\
     & -\gamma \bar q/2 & (\gamma |\bs{q}|)^2/3 + 1/2 & \ddots & \\
     & & \ddots & \ddots & -\gamma q/(N_{\ell}-1)  \\
     & & & -\gamma \bar q/(N_{\ell}-1) & (\gamma |\bs{q}|)^2/N_{\ell} + 1/(N_{\ell}-1) 
    ).
\end{align}
Dropping terms of $\mathcal{O}(u_D)$ and the last diagonal term $(\gamma |\bs{q}|)^2/N_\ell$, equivalently muting the intralayer $A_{N_\ell-1}$--$B_{N_\ell-1}$ hopping in the parent construction, gives
\begin{align}
    \tilde{h}^{(N_\ell)}_{A}(\bs{q}) = - \frac{t_1^2}{u_D} 
    \mqty( 
    (\gamma |\bs{q}|)^2& - \gamma q & & & \\
    - \gamma \bar q & (\gamma |\bs{q}|)^2/2 + 1 & - \gamma q/2 & & \\
     & -\gamma \bar q/2 & (\gamma |\bs{q}|)^2/3 + 1/2 & \ddots & \\
     & & \ddots & \ddots & -\gamma q/(N_{\ell}-1)  \\
     & & & -\gamma \bar q/(N_{\ell}-1) & 1/(N_{\ell}-1) 
    ).
\end{align}
This matrix has an exactly flat zero-energy band with eigenvector
\begin{align}
    \tilde{h}^{(N_\ell)}_{A}(\bs{q})
    \mqty(1 & \gamma \bar q & \cdots & (\gamma \bar q)^{N_\ell-1})^T=0.
\end{align}
The dropped term acts only on the component proportional to $(\gamma\bar q)^{N_\ell-1}$ and is therefore small when $|\gamma q|<1$ and/or $N_\ell$ is large. Thus, at the level of the local single-sublattice wavefunction, the parent construction is controlled by $|\gamma q|\ll1$, $u_D/t_1\ll1$, and sufficiently large $N_\ell$. As discussed below, for a fixed finite momentum window, there is an additional requirement: the displacement field must also be sufficiently large to isolate the surface band from its opposite-surface partner.

\subsection{Band properties in the ideal limit}

Here, we focus on the conduction band properties in the ideal model. To do so, we find it convenient to invert our layer labeling as $N_\ell-1 \to 0, N_\ell-2 \to 1, \ldots, 0\to N_{\ell}-1$. The orbital labeling is then the same as in the main text. From our above derivation, the continuum conduction band of interest in the ideal limit is described by single-particle states of the form
\begin{subequations}    
\begin{align}
    |\bs{q}\rangle = c_{\bs{q}}^\dagger |0\rangle, \qquad \braket{\bs{r},\ell}{\bs{q}}  = \bra{\bs{r},\ell} e^{i\bs{q}\cdot\bs{r}} \ket{s_{\bs{q}}} = e^{i\bs{q}\cdot\bs{r}} s_{\bs{q},\ell}, 
    \qquad \ell=0,1,\ldots,N_\ell-1, \label{eq:si_parent_band_single_particle_creation}
\end{align}
with 
\begin{align}
    s_{\bs{q},\ell} &= N_{\bs{q}} (\gamma q)^\ell, 
    \qquad 
    N_{\bs{q}}^{-2} = \sum_{\ell=0}^{N_{\ell}-1} |\gamma \bs{q}|^{2\ell} = \frac{1 - (\gamma |\bs{q}|)^{2N_\ell} }{1 - (\gamma |\bs{q}|)^2}. \label{eq:app_spinor_structure_parent_band}
\end{align}
\end{subequations}

The density operator projected into this continuum band is
\begin{align}
    \rho(\bs{q}) = \sum_{\bs{p}} F_P(\bs{p} + \bs{q}, \bs{p}) c_{\bs{p} + \bs{q}}^\dagger c_{\bs{p}},
\end{align}
with the parent band form factor 
\begin{align}
    F_P(\bs{q}, \bs{q}') &= \matrixel{\bs{q}}{ e^{i(\bs{q} - \bs{q}')\cdot\bs{r}} }{\bs{q}'} 
    = \braket{s_{\bs{q}}}{s_{\bs{q}'}}
    = \sum_{\ell=0}^{N_{\ell}-1} s^{*}_{\bs{q},\ell} s_{\bs{q}',\ell} = N_{\bs{q}} N_{\bs{q}'} \sum_{\ell=0}^{N_{\ell}-1} (\gamma^{2} \bar{q} q')^\ell .
\end{align}
Because the unnormalized spinor is holomorphic in $q$, the Berry curvature in the $\bs{K}'$-valley convention is
\begin{align}
    \Omega^{(N_\ell)}(\bs{q})
    &= -2\partial_q\partial_{\bar q}\log N_{\bs{q}}^{-2} 
    = -\frac{2 \gamma^2}{\left(\gamma^2 |\bs{q}|^2 -1 \right)^2}
    + \frac{2 N_{\ell}^2 \gamma^{2 N_{\ell}} |\bs{q}|^{2N_\ell-2}}{\left(\gamma^{2 N_{\ell}} |\bs{q}|^{2N_{\ell}}-1 \right)^2}. \label{eq:si_berry_curvature_ideal}
\end{align}
The continuum parent band thus encloses a total Berry flux of
\begin{align}
    \int d^2q \,\Omega^{(N_\ell)}(\bs{q}) = - 2 \pi (N_{\ell}-1),
\end{align}
with the sign reversed in the opposite valley. This ideal parent band Berry curvature is compared to the minimal model of R5G with $u_D=40~\mathrm{meV}$ in Fig.~\ref{si_fig:RNG_disp_summary}(c). We see that the overall agreement is reasonable, especially in the $\gamma |\bs{q}|\ll 1$ regime. From the holomorphic structure of the unnormalized wavefunction components in Eq.~\eqref{eq:app_spinor_structure_parent_band}, the Fubini--Study metric satisfies the ideal isotropic relation~\cite{wang2021exact}
\begin{align}
    g^{(N_\ell)}_{\alpha\beta}(\bs{q}) &= \frac{1}{2}|\Omega^{(N_\ell)}(\bs{q})|\delta_{\alpha\beta},
\end{align} 
which implies the band is ideal and saturates the trace condition~\cite{Parameswaran2012, parameswaran2013fractional, Roy2014, ledwith2020fractional, wang2021exact, ledwidth2023vortexability}
\begin{align}
    \Tr[g^{(N_\ell)}(\bs{q})] = |\Omega^{(N_\ell)}(\bs{q})|.
\end{align}

\subsection{Comparing band properties in the minimal model and ideal limit}
\label{si_subsec:comparison_ideal_limit_rng}

\begin{figure}
    \centering
    \includegraphics[width=1.00\linewidth]{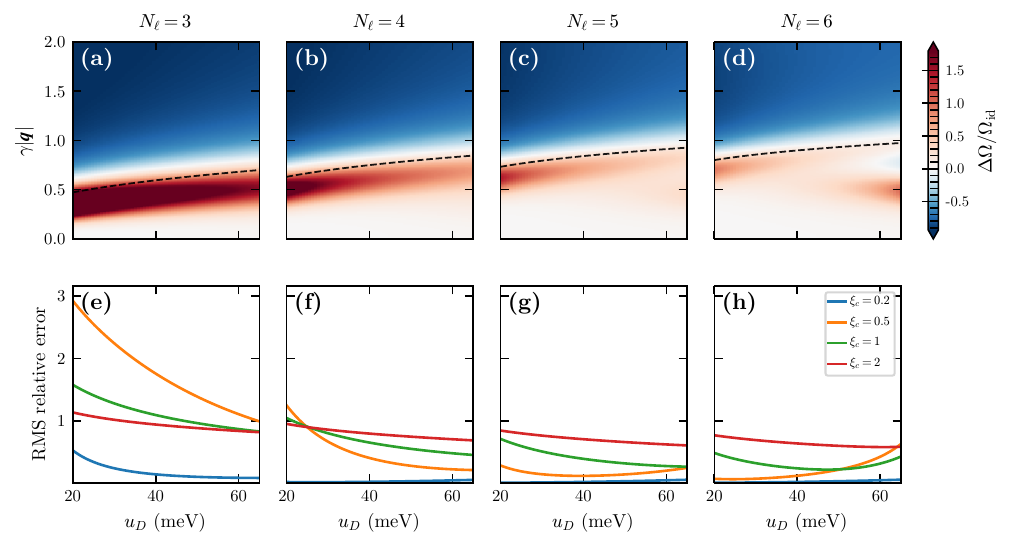}\vspace{-3mm}
    \caption{Validity of the ideal parent band geometry as a function of displacement field and number of layers. The comparison is performed for the minimal model with $t_0=3.1\,\mathrm{eV}$, $t_1=0.38\,\mathrm{eV}$, and $\gamma=1.73\,\mathrm{nm}$. Columns correspond to $N_\ell=3,4,5,6$. (a)--(d) Relative Berry curvature deviation for the first conduction band, $\Delta_\Omega=(\Omega_{\rm phys}-\Omega_{\rm id})/\Omega_{\rm id}$, as a function of displacement field $u_D$ and dimensionless momentum $\xi=\gamma|\bs q|$. The black dashed line marks the estimated surface-hybridization crossover $t_1\xi^{N_\ell}\sim (N_\ell-1)u_D$. (e)--(h) Radially integrated RMS relative deviation $\mathcal X_\Omega(u_D;\xi_c)$, defined in Eq.~\eqref{eq:si_rms_bc_error}, for several momentum cutoffs $\xi_c=0.2,0.5,1,2$.}
    \label{si_fig:comparison_ideal_limit_bc}
\end{figure}

We now compare the band geometry of the ideal parent band approximation with that of the minimal continuum model containing only $v_D$, $t_1$, and $u_D$. The comparison is performed for the first conduction band in a single valley. Since the minimal model is rotationally symmetric, the Berry curvature depends only on $\xi=\gamma|\bs{q}|$.

The previous subsection shows that the ideal single-surface description is controlled by two small parameters. The first is the local single-sublattice expansion parameter, $u_D/t_1$, and the second is the opposite-surface mixing parameter
\begin{align}
    \lambda_{\rm mix}(\xi,u_D)
    \equiv
    \frac{t_1\xi^{N_\ell}}{(N_\ell-1)u_D}.
\end{align}
Thus, for a fixed momentum window, the ideal parent band is expected to be accurate when
\begin{align}
    \lambda_{\rm mix}(\xi,u_D)\ll1,
    \qquad
    u_D/t_1\ll1,
\end{align}
or equivalently in the window stated in Eq.~\eqref{eq:ideal_parent_window}.

To quantify the comparison, we compute the relative Berry curvature deviation
\begin{align}
    \Delta_\Omega(\xi,u_D)
    =
    \frac{\Omega_{\rm phys}(\xi,u_D)-\Omega_{\rm id}(\xi)}
    {\Omega_{\rm id}(\xi)},
\end{align}
where $\Omega_{\rm phys}$ is obtained from the finite-difference quantum geometric tensor of the minimal model, and $\Omega_{\rm id}$ is the analytic ideal-parent result in Eq.~\eqref{eq:si_berry_curvature_ideal}. We also define the radially integrated RMS relative deviation
\begin{align}
    \mathcal X_\Omega(u_D;\xi_c)
    =
    \left[
    \frac{\int_0^{\xi_c}d\xi\,\xi\,|\Delta_\Omega(\xi,u_D)|^2}
    {\int_0^{\xi_c}d\xi\,\xi}
    \right]^{1/2}.
    \label{eq:si_rms_bc_error}
\end{align}
The factor $\xi d\xi$ is the radial momentum-space measure, up to an overall constant that cancels in the normalized RMS error.

The upper panels of Fig.~\ref{si_fig:comparison_ideal_limit_bc} show that the largest change in the Berry curvature error occurs near $\lambda_{\rm mix}\sim1$, shown by the black dashed line. For momenta above this crossover, the chiral inter-surface coupling is comparable to or larger than the displacement-field splitting, and the physical eigenstate is no longer well described by a single surface. Increasing $u_D$ shifts this crossover to larger $\xi$, thereby extending the momentum range over which the ideal single-surface description applies.

The lower panels show the same physics integrated over finite momentum windows. For larger cutoffs $\xi_c$, the RMS error is dominated by the region where inter-surface hybridization is important, so increasing $u_D$ generally improves the agreement. For smaller cutoffs, especially at larger $N_\ell$, the error is more sensitive to the local single-sublattice approximation and can develop a shallow optimum at intermediate $u_D$. Over the range $20~{\rm meV}\lesssim u_D\lesssim65~{\rm meV}$ shown here, however, $u_D/t_1$ remains small, and the dominant effect of increasing $u_D$ is to improve surface-band isolation.

\section{Lightning fast review of quantum Hall basics} \label{si_sec:review_qh_basics}

Throughout this Section, we take the electron charge to be $-e$, with $e>0$, and choose
\begin{equation}
    \bs B=-B\hat{\bs z},\qquad B>0,\qquad
    l_B=\sqrt{\frac{1}{eB}} .
\end{equation}
The single-particle Hamiltonian is
\begin{equation}
    H=\frac{(\bs p+e\bs A)^2}{2m}
    =\frac{\bs\pi^2}{2m},
    \qquad
    \bs\pi=\bs p+e\bs A,
\end{equation}
so that
\begin{equation}
    [\pi_x,\pi_y]=\frac{i}{l_B^2}.
\end{equation}
The guiding-center coordinate is
\begin{equation}
    \bs R=\bs r + l_B^2\wedge\bs\pi .
\end{equation}
It obeys
\begin{equation}
    [R_x,R_y]=-il_B^2,
    \qquad
    [R_\alpha,\pi_\beta]=0.
\end{equation}
Thus, the cyclotron and guiding-center degrees of freedom are independent. A convenient oscillator basis is
\begin{equation}
    a=\frac{l_B}{\sqrt2}(\pi_x+i\pi_y),
    \qquad
    b=\frac{1}{\sqrt2l_B}(R_x-iR_y),
\end{equation}
with $[a,a^\dagger]=[b,b^\dagger]=1$ and $[a,b]=[a,b^\dagger]=0$. The Hamiltonian depends only on the cyclotron oscillator,
\begin{equation}
    H=\hbar\omega_c\left(a^\dagger a+\frac12\right),
    \qquad
    \omega_c=\frac{eB}{m}.
\end{equation}
The eigenstates may be written as
\begin{equation}
    \ket{n,m}
    =\frac{(a^\dagger)^n(b^\dagger)^m}{\sqrt{n!m!}}\ket{0,0},
\end{equation}
where $n=0,1,2,\ldots$ labels the Landau level and $m=0,1,2,\ldots$ labels the degenerate guiding-center states within a Landau level.

\subsection{Torus geometry} \label{si_subsec:torus_magnetic_bloch}

\subsubsection{Magnetic translations} \label{si_subsubsec:magnetic_translations}

After projection to a fixed Landau level, magnetic translations on a particle $j$ act only on the guiding-center coordinate and are represented by
\begin{equation}
    T_j(\bs u)
    =
    \exp\left[-\frac{i}{l_B^2}\bs u\wedge\bs R_j\right]. \label{eq:def_magnetic_translation_operator}
\end{equation}
For a single-particle, these operators obey
\begin{equation}
    T_j(\bs u)T_j(\bs v)
    =\exp\left(\frac{i}{2l_B^2}\bs u\wedge\bs v\right)
      T_j(\bs u+\bs v),
\end{equation}
hence
\begin{equation}
    T_i(\bs u)T_j(\bs v)
    =T_j(\bs v)T_i(\bs u)
      \exp\left(\frac{i\delta_{ij}}{l_B^2}\bs u\wedge\bs v\right).
    \label{eq:single_particle_mag_translation_algebra}
\end{equation}

We place the system on a torus generated by two periods $\bs L_1$ and $\bs L_2$. We denote by $N_\Phi$ the Landau-level degeneracy,
\begin{equation}
    N_\Phi=\frac{\bs L_1\wedge\bs L_2}{2\pi l_B^2}\in\mathbb Z ,
\end{equation}
where the orientation is chosen so that $\bs L_1\wedge\bs L_2>0$. With our convention $\bs B=-B\hat{\bs z}$, this is the magnitude of the signed magnetic flux in units of $h/e$. In such a setting, it is useful to introduce a magnetic unit cell with primitive vectors $\bs a_1$ and $\bs a_2$ such that
\begin{equation}
    \bs L_\alpha=N_\alpha\bs a_\alpha,
    \qquad
    \bs a_1\wedge\bs a_2=2\pi l_B^2,
    \qquad
    N_1N_2=N_\Phi .
\end{equation}
Because a magnetic unit cell contains exactly one flux quantum, the magnetic translations $T(\bs a_1)$ and $T(\bs a_2)$ commute. Therefore, one may choose single-particle magnetic Bloch states satisfying
\begin{equation}
    T_j(\bs a_\alpha)\ket{\bs k}_{\mathrm{LL}}
    =e^{i\bs k\cdot\bs a_\alpha}\ket{ \bs k}_{\mathrm{LL}},
    \qquad \alpha=1,2 . \label{eq:definition_magnetic_bloch_state}
\end{equation}
The associated reciprocal vectors are 
\begin{equation}
    \bs b_1=\frac{\wedge\bs a_2}{l_B^2},
    \qquad
    \bs b_2=-\frac{\wedge\bs a_1}{l_B^2},
\end{equation}
such that $\bs b_i\cdot\bs a_j=2\pi\delta_{ij}$. The allowed magnetic momenta are fixed by $\bs k\cdot\bs L_i=2\pi m_i+\Phi_i$, so
\begin{equation}
    \bs k
    =\left(\frac{m_1}{N_1}+\frac{\Phi_1}{2\pi N_1}\right)\bs b_1
    +\left(\frac{m_2}{N_2}+\frac{\Phi_2}{2\pi N_2}\right)\bs b_2,
    \qquad
    m_\alpha=0,1,\ldots,N_\alpha-1,
    \label{eq:magnetic_momentum_grid_twisted}
\end{equation}
where $\Phi_\alpha$ are global twist angles (i.e., fluxes) defined by the torus boundary condition
\begin{equation}
    T_j(\bs L_\alpha)\ket{\phi_j}=e^{i\Phi_\alpha}\ket{\phi_j},
    \qquad \alpha=1,2 .
\end{equation}

\subsubsection{Flux-preserving translations} \label{si_subsubsec:flux_preserving_translations}

A generic magnetic translation changes the twist sector. Indeed,
\begin{equation}
    T_j(\bs L_\alpha)T_j(\bs u)\ket{\phi_j} =   \exp\left[i\Phi_\alpha+\frac{i}{l_B^2}\bs L_\alpha\wedge\bs u\right] T_j(\bs u)\ket{\phi_j}.
\end{equation}
Thus $T_j(\bs u)$ preserves fixed boundary conditions only if $\bs L_\alpha \wedge\bs u/l_B^2$ is an integer multiple of $2\pi$ for both $\alpha=1,2$. Modulo translations by the full periods $\bs L_\alpha$, the boundary-condition-preserving magnetic translations are generated by
\begin{equation}
    \bs d_\alpha=\frac{\bs L_\alpha}{N_\Phi},
    \qquad \alpha=1,2 .
\end{equation}
There are $N_\Phi^2$ such translations, and the generators satisfy
\begin{equation}
    T_j(\bs d_1)T_j(\bs d_2)
    =\exp\left(\frac{2\pi i}{N_\Phi}\right)
      T_j(\bs d_2)T_j(\bs d_1) .
    \label{eq:single_particle_d1_d2_algebra}
\end{equation}

\subsubsection{One-dimensional orbital labelling}
\label{si_subsubsec:one_dimensional_orbital_label}

A particularly convenient parametrization is obtained by choosing
\begin{equation}
    N_1=N_\Phi,
    \qquad
    N_2=1,
    \qquad
    \bs a_1=\bs d_1,
    \qquad
    \bs a_2=\bs L_2 .
\end{equation}
For zero twists, the $N_\Phi$ states in a Landau level are then labeled by
\begin{equation}
    \bs k_m=\frac{m}{N_\Phi}\bs b_1,
    \qquad
    m=0,1,\ldots,N_\Phi-1 .
\end{equation}
Twist angles simply shift this grid. In this basis, $T_j(\bs d_1)$ is a clock operator and $T_j(\bs d_2)$ is a shift operator. With a convenient phase convention,
\begin{equation}
    T_j(\bs d_1)\ket{\phi_m}
    =e^{2\pi i m/N_\Phi}\ket{\phi_m},
    \qquad
    T_j(\bs d_2)\ket{\phi_m}
    =\ket{\phi_{m+1}},
    \label{eq:clock_shift_single_particle}
\end{equation}
where $m$ is understood modulo $N_\Phi$ and $\ket{\phi_m}\equiv\ket{\phi_{\bs{k}_m}}$. If $c_m^\dagger\equiv c_{\bs{k}_m}^\dagger$ creates the orbital $\ket{\phi_m}$, the corresponding second-quantized action is
\begin{equation}
    \widehat T(\bs d_1)c_m^\dagger \widehat T(\bs d_1)^\dagger
    =e^{2\pi i m/N_\Phi}c_m^\dagger,
    \qquad
    \widehat T(\bs d_2)c_m^\dagger \widehat T(\bs d_2)^\dagger
    =c_{m+1}^\dagger .
    \label{eq:clock_shift_second_quantized}
\end{equation}
We shall use
\begin{equation}
    n_m=c_m^\dagger c_m
\end{equation}
for the orbital occupation operator.

\subsection{Many-body translations}
\label{si_subsec:many_body_translations}

For $N_e$ interacting electrons, define the center-of-mass magnetic translation
\begin{equation}
    T_{\rm MB}(\bs u)=\prod_{j=1}^{N_e}T_j(\bs u) .
\end{equation}
For simplicity, we set the twists to zero in the following discussion. Nonzero twists only shift the translation eigenvalues and do not modify the projective algebra discussed below. 

\subsubsection{Center-of-mass degeneracy}\label{si_subsubsec:center_of_mass_degeneracy}

For a translationally invariant interaction projected into a Landau level, $T_{\rm MB}(\bs u)$ commutes with the Hamiltonian whenever it preserves the chosen boundary conditions. In particular, the two elementary boundary-condition-preserving center-of-mass translations
\begin{equation}
    R\equiv T_{\rm MB}(\bs d_1),
    \qquad
    S\equiv T_{\rm MB}(\bs d_2)
\end{equation}
commute with $H$. Their algebra follows from Eq.~\eqref{eq:single_particle_d1_d2_algebra}:
\begin{equation}
    RS
    =\exp\left(\frac{2\pi iN_e}{N_\Phi}\right)SR .
\end{equation}
At filling
\begin{equation}
    \nu=\frac{N_e}{N_\Phi}=\frac{p}{q},
    \qquad \mathrm{GCD}(p,q)=1,
\end{equation}
this becomes
\begin{equation}
    RS=e^{2\pi i p/q}SR .
    \label{eq:many_body_projective_translation_algebra}
\end{equation}
$R$ and $S$ do not commute unless $q=1$. However, because $R$ and $S^q$ commute, the center-of-mass sectors may be labeled by the simultaneous eigenvalues of $H$, $R$, and $S^q$. Let $\ket{\Psi_{m,\sigma}}$ be such a many-body state:
\begin{equation}
    H\ket{\Psi_{m,\sigma}}=E\ket{\Psi_{m,\sigma}},
    \qquad
    R\ket{\Psi_{m,\sigma}}
    =e^{2\pi i m/N_\Phi}\ket{\Psi_{m,\sigma}},
    \qquad
    S^q\ket{\Psi_{m,\sigma}}
    =e^{i\sigma}\ket{\Psi_{m,\sigma}} .
\end{equation}
The additional label $\sigma$ denotes the phase of the $S^q$ eigenvalue. It is fixed within a given center-of-mass multiplet. Because $[H,S]=0$, the states $S^r\ket{\Psi_{m,\sigma}}$ have the same energy, and their $R$ eigenvalues are shifted according to
\begin{equation}
    R S^r\ket{\Psi_{m,\sigma}}
    =
    e^{2\pi i(m+rN_e)/N_\Phi}
    S^r\ket{\Psi_{m,\sigma}} .
\end{equation}
Writing $N_e=pN$ and $N_\Phi=qN$, with $\mathrm{GCD}(p,q)=1$, the $q$ states $r=0,1,\ldots,q-1$ have distinct $R$ eigenvalues, while $r=q$ returns to the same $R$ sector with the phase $e^{i\sigma}$. Thus, each fixed-$\sigma$ sector contains the usual $q$-fold center-of-mass multiplet. This is the torus magnetic-translation degeneracy reviewed in Ref.~\cite{haldane1985many}.

\subsubsection{One-body density correlations}
\label{si_subsubsec:one_body_density_matrix}

The distinction between a single pure ground state and a translation-invariant ground state density matrix is important. Let $\mathcal H_0$ be the full exactly degenerate ground state subspace of a translationally invariant Hamiltonian on the torus, and let
\begin{equation}
    P_0
    =\sum_{\mu=1}^{D}\dyad{\Psi_\mu}
\end{equation}
be the projector onto this subspace. Since $[H,R]=[H,S]=0$, both $R$ and $S$ map $\mathcal H_0$ to itself. Therefore,
\begin{equation}
    \rho_0=\frac{P_0}{D} \label{eq:uniform_density_matrix}
\end{equation}
is invariant under the magnetic translations,
\begin{equation}
    R\rho_0R^\dagger=\rho_0,
    \qquad
    S\rho_0S^\dagger=\rho_0 .
\end{equation}
The result is an invariant density matrix supported entirely within the ground state subspace.

Define the one-body density matrix in the one-dimensional orbital basis by
\begin{equation}
    \Gamma_{mn}
    =\mathrm{Tr}\left(\rho_0\,c_m^\dagger c_n\right) .
\end{equation}
Invariance under the clock translation $R$ gives
\begin{align}
    \Gamma_{mn}
    &=\mathrm{Tr}\left(\rho_0\,R c_m^\dagger c_n R^\dagger\right) =e^{2\pi i(m-n)/N_\Phi}\Gamma_{mn} .
\end{align}
Thus
\begin{equation}
    \Gamma_{mn}=0
    \qquad
    \text{unless }m=n .
\end{equation}
Invariance under the shift translation $S$ gives
\begin{equation}
    \Gamma_{m+1,m+1}=\Gamma_{m,m} .
\end{equation}
All diagonal occupations are therefore equal. Since
\begin{equation}
    \sum_{m=0}^{N_\Phi-1}\Gamma_{mm}=N_e,
\end{equation}
we obtain
\begin{equation}
    \Gamma_{mn}
    =\frac{N_e}{N_\Phi}\delta_{mn}
    =\nu\,\delta_{mn} .
    \label{eq:translation_invariant_density_matrix_gamma}
\end{equation}
The density matrix~\ref{eq:uniform_density_matrix} then has a uniform occupation of the magnetic Brillouin zone.

For a pure state, one must be more careful. Because $S$ is unitary and maps $\mathcal H_0$ into itself, its restriction to the finite-dimensional ground space can always be diagonalized. Hence there exists a basis of pure ground states satisfying
\begin{equation}
    S\ket{\Psi_\alpha}=e^{i\theta_\alpha}\ket{\Psi_\alpha} .
    \label{eq:S_eigenstate_ground_state}
\end{equation}
For any such state,
\begin{align}
    \mel{\Psi_\alpha}{n_{m+1}}{\Psi_\alpha}
    &=\mel{\Psi_\alpha}{S n_m S^\dagger}{\Psi_\alpha} = \mel{\Psi_\alpha}{n_m}{\Psi_\alpha},
\end{align}
where the phase in Eq.~\eqref{eq:S_eigenstate_ground_state} cancels. Therefore an $S$-eigenstate has
\begin{equation}
    \mel{\Psi_\alpha}{n_m}{\Psi_\alpha}=\nu
    \qquad
    \text{for all }m .
    \label{eq:S_eigenstate_uniform_diagonal_occupation}
\end{equation}
However, at fractional filling $R$ and $S$ obey the non-commuting algebra Eq.~\eqref{eq:many_body_projective_translation_algebra}. Consequently, a pure state cannot generally be an eigenstate of both $R$ and $S$. A pure $R$-eigenstate has a diagonal one-body density matrix in the $m$ basis, but its diagonal entries need not be uniform. A pure $S$-eigenstate has uniform orbital occupations, but symmetry alone does not force its off-diagonal one-body matrix elements to vanish. The fully constrained result Eq.~\eqref{eq:translation_invariant_density_matrix_gamma} is therefore symmetry-guaranteed for the translation-invariant density matrix $\rho_0 = P_0/D$, or for an explicit average over the center-of-mass multiplet, rather than for an arbitrary single pure ground state at fractional filling.

The same reasoning applies to Laughlin, Jain, hierarchy, and other homogeneous quantum Hall liquids on a torus. The conclusion follows from magnetic translation symmetry and fixed particle number, not from the detailed form of the many-body wave function. What changes from phase to phase are the full ground state degeneracy $D$ and the internal structure of the multiplet. In any case, since all states within $\mathcal{H}_0$ are degenerate, $\rho_0$ will have the same energy as any state in the multiplet. When interested in the ground state energy of the multiplet, one can use $\rho_0$, in which case the one-point correlator is diagonal and uniform in momentum space, simplifying calculations.

\subsection{Explicit form of the magnetic Bloch states} \label{si_subsec:magnetic_bloch_states}

We now provide explicit expressions for the magnetic Bloch states defined in~\eqref{eq:definition_magnetic_bloch_state}. A magnetic Bloch state can be constructed from a guiding center coherent state $\ket{n,m=0}$ (i.e., $ b\ket{n,m=0} = 0$) as
\begin{equation}
    \ket{\varphi_{\bs k}^{\rm nLL}}
    =\mathcal N_{\bs k}
    \sum_{l_1,l_2\in\mathbb Z}
    e^{-i\bs k\cdot\bs R_{l_1 l_2}}
    T(l_1\bs a_1)T(l_2\bs a_2)\ket{n,m=0} = \ket{n}\otimes\ket{\bs{k}}_{\mathrm{LL}},
    \qquad
    \bs R_{l_1l_2} = l_1\bs a_1 + l_2 \bs a_2 .
\end{equation}
In the lowest Landau level, this has the representative real-space form
\begin{align}
    \varphi_{\bs k}^{\rm LLL}(\bs r)
    &=\braket{\bs r}{\varphi_{\bs k}^{\rm LLL}} 
    =\mathcal N_{\bs k}
    \sum_{p,q\in\mathbb Z}
    \exp\left[
        -i\bs k\cdot\bs R_{pq}
        +i\pi pq
        +\frac{i}{2 l_B^2}\bs r\wedge\bs R_{pq}
        -\frac{|\bs r+\bs R_{pq}|^2}{4 l_B^2}
    \right], \label{eq:si_magnetic_bloch_states}
\end{align}
which can also be rewritten as~\cite{tan2025ideal, haldane1985many, haldane2018modular, haldane2018origin, haldane1985periodic}
\begin{align}
    \varphi_{\bs{k}}^{\text{LLL}}(\bs{r}) = \mathcal{N}_{\bs{k}} \exp \left(-\frac{z \bar{z}}{4 l_B^2}\right) e^{\frac{i}{2} \overline{k^{\prime}} z} \sigma\left(z+i l_B^2 k^{\prime}\right), \label{eq:si_magnetic_bloch_state_weirstrass_function}
\end{align}
where $k^{\prime}=k-\left(b_1+b_2\right) / 2$ and $\sigma\left(z\right)$ represents the modified Weierstrass $\sigma$-function~\cite{haldane2018modular} that satisfies the periodicity condition
\begin{align}
    \sigma\left(z+a_i\right)=-\exp \left(\frac{2 z \bar{a}_i+a_i \bar{a}_i}{4 l_B^2}\right) \sigma(z). \label{eq:si_periodicity_weirstrass_function}
\end{align}
Higher-Landau-level magnetic Bloch states are obtained by applying the cyclotron raising operator,
\begin{equation}
    \ket{\varphi_{\bs k}^{(n)}}
    =\frac{(a^\dagger)^n}{\sqrt{n!}}\ket{\varphi_{\bs k}^{\rm LLL}} .
\end{equation}

\subsubsection{Periodicity of magnetic Bloch states} \label{si_subsubsec:periodicity_bloch_states}

Using the magnetic Bloch states~\eqref{eq:si_magnetic_bloch_state_weirstrass_function} and the periodicity of the modified Weierstrass $\sigma$-function~\eqref{eq:si_periodicity_weirstrass_function}, it directly follows that the magnetic Bloch states obey
\begin{align}
    \varphi_{\bs{k}+\bs{b}_j}^{\text{LLL}}(\bs{r}) 
    &= \varphi_{\bs{k}}^{\text{LLL}}(\bs{r}) \exp( - i \frac{l_B^2}{2}  \bs{k} \wedge\bs{b}_j).
\end{align}
More generally, for a general shift by $\bs{g}=n_1\bs{b}_1 + n_2\bs{b}_2$, we can first apply all translations by $\bs{b}_1$ successively, which yields
\begin{align}
    \varphi_{\bs{k}+n_1\bs{b}_1}^{\text{LLL}}(\bs{r}) = \varphi_{\bs{k}}^{\text{LLL}}(\bs{r}) \exp(- i \frac{l_B^2}{2}  \bs{k} \wedge n_1\bs{b}_1),
\end{align}
since $\bs{b}_1\wedge \bs{b}_1=0$. Then applying translations by $\bs{b}_2$, we have
\begin{align}
    \varphi_{\bs{k}+n_1\bs{b}_1+\bs{b}_2}^{\text{LLL}}(\bs{r}) 
    &= \varphi_{\bs{k}}^{\text{LLL}}(\bs{r}) \exp(- i \frac{l_B^2}{2}  \bs{k} \wedge (n_1\bs{b}_1+\bs{b}_2)) \exp(- i \frac{l_B^2}{2}  n_1\bs{b}_1 \wedge \bs{b}_2) \nonumber \\
    &= \exp(i\pi n_1) \varphi_{\bs{k}}^{\text{LLL}}(\bs{r}) \exp( - i \frac{l_B^2}{2}  \bs{k} \wedge (n_1\bs{b}_1+\bs{b}_2)).
\end{align}
Repeating $n_2$ times, we find
\begin{align}
    \varphi_{\bs{k}+\bs{g}}^{\text{LLL}}(\bs{r}) 
    = \exp(i\pi n_1 n_2) \varphi_{\bs{k}}^{\text{LLL}}(\bs{r}) \exp( - i \frac{l_B^2}{2}  \bs{k} \wedge \bs{g}) = \exp(i\omega_{\bs{g}}) \varphi_{\bs{k}}^{\text{LLL}}(\bs{r}) \exp( -i \frac{l_B^2}{2}  \bs{k} \wedge \bs{g}),\label{eq:si_magnetic_bloch_state_periodicity}
\end{align}
where we introduced
\begin{align}
    \omega_{n_1\bs{b}_1 + n_2 \bs{b}_2} = \pi n_1 n_2. 
\end{align}

\subsubsection{Lowest Landau level form factor} \label{si_subsubsec:LLL_form_factor}

Using the periodicity relation~\eqref{eq:si_magnetic_bloch_state_periodicity} and denoting $\bs{g}_{\bs{k},\bs{q}}=\bs{k}+\bs{q}-\lceil\bs{k}+\bs{q}\rceil$ with $\lceil \bs{q} \rceil$ denoting folding back in the 1BZ, the lowest Landau level (LLL) form factor is 
\begin{align}
    \mathcal{F}^{\mathrm{LLL}}(\bs{k} + \bs{q} , \bs{k}) &= \bra{\varphi^{\mathrm{LLL}}_{\lceil \bs{k} + \bs{q} \rceil} }\exp(i\bs{q}\cdot\bs{r})\ket{\varphi^{\mathrm{LLL}}_{\bs{k}}} \nonumber \\
    &= \exp(i\omega_{\bs{g}_{\bs{k},\bs{q}}}) \exp\left[ - i \frac{l_B^2}{2}  (\bs{k}+\bs{q}) \wedge \bs{g}_{\bs{k},\bs{q}} \right] \bra{\varphi^{\mathrm{LLL}}_{ \bs{k} + \bs{q}} }\exp(i\bs{q}\cdot\bs{r})\ket{\varphi^{\mathrm{LLL}}_{\bs{k}}}  \nonumber \\
    &=  \exp(i\omega_{\bs{g}_{\bs{k},\bs{q}}}) \exp\left[ - i \frac{l_B^2}{2} (\bs{k}+\bs{q}) \wedge \bs{g}_{\bs{k},\bs{q}} \right] \exp\left[-\frac{l_B^2}{4}\left(|\bs{q}|^2 - 2i(\bs{k}+\bs{q})\wedge\bs{k} \right)\right] \nonumber \\
    &= \exp(i\omega_{\bs{g}_{\bs{k},\bs{q}}}) \exp\left[-\frac{l_B^2}{4}\left(|\bs{q}|^2 - 2i\bs{q}\wedge\bs{k} + 2i  (\bs{k}+\bs{q}) \wedge \bs{g}_{\bs{k},\bs{q}} \right)\right].
\end{align}

\section{AHC single-particle basis} \label{si_sec:ahc_single_particle_basis}

\subsection{General considerations}

As explained in the main text and laid out in greater detail in Ref.~\cite{tan2025ideal}, restriction to the first conduction band in the ideal limit (see Sec.~\ref{si_sec:validity_ideal_limit_rng}), enforces the following relation on the many-body state
\begin{align}
    \Psi_{\{\ell\}}(\{\bs{r}\}) = \left[ \prod_{j=1}^{N_e} (-2i\gamma\partial_{\bar{z}_j})^{\ell_j}\right] \Psi_{\{0\}}(\{\bs{r}\}), \label{eq:si_antiholormphic_condition_mb_wavefunction}
\end{align}
where the wavefunction with electrons on higher layers can be generated by acting with successive antiholomorphic derivatives on the lowest layer wavefunction. Here we denote positions and layer indices as $\{\bs{r}\}\equiv(\bs{r}_1,\bs{r}_2,\ldots,\bs{r}_{N_e})$ and $\{\ell\}\equiv(\ell_1,\ell_2,\ldots,\ell_{N_e})$, respectively. Minimization of the contact interaction $H_{\mathrm{int}} = \sum_{i<j}\delta(\bs{r}_i - \bs{r}_j)$, further enforces the presence of chiral nodes in the wavefunction as
\begin{align}
    \Psi_{\{\ell\}}(\{\bs{r}\}) = \left[ \prod_{i<j} (z_i-z_j)\right] \Xi_{\{\ell\}}(\{\bs{r}\}),
\end{align}
where the layer-dependent component $\Xi_{\{\ell\}}(\{\bs{r}\})$ is symmetric under particle exchange and must satisfy
\begin{align}
    \Xi_{\{\ell\}}(\{\bs{r}\}) = \left[ \prod_{j=1}^{N_e} (-2i\gamma\partial_{\bar{z}_j})^{\ell_j}\right] \Xi_{\{0\}}(\{\bs{r}\})
\end{align}
as follows from Eq.~\eqref{eq:si_antiholormphic_condition_mb_wavefunction}. 

From this manifold of states, we restrict our attention in this work to anomalous Hall crystal (AHC) wavefunctions of the form
\begin{align}
    \Psi_{\{\ell\}}(\{\bs{r}\})
    =\Phi_{\mathrm{QH}}(\{\bs{r}\}) \prod_{i=1}^{N_e}\chi^{(\bs{q})}_{\ell_i}(\bs{r}_i),
\end{align}
where $\Phi_{\mathrm{QH}}(\{\bs{r}\})$ is a many-body quantum Hall state in the LLL and $\prod_{i=1}^{N_e}\chi^{(\bs{q})}_{n,\ell_i}(\bs{r}_i)$ describe a boson condensate in an Abrikosov lattice subjected to an opposite magnetic field. Such wavefunctions describe quantum Hall crystals that spontaneously break translation symmetry, whose periodicity is set by the Abrikosov lattice. Here, we consider general crystals with one ``emergent'' flux quantum and $\nu$ electrons per unit cell. In the $\nu=1$ case, the lattice periodicity is set by the electronic density $A_{\mathrm{u.c.}}=1/n_{e}$ and we have a conventional integer anomalous Hall crystal (IAHC). In contrast, for $\nu=p/q$, $\Phi_{\mathrm{QH}}(\{\bs{r}\})$ corresponds to a general hierarchy incompressible quantum Hall state such that $\Psi_{\{\ell\}}(\{\bs{r}\})$ would display a fractional quantum anomalous Hall effect (when pinning the crystal). We refer to such a state as a fractional anomalous Hall crystal (FAHC).

Given that a QH wavefunction admits a general expansion in terms of magnetic Bloch states Slater determinants
\begin{align}
    \Phi_{\mathrm{QH}}(\{\bs{r}\}) = \sum_{\{\bs{k}\}} C_{\{\bs{k}\}} \det_{j,\bs{k}}[ \varphi_{\bs{k}}^{\mathrm{LLL}}(\bs{r}_j) ], \label{eq:si_expansion_qh_first_quantized}
\end{align}
our AHC wavefunction of interest admits a similar expansion as
\begin{align}
    \Psi_{\{\ell\}}(\{\bs{r}\}) 
    = \sum_{\{ \bs{k} \}} \frac{C_{\{\bs{k}\}}}{\mathfrak{N}_{\{\bs{k}\}}} \det_{j,\bs{k}}[ \mathfrak{N}_{\bs{k}} \varphi_{\bs{k}}^{\mathrm{LLL}}(\bs{r}_j) \chi_{\ell_j}^{(\bs{q})}(\bs{r}_j) ] 
    = \sum_{\{ \bs{k} \}} C_{\{\bs{k}\}} \left( \det_{j,\bs{k}}[ \varphi_{\bs{k}}^{\mathrm{LLL}}(\bs{r}_j)] \prod_j \chi_{\ell_j}^{(\bs{q})}(\bs{r}_j) \right)
    = \Phi_{\mathrm{QH}}(\{\bs{r}\}) \prod_{i=1}^{N_e}\chi^{(\bs{q})}_{\ell_i}(\bs{r}_i), \label{eq:si_expansion_ahc_first_quantized}
\end{align}
where $\mathfrak{N}_{\bs{k}}$ is a normalization constant and $\mathfrak{N}_{\{\bs{k}\}}= \prod_{\bs{k} \in \{\bs{k}\}} \mathfrak{N}_{\bs{k}}$. 

From the above, we see that integer and fractional anomalous Hall crystals (IAHC and AHC) wavefunctions can then be expanded as a superposition over Slater determinants of single-particle states of the form
\begin{align}
    \psi_{\bs{k},\bs{q},\ell}(\bs{r}) = \mathfrak{N}_{\bs{k}} \varphi_{\bs{k}}^{\mathrm{LLL}}(\bs{r}) \chi_{\ell}^{(\bs{q})}(\bs{r}). \label{eq:si_ahc_single-particle_basis_general}
\end{align}
For this single-particle state to be Bloch periodic
\begin{align}
    \psi_{\bs{k},\bs{q},\ell}(\bs{r} + \bs{a}_\alpha) = \exp(i (\bs{k}+\bs{q}) \cdot \bs{a}_\alpha) \psi_{\bs{k},\bs{q},\ell}(\bs{r}), 
\end{align}
we require that 
\begin{align}
    \chi_{\ell}^{(\bs{q})}(\bs{r} + \bs{a}_\alpha)  = \chi_{\ell}^{(\bs{q})}(\bs{r}) \exp \left(i \bs{q} \cdot \bs{a}_\alpha\right) \exp \left( i \frac{\bs{r} \wedge \bs{a}_\alpha}{2 l_B^2}\right)
\end{align}
since
\begin{align}
    \varphi_{\bs{k}}^{\mathrm{LLL}}\left(\bs{r}+\bs{a}_\alpha\right)
    = \varphi_{\bs{k}}^{\mathrm{LLL}}(\bs{r}) \exp \left(i \bs{k} \cdot \bs{a}_\alpha\right) \exp \left(-i \frac{\bs{r} \wedge \bs{a}_\alpha}{2 l_B^2}\right). \label{eq:periodicity_real_space_magnetic_bloch_state}
\end{align}
Assuming a single emergent flux quantum per unit cell as above, the above periodic boundary condition implies that $\chi_{0}^{(\bs{q})}(\bs{r})$ can be expanded in terms of the complex conjugate of the LL magnetic Bloch states
\begin{align}
    \chi_{0}^{(\bs{q})}(\bs{r}) = \sum_{n} c_n \bar{\varphi}_{-\bs{q}}^{\mathrm{nLL}}(\bs{r}),
\end{align}
with the overline indicating complex conjugation. Here, $\bs{q}$ determines magnetic Bloch momenta and the position of the vortices. In what follows, we will always set $\bs{q}=\bs{0}$ and will then suppress the index when it is not relevant to the discussion. For a triangular lattice, discrete rotational symmetry enforces that one can only mix $\bar{\varphi}_{\bs{0}}^{\mathrm{nLL}}$ with $n$ differing by a multiple of six~\cite{Fang2012, kruthoff2017topological, po2017symmetry}. In what follows, we will neglect such a mixing and assume that 
\begin{align}
    \chi_{n,0}^{(\bs{q})}(\bs{r}) 
    = \bar{\varphi}_{-\bs{q}}^{\mathrm{nLL}}(\bs{r})
    = \frac{(\bar{A}^\dagger)^n}{\sqrt{n!}} \bar{\varphi}_{-\bs{q}}^{0\mathrm{LL}},(\bs{r}),
\end{align}
where $\bar{A}^{\dagger}=i\left(2 l_B \partial_{\bar{z}}-z / 2 l_B\right) / \sqrt{2}$ is a conjugate LL raising operator. After fixing this, single-particle states on other layers can be obtained by successive anti-holomorphic derivatives as
\begin{align}
    \psi_{\bs{k},n,\ell}(\bs{r}) 
    &= (-2 i \gamma \partial_{\bar{z}})^\ell \psi_{\bs{k},n,0}(\bs{r}) 
    =  \mathfrak{N}_{\bs{k},n} (-2 i \gamma \partial_{\bar{z}})^\ell \varphi_{\bs{k}}^{\mathrm{LLL}}(\bs{r}) \chi_{n,0}^{(\bs{0})}(\bs{r}) 
    = \mathfrak{N}_{\bs{k},n} \sqrt{\frac{(n + \ell)!}{n!}} \left( -\frac{\sqrt{2} \gamma}{l_B} \right)^\ell \varphi_{\bs{k}}^{\mathrm{LLL}}(\bs{r})  \bar{\varphi}_{ \bs{0}}^{(n+\ell)\mathrm{LL}}(\bs{r}),
\end{align}
such that
\begin{align}
    \chi_{n,\ell}^{(\bs{0})}(\bs{r}) =  \sqrt{\frac{(n + \ell)!}{n!}} \left( -\frac{\sqrt{2} \gamma}{l_B} \right)^\ell \bar{\varphi}_{\bs{0}}^{(n+\ell)\mathrm{LL}}(\bs{r}). \label{eq:si_rewrite_chi_with_magnetic_bloch_states}
\end{align}
We note that our treatment in the Supplemental Material is slightly more general than in the main text, as we allow $n\ne 0$.

\subsection{Generalities about ideal bands}
\label{subsec:si_general_ideal_bands}

Before proceeding with further characterization of the AHC single-particle basis, we find it convenient to first gather a few general facts about ideal Chern bands. This also fixes the geometric conventions used in the rest of the Supplemental Material. A local quantity that characterizes the geometry of a band is the quantum geometric tensor (QGT)~\cite{Parameswaran2012, parameswaran2013fractional, Roy2014, yu2025quantum, wang2021exact, ledwidth2023vortexability, ledwith2020fractional, estienne2023ideal}
\begin{align}
    \mathcal Q_{\alpha\beta}(\bs k)
    =
    \braket{\mathcal D_{k_\alpha}u_{\bs k}}{\mathcal D_{k_\beta}u_{\bs k}}
    =
    g_{\alpha\beta}(\bs k)
    -
    \frac{i}{2}\varepsilon_{\alpha\beta}\Omega(\bs k),
    \label{eq:si_general_QGT_convention}
\end{align}
where $\ket{u_{\bs{k}}}$ is the cell periodic part of the Bloch wavefunction (i.e., $\ket{\psi_{\bs{k}}} = e^{i\bs{k}\cdot\bs{r}}\ket{u_{\bs{k}}}$) and the covariant derivative is
\begin{align}
    \mathcal D_{k_\alpha}\ket{u_{\bs k}}
    =
    \left(\partial_{k_\alpha}+i\mathcal A_\alpha(\bs k)\right)
    \ket{u_{\bs k}}, \qquad  \mathcal A_\alpha(\bs k)
    =
    i\braket{u_{\bs k}}{\partial_{k_\alpha}u_{\bs k}}.
\end{align}
The real part of the QGT is the Fubini-Study metric, which is defined by 
\begin{align}
    1- |\braket{u_{\bs{k}}}{u_{\bs{k} + \delta \bs{k}}}|^2 = \sum_{\alpha\beta} g_{\alpha\beta}(\bs{k}) \delta k_\alpha \delta k_\beta + \ldots ,
\end{align}
and the imaginary part is the Berry curvature
\begin{align}
    \Omega(\bs k)
    =
    \partial_{k_x}\mathcal A_y(\bs k)
    -
    \partial_{k_y}\mathcal A_x(\bs k).
\end{align}
The Berry curvature always lower bounds the metric as~\cite{wang2021exact}
\begin{align}
    \Tr[g_{\alpha\beta}(\bs{k})] \ge |\Omega(\bs{k})|.
\end{align}

A band is said to be ideal if it saturates the above trace condition for all momenta
\begin{align}
    \Tr[g_{\alpha\beta}(\bs{k})] = |\Omega(\bs{k})|.
\end{align}
A convenient local characterization of an ideal band is the existence of a holomorphic frame. In a smooth patch of the Brillouin zone, one can write
\begin{align}
    \ket{u_{\bs k}}
    =
    \mathcal N_{\bs k}
    \ket{\widetilde u_{\bs k}},
    \qquad
    \partial_{\bar k}\ket{\widetilde u_{\bs k}}=0,
    \label{eq:si_holomorphic_frame_general}
\end{align}
where $(\partial_{k_x}+i\partial_{k_y})/2$, and $\mathcal{N}_{\bs{k}} = 1/\sqrt{ \braket{\widetilde u_{\bs k}}{\widetilde u_{\bs k}}}$. The associated K\"ahler potential is
\begin{align}
    \mathcal K(\bs k)
    =
    \log( \braket{\widetilde u_{\bs k}}{\widetilde u_{\bs k}} )
    =
    -2\log \mathcal N_{\bs k}.
    \label{eq:si_kahler_potential_general}
\end{align}
For an ideal band, the momentum dependence of the Kähler potential controls the Berry curvature (and Fubini-Study metric) distribution as
\begin{align}
    \Omega(\bs k)
    =
    -\frac12\nabla_{\bs k}^2\mathcal K(\bs k).\label{eq:si_curvature_from_kahler_general}
\end{align}
Under a holomorphic change of frame,
\begin{align}
    \ket{\widetilde u_{\bs k}}
    \rightarrow
    e^{f(k)}\ket{\widetilde u_{\bs k}}, \qquad
    \partial_{\bar k} f(k)=0, \label{eq:si_holomorphic_transformation}
\end{align}
the normalized state changes only by a phase, while the K\"ahler potential undergoes
\begin{align}
    \mathcal K(\bs k)
    \rightarrow
    \mathcal K(\bs k)+f(k)+\bar f(\bar k).
    \label{eq:si_kahler_transformation_general}
\end{align}
The Berry curvature~\eqref{eq:si_curvature_from_kahler_general} remains invariant under such transformations. 

Let us briefly recall why Eq.~\eqref{eq:si_holomorphic_frame_general} implies ideal quantum geometry. Since $\partial_{\bar k}\ket{\widetilde u_{\bs k}}=0$, the anti-holomorphic derivative of the normalized state is purely parallel to the state itself
\begin{align}
    \partial_{\bar k}\ket{u_{\bs k}}
    =
    (\partial_{\bar k}\log\mathcal N_{\bs k})
    \ket{u_{\bs k}} .
\end{align}
After projection to a subspace orthogonal to $\ket{u_{\bs k}}$, we have
\begin{align}
    \mathcal D_{\bar k}\ket{u_{\bs k}}=0,
    \qquad
    \mathcal D_{\bar k}
    =
    \frac12(\mathcal D_{k_x}+i\mathcal D_{k_y}).
\end{align}
Equivalently,
\begin{align}
    \mathcal D_{k_y}\ket{u_{\bs k}}
    =
    i\mathcal D_{k_x}\ket{u_{\bs k}} .
\end{align}
Denoting $\ket{v}=\mathcal D_{k_x}\ket{u_{\bs k}}$ this gives
\begin{align}
    g_{xx}(\bs k) = g_{yy}(\bs k) =\braket{v}{v},
    \qquad
    g_{xy}(\bs k)=0,
    \qquad
    \Omega(\bs k)=-2\braket{v}{v},
\end{align}
such that $\mathrm{Tr}[g(\bs k)]=|\Omega(\bs k)|$, as claimed.

For a Chern band, the holomorphic frame cannot be chosen to be globally smooth and periodic. Instead, momenta related by reciprocal lattice vectors are connected by transition functions. For a $|C|=1$ ideal band, one may choose the same line-bundle structure as in a LLL magnetic Bloch band. In this representation, the normalized Bloch wavefunction can be written as~\cite{wang2021exact, ledwidth2023vortexability, cano2026ideal}
\begin{align}
    \psi_{\bs k}(\bs r)
    =
    \mathfrak M_{\bs k}\,
    \mathcal B(\bs r)
    \varphi^{\rm LLL}_{\bs k}(\bs r),
    \label{eq:si_generic_C1_ideal_wavefunction}
\end{align}
where $\varphi^{\rm LLL}_{\bs k}$ is the normalized LLL magnetic Bloch state with one flux quantum per unit cell, $\mathcal B(\bs r)$ is a momentum-independent real-space factor, and $\mathfrak M_{\bs k}$ normalizes the state. The factor $\mathcal B(\bs r)$ is quasi-periodic in real space, so the product in Eq.~\eqref{eq:si_generic_C1_ideal_wavefunction} obeys ordinary Bloch boundary conditions, given the periodicity of the magnetic Bloch state~\eqref{eq:periodicity_real_space_magnetic_bloch_state}. 

The connection between Eq.~\eqref{eq:si_generic_C1_ideal_wavefunction} and the holomorphic frame in Eq.~\eqref{eq:si_holomorphic_frame_general} is as follows. In a local gauge, the LLL magnetic Bloch state contains a universal non-holomorphic normalization factor. Using the definition~\eqref{eq:si_magnetic_bloch_state_weirstrass_function}, we see that one can define a holomorphic LLL cell-periodic section by
\begin{align}
    \varphi^{\rm LLL}_{\bs k}(\bs r) = e^{i\bs{k}\cdot\bs{r}} u_{\bs{k}}^{\rm LLL}(\bs{r}) = e^{i\bs{k}\cdot\bs{r}} \mathcal{N}_{\bs{k}}^{\rm LLL} \tilde{u}_{\bs{k}}^{\rm LLL}(\bs{r}), 
\end{align}
where $\partial_{\bar{k}} \tilde{u}_{\bs{k}}^{\rm LLL}(\bs{r}) = 0$ and $\mathcal{N}_{\bs{k}}^{\rm LLL} = \exp(-\mathcal{K}^{\rm LLL}(\bs{k})/2) = \exp(-l_B^2 |\bs{k}|^2/4)$. It then directly follows that $\Omega^{\rm LLL}_{\bs{k}} = -\nabla_{\bs{k}}^2 \mathcal{K}^{\rm LLL}(\bs{k})/2 = -l_B^2$. Given this, we can relate the ideal $|C|=1$ band~\eqref{eq:si_generic_C1_ideal_wavefunction} and \eqref{eq:si_holomorphic_frame_general} by using the rewriting
\begin{align}
    \psi_{\bs k}(\bs r)
    =
    \mathfrak{M}_{\bs k}  e^{i\bs{k}\cdot\bs{r}} \mathcal{N}_{\bs{k}}^{\rm LLL} \tilde{u}^{\rm LLL}_{\bs k}(\bs r)
    \mathcal B(\bs r)
    = \mathcal{N}_{\bs{k}} e^{i\bs{k}\cdot\bs{r}} \tilde{u}_{\bs{k}}(\bs{r}),
\end{align}
from which we can identify (up to an arbitrary transformation~\eqref{eq:si_holomorphic_transformation})
\begin{align}
    \mathcal{N}_{\bs{k}} = \mathfrak{M}_{\bs{k}} \mathcal{N}^{\rm LLL}_{\bs{k}}, \quad \text{and} \quad 
    \tilde{u}_{\bs{k}}(\bs{r}) = \mathcal B(\bs r) \tilde{u}^{\rm LLL}_{\bs k}(\bs r).
\end{align}
Such a band then has a Berry curvature distribution 
\begin{align}
    \Omega(\bs{k})
    = \nabla_{\bs k}^2 \log( \mathcal{N}_{\bs{k}}^{\rm LLL} \mathfrak{M}_{\bs{k}} ) 
    = - l_B^2 + \nabla_{\bs k}^2 \log( \mathfrak{M}_{\bs{k}} ) . 
\end{align}
For the Chern number of the band to be $\int d^2k \Omega(\bs{k})/2\pi = -1$, we see $\nabla_{\bs k}^2 \log( \mathfrak{M}_{\bs{k}} )$ must integrate to zero and instead controls the momentum dependence of the curvature. 

The form factor for such a band can be generally written as
\begin{align}
    \mathcal F(\bs k+\bs q,\bs k)
    =
    \mathcal F^{\rm LLL}(\bs k+\bs q,\bs k)
    \frac{
    \mathcal Z_{\bs q}(\bs k)
    }{
    \sqrt{\mathcal Z_{\bs 0}(\bs k)\mathcal Z_{\bs 0}(\lceil\bs k+\bs q\rceil)}
    },
    \label{eq:si_general_ideal_form_factor_factorization}
\end{align}
where we have introduced
\begin{align}
    \mathcal Z_{\bs q}(\bs k)
    \equiv
    \frac{
    \matrixel{\mathcal B\varphi^{\rm LLL}_{\lceil\bs k+\bs q\rceil}}{e^{i\bs q\cdot\bs r}}{\mathcal B\varphi^{\rm LLL}_{\bs k}}
    }{
    \mathcal F^{\rm LLL}(\bs k+\bs q,\bs k)
    }, \qquad \braket{\bs{r}}{\mathcal B\varphi^{\rm LLL}_{\bs k}} = \mathcal B(\bs{r})\varphi^{\rm LLL}_{\bs k}(\bs{r}),
    \label{eq:si_reduced_kernel_general}
\end{align}
such that 
\begin{align}
    \mathcal Z_{\bs 0}(\bs k)
    =
    \int d^2r\, |\mathcal B(\bs r)|^2 |\varphi^{\rm LLL}_{\bs k}(\bs r)|^2
    =
    \mathfrak M_{\bs k}^{-2}.
    \label{eq:si_reduced_kernel_diagonal_general}
\end{align}
The LLL factor carries the nontrivial reciprocal-lattice sewing phases of the $|C|=1$ bundle, while $\mathcal Z_{\bs q}(\bs k)$ is periodic in $\bs k$ and encodes the Berry curvature inhomogeneity.

\subsection{Explicit form of the normalization}

From the above general discussion of ideal bands, we see that the momentum dependence of the normalization factor plays a crucial role in determining the underlying band geometry. Given its importance, we thus start by providing an explicit evaluation of the AHC single-particle state normalization in Eq.~\eqref{eq:si_ahc_single-particle_basis_general}
\begin{align}
    \mathfrak{N}_{\bs{k},n}^{-2} = \int d^{2}r \sum_{\ell} \left| \varphi^{\mathrm{LLL}}_{\bs{k}}(\bs{r}) \chi_{n,\ell}(\bs{r}) \right|^2.
\end{align}
Using Eq.~\eqref{eq:si_rewrite_chi_with_magnetic_bloch_states}, the normalization constant can be rewritten as
\begin{align}
     \mathfrak{N}_{\bs{k},n}^{-2}  = \sum_{\ell} \frac{(n+\ell)!}{n!} \left(\frac{2 \gamma^2}{l_B^2}\right)^{\ell} \mathfrak{I}_{n+\ell}(\bs{k}), \label{eq:si_normalization_initial_form}
\end{align}
where 
\begin{align}
    \mathfrak{I}_{m}(\bs{k}) &= \int d^2r |\varphi^{\mathrm{LLL}}_{\bs{k}}(\bs{r})|^2  |\varphi^{\mathrm{mLL}}_{\bs{0}}(\bs{r})|^2.
\end{align}
Since $|\varphi^{\mathrm{mLL}}_{\bs{k}}(\bs{r}+\bs{a}_\alpha)|^2 = |\varphi^{\mathrm{mLL}}_{\bs{k}}(\bs{r})|^2$ is periodic, we can expand them in lattice harmonics as 
\begin{align}
    |\varphi^{\mathrm{mLL}}_{\bs{k}}(\bs{r})|^2 = \frac{1}{A_{\rm u.c.}} \sum_{\bs{G}}  \mathfrak{p}_{\bs{k}}^{(m)}(\bs{G}) e^{-i\bs{G}\cdot\bs{r}},
\end{align}
where 
\begin{align}
    \mathfrak{p}_{\bs{k}}^{(m)}(\bs{G})  
    = \int_{\mathrm{u.c.}} d^2r |\varphi^{\mathrm{mLL}}_{\bs{k}}(\bs{r})|^2 e^{i\bs{G}\cdot\bs{r}}
    = \bra{\varphi_{\bs{k}}^{\mathrm{mLL}}} e^{i\bs{G}\cdot\bs{r}} \ket{\varphi_{\bs{k}}^{\mathrm{mLL}}} = \bra{m}e^{-i l_B^2\bs{G}\wedge\bs{\pi}}\ket{m} \bra{\bs{k}} e^{i\bs{G}\cdot\bs{R}} \ket{\bs{k}}_{\mathrm{LL}}. 
\end{align}
In this last equality we split the magnetic Bloch state into its guiding center and cyclotron parts $\ket{\varphi^{mLL}_{\bs{k}}}=\ket{m}\otimes\ket{\bs{k}}_{\mathrm{LL}}$ and similarly split the position operator as $\bs{r}=\bs{R} - l_B^2 \wedge \bs{\pi}$. For the guiding center part, we can identify
\begin{align}
    e^{i\bs{G}\cdot\bs{R}} = T(-l_B^2 \wedge \bs{G}),
\end{align}
with $T(\bs{u})$ in Eq.~\eqref{eq:def_magnetic_translation_operator}, from which it directly follows that
\begin{align}
    \bra{\bs{k}} e^{i\bs{G}\cdot\bs{R}} \ket{\bs{k}}_{\mathrm{LL}} 
    = \bra{\bs{k}} T(-l_B^2 \wedge \bs{G}) \ket{\bs{k}}_{\mathrm{LL}} 
    = \exp(i l_B^2 \bs{G} \wedge \bs{k}).
\end{align}
The cyclotron matrix element can be evaluated using 
\begin{align}
    i l_B^2 \mathbf{G} \wedge \bs{\pi}=\alpha a^{\dagger}-\alpha^* a, \qquad \alpha = \frac{l_B}{\sqrt{2}} (G_x - i G_y),
\end{align}
such that
\begin{align}
     \bra{m}e^{-i l_B^2\bs{G}\wedge\bs{\pi}}\ket{m} 
     = e^{-|\alpha|^2 / 2} L_m\left(|\alpha|^2\right) 
     = e^{-l_B^2 |\bs{G}|^2 / 4} L_m\left( \frac{l_B^2 |\bs{G}|^2}{2} \right),
\end{align}
where $L_m(x)$ are Laguerre polynomials. Putting everything together, we have 
\begin{align}
    \mathfrak{p}_{\bs{k}}^{(m)}(\bs{G}) = e^{-l_B^2 |\bs{G}|^2 / 4} L_m\left( \frac{l_B^2 |\bs{G}|^2}{2} \right) \exp(i l_B^2 \bs{G} \wedge \bs{k}). 
\end{align}
Using Parseval's identity, this yields
\begin{align}
    \mathfrak{I}_{m}(\bs{k}) 
    = \frac{1}{A_{\mathrm{u.c.}}} \sum_{\bs{G}} \mathfrak{p}_{\bs{k}}^{(0)}(\bs{G})  \mathfrak{p}_{\bs{0}}^{(m)}(-\bs{G})
    = \frac{1}{A_{\mathrm{u.c.}}} \sum_{\bs{G}}  e^{-l_B^2 |\bs{G}|^2 / 2}  L_m\left( \frac{l_B^2 |\bs{G}|^2}{2} \right) \exp(i l_B^2 \bs{G} \wedge \bs{k}).
\end{align}
We can further use the map
\begin{align}
    l_B^2 \bs{G} \wedge \bs{k} = \bs{k}\cdot \bs{R}_{\bs{G}}, \qquad \bs{R}_{\bs{G}} = -l_B^2 \wedge \bs{G}, 
\end{align}
where $\bs{R}_{G}=n_1\bs{a}_1 + n_2 \bs{a}_2$ is a real space lattice vector, to rewrite our integral as
\begin{align}
    \mathfrak{I}_{m}(\bs{k}) 
    = \frac{1}{A_{\mathrm{u.c.}}} \sum_{\bs{R}}  \exp(-\frac{|\bs{R}|^2}{2 l_B^2})  L_m\left( \frac{ |\bs{R}|^2}{2 l_B^2} \right) \exp(i \bs{R} \cdot \bs{k}).
\end{align}
Replacing into~\eqref{eq:si_normalization_initial_form}, we finally get that
\begin{subequations} \label{eq:si_explicit_form_normalization_ahc_single-particle_basis}
\begin{align}
    \mathfrak{N}_{\bs{k},n}^{-2} = 
    \frac{1}{A_{\mathrm{u.c.}}} \sum_{\bs{R}}  \exp(i \bs{R} \cdot \bs{k}) \mathfrak{B}_{n}(\bs{R})  
\end{align}
with
\begin{align}
    \mathfrak{B}_{n}(\bs{R}) = \exp(-\frac{|\bs{R}|^2}{2 l_B^2}) \sum_{\ell} \frac{(n+\ell)!}{n!} \left(\frac{2 \gamma^2}{l_B^2}\right)^{\ell} L_{n+\ell}\left( \frac{ |\bs{R}|^2}{2 l_B^2} \right). 
\end{align}
\end{subequations}

\subsection{Momentum space representation}

The above AHC single-particle wavefunctions $\psi_{\bs{k},n,\ell}(\bs{r})$ admit a plane-wave expansion that was derived in Ref.~\cite{tan2025ideal}. For brevity, we will quote only the result here, but refer interested readers to Appendix~B.3 of Ref.~\cite{tan2025ideal} for the full derivation. The product of a magnetic Bloch state and the conjugate Landau-level factor can be written as
\begin{subequations}
\begin{align}
    \psi_{\bs{k}, n, \ell}(\bs{r})
    &= \mathfrak{N}_{\bs{k},n} \phi_{\bs{k}}^{\mathrm{LLL}}(\bs{r}) \chi_{n,\ell}^{(\bs{0})}(\bs r) \nonumber \\
    &= \mathfrak{N}_{\bs{k},n} (-2i\gamma\partial_{\bar{z}})^{\ell} \phi_{\bs{k}}^{\mathrm{LLL}}(\bs{r}) (\phi_{\bs{0}}^{\mathrm{nLL}}(\bs r))^* \nonumber \\
    &= \mathfrak{N}_{\bs{k},n} \frac{1}{\sqrt{A_{\rm u.c.}}}\sum_{\bs{g}} e^{i\omega_{\bs{g}}} Q_n(\bs{k} + \bs{g}) \exp(-\frac{l_B^2}{4}|\bs{k}+\bs{g}|^2) \exp(\frac{il_B^2}{2} \bs{k} \wedge\bs{g}) \frac{e^{i(\bs{k}+\bs{g}) \cdot \bs{r}}}{N_{\bs{k}+\bs{g}}} s_{\bs{k}+\bs{g},\ell}
\end{align}
where
\begin{align}
    Q_{n}(\bs{k}+\bs{g}) &= \left(-\frac{k+g}{\sqrt{2}}\right)^n \frac{l_B^n}{\sqrt{n!}}.
\end{align}
\end{subequations}
The above further implies an equivalent definition of the normalization factor as
\begin{align}
    \mathfrak{N}_{\bs{k},n}^{-2} &= \frac{1}{A_{\rm u.c.}} \sum_{\bs{g}} |Q_n(\bs{k} + \bs{g})|^2  \exp( - \frac{l_B^2}{2}|\bs{k} +\bs{g}|^2 ) N_{\bs{k}+\bs{g}}^{-2}. \label{eq:si_second_definition_normalization_ahc_single-particle_basis}
\end{align}

As is clear from comparing Eq.~\eqref{eq:si_ahc_single-particle_basis_general} and \eqref{eq:si_generic_C1_ideal_wavefunction}, the AHC single-particle basis describes an ideal $C=-1$ mini-band. One can also see this clearly by rewriting it as
\begin{align}
    \psi_{\bs{k},n,\ell} &= e^{i\bs{k}\cdot\bs{r}} \mathfrak{N}_{\bs{k},n} e^{-l_B^2|\bs{k}|^2/4} \tilde{u}_{\bs{k},n,\ell}(\bs{r}),
\end{align}
with the holomorphic component
\begin{align}
    \tilde{u}_{\bs{k},n,\ell}(\bs{r}) &= \sum_{\bs{g}} Q_{n}(\bs{k}+\bs{g}) e^{i\omega_{\bs{g}}} (\gamma (k+g))^\ell \exp(-\frac{l_B^2}{4} (|\bs{g}|^2 + 2 k \overline{g})) e^{i\bs{g}\cdot\bs{r}}.
\end{align}
Using Eq.~\eqref{eq:si_curvature_from_kahler_general}, the Berry curvature of the AHC band is then
\begin{align}
    \Omega(\bs{k}) 
    &= -l_B^2  + \nabla_{\bs k}^2 \log( \mathfrak{N}_{\bs{k}} )
    = - l_B^2 - \frac{1}{2} \nabla_{\bs k}^2 \log \left[ \sum_{\bs{R}}  \exp(-i \bs{R} \cdot \bs{k}) \exp(-\frac{|\bs{R}|^2}{2 l_B^2}) \sum_{\ell} \frac{(n+\ell)!}{n!} \left(\frac{2 \gamma^2}{l_B^2}\right)^{\ell} L_{n+\ell}\left( \frac{ |\bs{R}|^2}{2 l_B^2} \right)\right].
\end{align}

We can thus define the following AHC single-particle basis creation operator 
\begin{align}
    \psi^{\dagger}_{\bs{k},n} = \sum_{\bs{g}} a_{\bs{g}}(\bs{k}) c^\dagger_{\bs{k}+\bs{g}}, \qquad \psi_{\bs{k},n,\ell}(\bs{r}) = \bra{\bs{r},\ell} \psi^{\dagger}_{\bs{k},n} \ket{0} = \braket{\bs{r},\ell}{\psi_{\bs{k},n}}.
\end{align}
Based on our previous definition for the parent band creation operator~\eqref{eq:si_parent_band_single_particle_creation}, we then have 
\begin{align}
    \psi_{\bs{k},n,\ell}(\bs{r}) = \sum_{\bs{g}} a_{\bs{g},n}(\bs{k}) s_{\bs{k} + \bs{g},\ell}e^{i(\bs{k} + \bs{g}) \cdot \bs{r}}
\end{align}
and
\begin{align}
    a_{\bs{g},n}(\bs{k}) 
    &= \frac{1}{\sqrt{A_{\rm u.c.}}}\mathfrak{N}_{\bs{k},n} Q_{n}(\bs{k}+\bs{g}) e^{i \omega_{\bs{g}}} \exp \left(-\frac{l_B^2}{4}|\bs{k}+\bs{g}|^2\right) \exp \left(\frac{i l_B^2}{2} \bs{k} \wedge \bs{g}\right) N^{-1}_{\bs{k}+\bs{g}}.
\end{align}

\subsection{Form factor}

\subsubsection{General form}

The form factor for the AHC basis states is given by 
\begin{align}
    \mathcal{F}_{n}^{\rm AHC}(\bs{k} + \bs{q}, \bs{k}) = \matrixel{\psi_{\lceil \bs{k} + \bs{q} \rceil,n}}{\exp(i\bs{q} \cdot \bs{r})}{\psi_{\bs{k},n}}.
\end{align}
Based on our previous discussion in Sec.~\ref{subsec:si_general_ideal_bands}, one should be able to rewrite it following the general form~\eqref{eq:si_general_ideal_form_factor_factorization}. Using our previous expression, we have
\begin{align}
    \mathcal{F}_n^{\rm AHC}(\bs{k} + \bs{q}, \bs{k}) 
    &= \sum_{\bs{g}_1, \bs{g}_2} a_{\bs{g}_1,n}^*(\lceil \bs{k} + \bs{q} \rceil) a_{\bs{g}_2,n}(\bs{k}) \matrixel{ \lceil \bs{k} + \bs{q}\rceil }{\exp(i\bs{q} \cdot \bs{r})}{\bs{k}} \nonumber \\
    &= \sum_{\bs{g}_1, \bs{g}_2} a_{\bs{g}_1,n}^*(\lceil \bs{k} + \bs{q} \rceil) a_{\bs{g}_2,n}(\bs{k}) \matrixel{s_{\lceil \bs{k} + \bs{q} \rceil + \bs{g}_1}}{e^{i( - \bs{k} - \bs{q} + \bs{g}_{\bs{k},\bs{q}} - \bs{g}_1 + \bs{k} + \bs{g}_2 + \bs{q})\cdot\bs{r}}}{ s_{\bs{k} +\bs{g}_2} } \nonumber \\
    &= \sum_{\bs{g}} a_{\bs{g} + \bs{g}_{\bs{k},\bs{q}},n}^*(\lceil \bs{k} + \bs{q} \rceil) a_{\bs{g},n}(\bs{k}) F_{P}(\bs{k} + \bs{q} + \bs{g}, \bs{k} + \bs{g}) \nonumber \\
    &= \frac{1}{A_{\rm u.c.}}\sum_{\bs{g}} F_P(\bs{k} + \bs{q} + \bs{g}, \bs{k} + \bs{g}) 
    \mathfrak{N}_{\bs{k},n} Q_n(\bs{k}+\bs{g}) e^{i \omega_{\bs{g}}} \exp \left(-\frac{l_B^2}{4}|\bs{k}+\bs{g}|^2\right) \exp \left(\frac{i l_B^2}{2} \bs{k} \wedge \bs{g}\right) N^{-1}_{\bs{k}+\bs{g}} N^{-1}_{ \bs{k} + \bs{q}  + \bs{g}} \nonumber \\
    &\quad \times \mathfrak{N}_{\lceil \bs{k} + \bs{q} \rceil, n} Q_n^*( \bs{k} + \bs{q} + \bs{g}) e^{-i \omega_{\bs{g}+\bs{g}_{\bs{k},\bs{q}}}} \exp \left(-\frac{l_B^2}{4}| \bs{k} + \bs{q} + \bs{g}|^2\right) \exp \left(- \frac{i l_B^2}{2} ( \bs{k} + \bs{q} - \bs{g}_{\bs{k},\bs{q}} ) \wedge (\bs{g}_{\bs{k},\bs{q}}+ \bs{g})\right),
\end{align}
To simplify, we can use
\begin{align}
    |\bs{k} + \bs{q} + \bs{g}|^2 = |\bs{k} + \bs{g}|^2 + |\bs{q}|^2 + 2 (\bs{k} + \bs{g})\cdot\bs{q}
\end{align}
and
\begin{align}
    \omega_{\bs{g}} - \omega_{\bs{g}+\bs{g}_0} + \frac{l_B^2}{2} \bs{g}_0 \wedge \bs{g} 
    &= (n_1n_2)\pi - (n_1 + n_{0,1})(n_2 + n_{0,2})\pi + \frac{l_B^2}{2} \left( n_{0,1}\bs{b}_1 + n_{0,2}\bs{b}_2 \right) \wedge \left( n_{1}\bs{b}_1 + n_{2}\bs{b}_2 \right) \nonumber \\
    &= 2\pi n_1 n_{0,2} - n_{0,1}n_{0,2}\pi \nonumber \\
    &= \omega_{\bs{g}_0} ~(\mathrm{mod}~2\pi)
\end{align}
such that 
\begin{align}
    \mathcal{F}_{n}^{\rm AHC}(\bs{k} + \bs{q}, \bs{k}) 
    &= \frac{ \mathcal{R}_{n,\bs{q}}(\bs{k}) }{\sqrt{\mathcal{R}_{n,\bs{0}}(\bs{k}) \mathcal{R}_{n,\bs{0}}(\lceil \bs{k} + \bs{q} \rceil) }} e^{i\omega_{\bs{g}_{\bs{k},\bs{q}}}} \exp(-\frac{l_B^2}{4}\left(|\bs{q}|^2 - 2 i \bs{q}\wedge\bs{k} + 2i(\bs{k}+\bs{q})\wedge\bs{g}_{\bs{k},\bs{q}} \right)),
\end{align}
with
\begin{align}
    \mathcal{R}_{n,\bs{q}}(\bs{k})
    &= \frac{1}{A_{\rm u.c.}} \sum_{\bs{g}} \mathcal{P}_{n,\bs{q}}(\bs{k}+\bs{g})
\end{align}
and
\begin{align}
     \mathcal{P}_{n,\bs{q}}(\bs{k}+\bs{g}) &= F_P(\bs{k} + \bs{q} + \bs{g}, \bs{k} + \bs{g})  Q_n(\bs{k}+\bs{g})  Q_n^*( \bs{k} + \bs{q} + \bs{g}) N_{\bs{k}+\bs{g}}^{-1} N_{\bs{k}+\bs{q}+\bs{g}}^{-1} \nonumber \\
     &\qquad \times \exp(-\frac{l_B^2}{2}\left( |\bs{k}+\bs{g}|^2 +  (\bs{k}+\bs{g})\cdot\bs{q} + i \bs{q}\wedge(\bs{k}+\bs{g})\right)).
\end{align}
Using the LLL form factor derived in Sec.~\ref{si_subsubsec:LLL_form_factor}, the above AHC form factor can be rewritten as
\begin{align}
    \mathcal{F}_n^{\rm AHC}(\bs{k} + \bs{q}, \bs{k}) &=  \frac{ \mathcal{R}_{n,\bs{q}}(\bs{k}) }{\sqrt{\mathcal{R}_{n,\bs{0}}(\bs{k}) \mathcal{R}_{n,\bs{0}}(\lceil \bs{k} + \bs{q} \rceil) }}  \mathcal{F}^{\mathrm{LLL}}(\bs{k} + \bs{q}, \bs{k}) , \label{eq:si_dressed_lll_orm_factor}
\end{align}
We note here that 
\begin{align}
    \mathcal{R}_{n,\bs{0}}(\bs{k})
    &= \frac{1}{A_{\rm u.c.}} \sum_{\bs{g}}  |Q_n(\bs{k}+\bs{g})|^2 \exp(-\frac{l_B^2}{2}\left( |\bs{k}+\bs{g}|^2 \right)) N_{\bs{k}+\bs{g}}^{-2} 
\end{align}
which is equal to $\mathfrak{N}_{\bs{k},n}^{-2}$ given~\eqref{eq:si_second_definition_normalization_ahc_single-particle_basis}, as consistent with our convention in Eq.~\eqref{eq:si_reduced_kernel_diagonal_general}.

\subsubsection{Poisson resummation for $\mathcal{R}_{n,\mathbf{q}}(\mathbf{k})$} \label{app:approx_R_k}

We now derive a Poisson resummed form for $\mathcal{R}_{n,\mathbf{q}}(\bs{k})$. This form is particularly useful since the Poisson resummation can be truncated to yield some useful approximations for $\mathcal{R}_{n,\mathbf{q}}(\bs{k})$. Since $\mathcal{R}_{n,\mathbf{q}}(\bs{k})$ is periodic under $\bs{k}\to\bs{k}+\bs{g}$, we can reexpress it as 
\begin{align}
    \mathcal{R}_{n,\bs{q}}(\bs{k}) = \sum_{\bs{R}} b_{n,\bs{q}}(\bs{R}) e^{i\bs{k}\cdot\bs{R}},
\end{align}
where 
\begin{align}
    b_{n,\bs{q}}(\bs{R}) &= \frac{1}{\Omega_{\mathrm{BZ}}} \int_{\mathrm{BZ}} d^2p \mathcal{R}_{n,\bs{q}}(\bs{p}) e^{-i\bs{p}\cdot\bs{R}} = \frac{1}{\Omega_{\mathrm{BZ}}} \int_{\mathbb{R}^2} d^2p \mathcal{P}_{n,\bs{q}}(\bs{p}) e^{-i\bs{p}\cdot\bs{R}}
\end{align}

These Fourier coefficients are
\begin{align}
    b_{n,\bs{q}}(\bs{R}) &= \frac{1}{\Omega_{\rm BZ} A_{\rm u.c.}} \int_{\mathbb{R}^2} d^2p F_P(\bs{p}+ \bs{q}, \bs{p}) N_{\bs{p}}^{-1} N_{\bs{p}+\bs{q}}^{-1} Q_n^*( \bs{p} + \bs{q} ) Q_n(\bs{p})  \exp(-\frac{l_B^2}{2}\left( |\bs{p}|^2 +  \bs{p}\cdot\bs{q} + i \bs{q}\wedge \bs{p}\right) ) e^{-i\bs{p}\cdot\bs{R}} \nonumber \\
    &= \frac{1}{\Omega_{\rm BZ} A_{\rm u.c.}}\int_{\mathbb{R}^2} d^2p \left( \sum_{\ell=0}^{N_{\ell}-1}(\gamma^2 (\bar{p}+\bar{q})p)^\ell \right) \left( \left( (\bar{p}+\bar{q}) p \right)^n \frac{l_B^{2n}}{2^n n!} \right)  \exp(-\frac{l_B^2}{2}\left( |\bs{p}|^2 +  \bs{p}\cdot\bs{q} + i \bs{q}\wedge \bs{p}\right) ) e^{-i\bs{p}\cdot\bs{R}} \nonumber \\
    &= \frac{1}{\Omega_{\rm BZ} A_{\rm u.c.}}  \frac{ l_B^{2n} }{ 2^n n! } \sum_{\ell=0}^{N_{\ell}-1}\gamma^{2\ell} \int_{\mathbb{R}^2} d^2p \left((\bar{p}+\bar{q}) p \right)^{\ell + n} \exp(-\frac{l_B^2}{2}\left( |\bs{p}|^2 + p \bar{q} \right) ) e^{-i\bs{p}\cdot\bs{R}} \nonumber \\
    &=  \frac{ a^{n+1} }{ n! \pi  A_{\rm u.c.} } \sum_{\ell=0}^{N_{\ell}-1}\gamma^{2\ell} I_{\ell+n}(\bs{q}), \label{eq:general_expression_bnqR}
\end{align}
where we have defined $a\equiv l_B^2/2$ and
\begin{align}
    I_{m}(\bs{q}) \equiv \int_{\mathbb{R}^2} d^2p \left((\bar{p}+\bar{q}) p \right)^{m} \exp(-\frac{l_B^2}{2}\left( |\bs{p}|^2 +  p \bar{q}\right) ) e^{-i\bs{p}\cdot\bs{R}}.
\end{align}
Let's give some compact closed form expression for $b_{n,\bs{q}}(\bs{R})$ by evaluating $I_{m}(\bs{q})$. To do so, it is convenient to further introduce the shorthand notation
\begin{align}
    x_{\bs{R}} \equiv \frac{|\bs{R}|^2}{2l_B^2},
    \qquad
    \lambda_\gamma \equiv \frac{2\gamma^2}{l_B^2} = \frac{\gamma^2}{a}.
\end{align}
To evaluate $I_{m}(\bs{q})$, we introduce the generating parameter $t$
\begin{align}
    \sum_{m=0}^{\infty} \frac{t^m}{m!} I_m
    &= \int_{\mathbb{R}^2} d^2p \,
    \exp\!\left[-(a-t)\bar{p}p-(a-t)\bar{q}p-i\bs{p}\cdot\bs{R}\right] 
    = \frac{\pi}{a-t}
    \exp\!\left(
        \frac{i}{2}\bar{q}R
        -
        \frac{|R|^2}{4(a-t)}
    \right).
\end{align}
Given the generating function of Laguerre polynomials~\cite{thompson2011nist}
\begin{align}
    \frac{1}{1-s}\exp\!\left(-\frac{x\,s}{1-s}\right)
    =
    \sum_{m=0}^{\infty} L_m(x)s^m
\end{align}
and the identification $s=t/a$ and $x=x_{\bs{R}}$, this gives the exact closed form
\begin{align}
    I_m(\bs{q})
    =
    \pi m!\left(\frac{1}{a}\right)^{m+1}
    e^{\frac{i}{2}\bar{q}R-x_{\bs{R}}}
    L_m(x_{\bs{R}}).
\end{align}
Substituting back into Eq.~\eqref{eq:general_expression_bnqR} yields
\begin{align}
    b_{n,\bs{q}}(\bs{R})
    =
    \frac{e^{\frac{i}{2}\bar{q}R-x_{\bs{R}}}}{n! A_{\rm u.c.}}
    \sum_{\ell=0}^{N_{\ell}-1}
    \lambda_{\gamma}^{\ell}
    (\ell+n)!
    L_{\ell+n}(x_{\bs{R}}),
    \label{eq:bnq_closed_form}
\end{align}
such that 
\begin{align}
     \mathcal{R}_{n,\bs{q}}(\bs{k}) &= \sum_{\bs{R}}   \frac{e^{\frac{i}{2}\bar{q}R-x_{\bs{R}}}}{n! A_{\rm u.c.} } \sum_{\ell=0}^{N_{\ell}-1} \lambda_{\gamma}^{\ell} (\ell+n)! L_{\ell+n}(x_{\bs{R}}) e^{i\bs{k}\cdot\bs{R}}. \label{eq:Rnq_closed_form}
\end{align}
Denoting the average by
\begin{align}
    b_{n,\bs{q}}(\bs{0})
    =
    \overline{\mathcal{R}}_n
    \equiv
    \frac{1}{n! A_{\rm u.c.}}
    \sum_{\ell=0}^{N_{\ell}-1}
    \lambda_{\gamma}^{\ell}
    (\ell+n)!,
    \label{eq:Rbar_closed_form}
\end{align}
and the normalized shell amplitude by
\begin{align}
    \Lambda_n(x)
    \equiv
    e^{-x}
    \frac{\sum_{\ell=0}^{N_{\ell}-1}\lambda_{\gamma}^{\ell}(\ell+n)!L_{\ell+n}(x)}
    {\sum_{\ell=0}^{N_{\ell}-1}\lambda_{\gamma}^{\ell}(\ell+n)!},
    \label{eq:Lambda_n_def}
\end{align}
Equation~\eqref{eq:bnq_closed_form} can then be rewritten as
\begin{align}
    b_{n,\bs{q}}(\bs{R})
    =
    \overline{\mathcal{R}}_n\,
    \Lambda_n(x_{\bs{R}})\,
    e^{\frac{i}{2}\bar{q}R}.
\end{align}
We then conclude that
\begin{align}
    \mathcal{R}_{n,\bs{q}}(\bs{k}) 
    = \sum_{\bs{R}} \overline{\mathcal{R}}_n \Lambda_n\left(\frac{|\bs{R}|^2}{2l_B^2}\right) e^{\frac{i}{2}\bar{q}R} e^{i\bs{k}\cdot\bs{R}} \label{eq:si_poisson_resummed_R}.
\end{align}

We finally note that 
\begin{align}
    \mathcal{R}_{n,\bs{0}}(\bs{k}) &= \frac{1}{A_{\rm u..c.}} \sum_{\bs{R}} e^{i\bs{k}\cdot\bs{R}} e^{-|\bs{R}|^2/(2l_B^2)} \sum_{\ell=0}^{N_{\ell}-1} \frac{(\ell + n)!}{n!} \left( \frac{2\gamma^2}{l_B^2} \right)^{\ell} L_{\ell+n}\left(\frac{|\bs{R}|^2}{2l_B^2}\right)  
    = \mathfrak{N}_{\bs{k},n}^{-2},
\end{align}
which is once again consistent with~\eqref{eq:si_explicit_form_normalization_ahc_single-particle_basis}, given our identification~\eqref{eq:si_reduced_kernel_diagonal_general}

\subsubsection{$0^{\mathrm{th}}$ harmonic approximation} \label{app:zeroth_harmonic_approx_R}

If we consider the $0^{\mathrm{th}}$ harmonic approximation by restricting the above Poisson resummation to the leading-order term $b_{n,\bs{q}}(\bs{0})$, the approximation gives
\begin{align}
    \mathcal{R}_{n,\bs{q}}(\bs{k}) 
    \approx \overline{\mathcal{R}_{n,\bs{q}}} =  \frac{1}{\Omega_{\mathrm{BZ}}} \int_{\mathrm{BZ}}d^2k \mathcal{R}_{n,\bs{q}}(\bs{k}) 
    = \frac{1}{A_{\rm u.c.} \Omega_{\mathrm{BZ}}} \int d^2p \mathcal{P}_{n,\bs{q}}(\bs{p})  
    = \frac{1}{n! A_{\rm u.c.} } \sum_{\ell=0}^{N_{\ell}-1} \lambda_{\gamma}^{\ell}(\ell+n)! 
    = \overline{\mathcal{R}_{n,\bs{0}}}. \label{eq:si_poisson_resummed_R_0_harmonics}
\end{align}
Within such approximations, the AHC form factor is approximated by the LLL one
\begin{align}
    \mathcal{F}(\bs{k} + \bs{q}, \bs{k}) 
    &\approx \left( \frac{\overline{\mathcal{R}_{n,\bs{q}}}}{\overline{\mathcal{R}_{n,\bs{0}}}} \right) \mathcal{F}^{\mathrm{LLL}}(\bs{k} + \bs{q}, \bs{k}) 
    = \mathcal{F}^{\mathrm{LLL}}(\bs{k} + \bs{q}, \bs{k}). \label{eq:si_approx_form_factor_zeroth_harmonic}
\end{align}

Using $2\pi l_B^2=A_{\mathrm{u.c.}}$, the Gaussian suppression of a lattice vector $\bs{R}$ is $\exp(-x_{\bs{R}}) = \exp(-\pi |\bs{R}|^2/A_{\mathrm{u.c.}})$. This small parameter controls the truncation of the Poisson resummation. It is already on the order of $2.7\cdot 10^{-2}$ for the smallest lattice vectors on the triangular lattice. The zeroth-harmonic truncation is therefore a reasonable baseline whenever the Laguerre prefactor does not overcome the Gaussian decay.

\section{AHC many-body state and quantum Hall correspondence} \label{si_sec:ahc_many_body_and_ahc_correspondence}

Given Eqs.~\eqref{eq:si_expansion_qh_first_quantized} and~\eqref{eq:si_expansion_ahc_first_quantized}, it directly follows that the second-quantized form of the QH wavefunction
\begin{equation}
    \ket{\Phi^{\rm QH}}
    =\sum_{\{\bs k\}}C_{\{\bs k\}} \ket{\{ \varphi_{\bs k}^{\mathrm{LLL}} \}}, \qquad \ket{\{ \varphi_{\bs k}^{\mathrm{LLL}} \}} = \prod_{i=1}^{N_e}c_{\bs k_i}^{\rm LLL\dagger}\ket{0}
\end{equation}
is related to the AHC wavefunction by
\begin{equation}
    \ket{\Psi^{\rm AHC}}
    =  \sum_{\{\bs k\}} \frac{ C_{\{\bs k\}}}{\mathfrak{N}_{\{\bs{k}\},n}}
      \ket{\{\psi_{\bs k,n}\}}, \qquad \ket{\{\psi_{\bs k,n }\}} = \prod_{i=1}^{N_e}\psi_{\bs k_i,n}^\dagger\ket{0},
\end{equation}
where $\mathfrak{N}_{\{\bs{k}\},n} = \prod_{\bs{k}\in\{\bs{k}\}} \mathfrak{N}_{\bs{k},n}$.

\subsection{Bounds on the momentum dependence of the normalization}

Since the coefficients in the second-quantized AHC wavefunctions differ from the corresponding QH wavefunction by products of the single-particle normalization factors, we now estimate the momentum dependence of $\mathfrak N_{\bs k,n}$. If this dependence is weak, then $\mathfrak N_{\{\bs k\},n}=\prod_{\bs k\in\{\bs k\}}\mathfrak N_{\bs k,n} $ is approximately an overall constant at fixed particle number, and the AHC and QH Slater coefficients are simply related. Given this correspondence, one may be able to relate AHC correlations to known QH ones.

To bound the momentum dependence of $\mathfrak N_{\bs k,n}$, we rewrite the explicit expression~\eqref{eq:si_explicit_form_normalization_ahc_single-particle_basis} as
\begin{align}
    \mathfrak{N}_{\bs{k},n}^{-2} 
    = \frac{\mathfrak{B}_{n}(\bs{0})}{A_{\mathrm{u.c.}}} \left( 1 + \sum_{\bs{R}\ne \bs{0}} \frac{\mathfrak{B}_{n}(\bs{R})}{\mathfrak{B}_{n}(\bs{0})} e^{-i\bs{k}\cdot\bs{R}}\right) 
    = \overline{\mathfrak{N}}_{n}^{-2} \left( 1 + \mathfrak{D}_{n}(\bs{k}) \right), \qquad  \overline{\mathfrak{N}}_{n} = \sqrt{\frac{A_{\mathrm{u.c.}} }{\mathfrak{B}_{n}(\bs{0})}}. \label{eq:si_normalization_after_simplification}
\end{align}
We see that the momentum dependence, given by 
\begin{align}
    \mathfrak{D}_{n}(\bs{k}) = \sum_{\bs{R}\ne \bs{0}} \frac{\mathfrak{B}_{n}(\bs{R})}{\mathfrak{B}_{n}(\bs{0})} e^{-i\bs{k}\cdot\bs{R}},
\end{align}
is suppressed by factors of $\mathfrak{B}_{n}(\bs{R})/\mathfrak{B}_{n}(\bs{0})$, which decay rapidly because of the exponential factor $\exp(- \pi |\bs{R}|^2/A_{\mathrm{u.c.}})$.

Let us try to make these considerations more robust and formal. We first introduce once again $\lambda_{\gamma}\equiv 2\gamma^2/l_B^2$ and $x_{\bs R}\equiv |\bs R|^2/(2 l_B^2)$ which we group as 
\begin{align}
    \mathcal Q\equiv
    \{x_{\bs R}\,|\,\bs R\in\Lambda,\ \bs R\neq\bs0\},
\end{align}
where $\Lambda$ is the direct lattice. For each $Q\in\mathcal{Q}$, we can then introduce a shell of vectors sharing the same magnitude
\begin{align}
    \mathcal S_Q
    \equiv
    \{\bs R\in\Lambda\setminus\{\bs0\} | x_{\bs R}=Q\}.
\end{align}
The magnitude of the momentum-dependent component of the normalization is then bounded by
\begin{align}
    \left|\mathfrak{D}_{n}(\bs{k})\right|
    \le \mathfrak{E}_n^{\infty} =  \sum_{Q\in\mathcal{Q}}  \mathfrak{E}_n(Q)
    = \sum_{\bs{R}\ne \bs{0}} e^{-x_{\bs{R}}} \frac{ \left|\sum_{\ell=0}^{N_{\ell}-1} (n+\ell)! \lambda_\gamma^{\ell}     L_{n+\ell}\left( x_{\bs{R}} \right) \right| }{ \left| \sum_{\ell=0}^{N_{\ell}-1} (n+\ell)! \lambda_\gamma^{\ell} \right|}, \label{eq:si_bound_norm_momentum_dep}
\end{align}
with 
\begin{align}
    \mathfrak E_n(Q)
    =
    \sum_{\bs R\in\mathcal S_Q}
    \left|
    \frac{\mathfrak B_n(\bs R)}{\mathfrak B_n(\bs0)}
    \right|.
\end{align}
This bound and the definition~\eqref{eq:si_normalization_after_simplification} directly imply that
\begin{align}
    (1 + \mathfrak{E}_{n}^{\infty})^{-1/2} \le \frac{\mathfrak{N}_{\bs{k},n}}{ \overline{\mathfrak{N}}_{n} } \le (1 - \mathfrak{E}_{n}^{\infty})^{-1/2}, 
\end{align}
provided that $\mathfrak E_n^\infty<1$.

\begin{figure}
    \centering
    \includegraphics[width=1.00\linewidth]{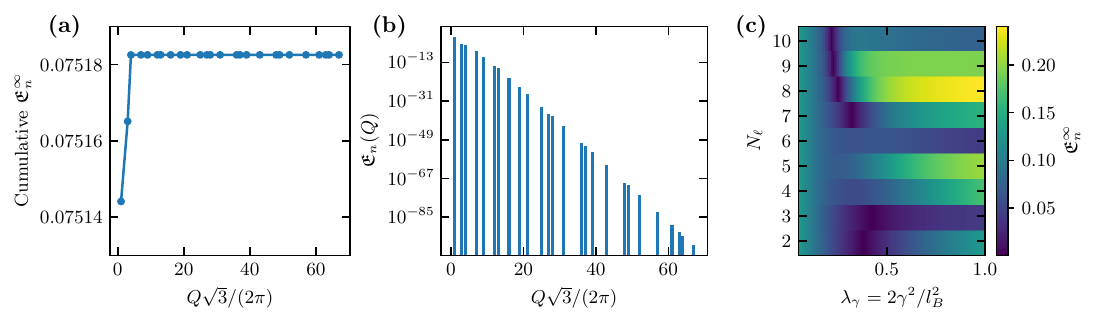}\vspace{-3mm}
    \caption{ Bound on the momentum dependence of the normalization prefactor. (a) Cumulative value of $\mathfrak E_n^\infty$ as direct-lattice shells are included, for $\lambda_\gamma=0.2$, $N_\ell=5$, and $n=0$. (b) Individual shell contributions $\mathfrak E_n(Q)$ for the same parameters as (a), shown on a logarithmic scale. (c) $\mathfrak E_n^\infty$ as a function of $\lambda_\gamma=2\gamma^2/l_B^2$ and $N_\ell$ for $n=0$.
}
\label{si_fig:bound_normalization}
\end{figure}

We evaluate the magnitude of this bound by specializing to the triangular lattice. In this case, we have
\begin{align}
    x_{\bs R}
    =
    \frac{2\pi}{\sqrt3}
    \left(m_1^2+m_1 m_2 + m_2^2\right).
\end{align}
with $m_1,m_2\in\mathbb{Z}$ and
\begin{align}
    \mathcal S_Q
    =
    \left\{
    m_1\bs{a}_1 + m_2 \bs{a}_2 \Big| (m_1,m_2)\in\mathbb Z^2\setminus(0,0) \quad \mathrm{and}\quad
    m_1^2 + m_1 m_2 + m_2^2=Q \sqrt{3}/(2\pi)
    \right\}.
\end{align}
We show in Fig.~\ref{si_fig:bound_normalization}(a) the evolution of $\mathfrak{E}^{\infty}_n$ as more shells are included for some relevant parameters. The bound converges very quickly due to the rapid exponential decay of the contributions from different shells (see Fig.~\ref{si_fig:bound_normalization}(b)) and ultimately saturates to a small value. Fig.~\ref{si_fig:bound_normalization}(c) shows a more detailed scan of $\mathfrak{E}^{\infty}_n$ as a function of $N_\ell$ and $\lambda_\gamma$. $\mathfrak{E}^{\infty}_n$ remains small, especially in the range $\lambda_\gamma<0.5$. For the density range discussed in the main text for R5G, we note that $\gamma/l_B$ is on the order of $0.3$, yielding $\lambda_\gamma\approx 0.18$.

\subsection{Matching expectation values}

Given that the momentum dependence of the normalization constant $\mathfrak{N}_{\bs{k},n}$ is parametrically small ($\mathfrak{N}_{\bs{k},n} \approx \overline{\mathfrak{N}}_{n}$), we hereafter assume that the AHC and QH wavefunctions are in direct correspondence
\begin{equation}
    \ket{\Phi^{\rm QH}}
    =\sum_{\{\bs k\}}C_{\{\bs k\}} \ket{\{ \varphi_{\bs k}^{\mathrm{LLL}} \}}
    \longleftrightarrow
     \ket{\Psi^{\rm AHC}}
    = \sum_{\{\bs k\}}  C_{\{\bs k\}}
      \ket{\{\psi_{\bs k,n}\}},
\end{equation}
Equivalently, there is a natural isometry on the band-projected Fock spaces under which
\begin{equation}
    c_{\bs k}^{\rm LLL\dagger}
    \longleftrightarrow
    \psi_{\bs k,n}^\dagger,
\end{equation}
which implies any correlator constructed only from these creation and annihilation operators is preserved by the map. In particular,
\begin{equation}
    \left\langle
    c_{\bs k_1}^{\rm LLL\dagger}c_{\bs k_2}^{\rm LLL}
    \right\rangle_{\rm QH}
    =
    \left\langle
    \psi_{\bs k_1,n}^\dagger\psi_{\bs k_2,n}
    \right\rangle_{\rm AHC},
    \label{eq:ahc_qh_mapping_one_body_corrected}
\end{equation}
and similarly
\begin{align}
    \left\langle
    c_{\bs k_1}^{\rm LLL\dagger}
    c_{\bs k_2}^{\rm LLL\dagger}
    c_{\lceil\bs k_2+\bs q_1\rceil}^{\rm LLL}
    c_{\lceil\bs k_1-\bs q_1\rceil}^{\rm LLL}
    \right\rangle_{\rm QH}
    =
    \left\langle
    \psi_{\bs k_1,n}^\dagger
    \psi_{\bs k_2,n}^\dagger
    \psi_{\lceil\bs k_2+\bs q_1\rceil, n}
    \psi_{\lceil\bs k_1-\bs q_1\rceil, n}
    \right\rangle_{\rm AHC} .
\end{align}
Here $\expval{\ldots}_{\rm QH}$ denotes an average with respect to the QH wavefunction~\eqref{eq:si_expansion_qh_first_quantized} and $\expval{\ldots}_{\rm AHC}$ with respect to~\eqref{eq:si_expansion_ahc_first_quantized}. For the IAHC, the corresponding QH reference state is the filled LLL. For the FAHC at filling $\nu$ per unit cell, the corresponding QH state is a representative of the ground state multiplet (or density matrix) at filling $\nu$ of the LLL (see Sec.~\ref{si_sec:review_qh_basics}). Such a correspondence was already pointed out in Appendix D of Ref.~\cite{tan2024wavefunction}. 

To illustrate how the correspondence~\eqref{eq:ahc_qh_mapping_one_body_corrected} can be used, let's consider the diagonal one-body expectation values that enter the kinetic and periodic potential energies. We have
\begin{align}
    \expval{\psi_{\bs{k}}^\dagger \psi_{\bs{k}}}_{\rm AHC}
    =
    \expval{ c_{\bs k}^{\rm LLL\dagger}c_{\bs k}^{\rm LLL} }_{\rm QH}.
\end{align}
Following the translation-symmetry argument in Sec.~\ref{si_subsubsec:one_body_density_matrix}, one may choose a representative QH state within the center-of-mass multiplet, or use the translation-invariant ground state density matrix, for which the momentum space occupation is uniform
\begin{align}
    \expval{\psi_{\bs{k}}^\dagger \psi_{\bs{k}}}_{\rm AHC}
    =
    \expval{ c_{\bs k}^{\rm LLL\dagger}c_{\bs k}^{\rm LLL} }_{\rm QH}
    = \nu . \label{eq:si_qh_ahc_uniform_occupation}
\end{align}
We note that although we neglect the momentum dependence of the normalization in evaluating matrix elements, we still preserve all momentum dependence that arises from, e.g., the ratio of $\mathcal{R}_{n,\bs{q}}(\bs{k})$ in the form factor.

\section{Energetics of integer and fractional anomalous Hall crystals} \label{si_sec:energetics_of_ahcs}

We finally move on to our central results: an explicit evaluation of the kinetic, interaction, and periodic potential energy of our IAHC and FAHC candidate wavefunctions. 

\subsection{Kinetic energy}

The kinetic energy can be written in the AHC basis as 
\begin{align}
    \expval{(-\nabla^{2})^{\alpha}} 
    &= \frac{1}{N_{e}} \sum_{\bs{k}_1,\bs{k}_2} \matrixel{\psi_{\bs{k}_1,n}}{(-\nabla^{2})^{\alpha}}{\psi_{\bs{k}_2,n}} \expval{ \psi^\dagger_{\bs{k}_1,n} \psi_{\bs{k}_2,n}},
\end{align}
where the matrix element is
\begin{align}
    \langle \psi_{\bs{k}_1,n} | (-\nabla^{2})^{\alpha} | \psi_{\bs{k}_2,n} \rangle
    &= \delta_{\bs{k}_1,\bs{k}_2} \frac{\sum_{\bs{g}}  \left| Q_n(\bs{k}+\bs{g}) \right|^2 N_{\bs{k}_1+\bs{g}}^{-2} e^{-\frac{l_B^2}{2} |\bs{k}_1 + \bs{g}|^2} |\bs{k}_1 + \bs{g}|^{2\alpha}}{ \sum_{\bs{g}} \left| Q_n(\bs{k}+\bs{g}) \right|^2 N_{\bs{k}_1+\bs{g}}^{-2} e^{-\frac{l_B^2}{2} |\bs{k}_1 + \bs{g}|^2} } \nonumber \\
    &= \delta_{\bs{k}_1,\bs{k}_2} \frac{\sum_{\bs{g}} N_{\bs{k}_1+\bs{g}}^{-2} e^{-\frac{l_B^2}{2} |\bs{k}_1 + \bs{g}|^2} |\bs{k}_1 + \bs{g}|^{2\alpha + 2 n}}{ \sum_{\bs{g}} |\bs{k}+\bs{g}|^{2n} N_{\bs{k}_1+\bs{g}}^{-2} e^{-\frac{l_B^2}{2} |\bs{k}_1 + \bs{g}|^2} } \nonumber \\
    &= \delta_{\bs{k}_1,\bs{k}_2} \frac{ \sum_{\bs{g}} \mathcal{P}_{n,\bs{0}}(\bs{k}_1+\bs{g}) |\bs{k}_1+\bs{g}|^{2\alpha}}{ \sum_{\bs{g}} \mathcal{P}_{n,\bs{0}}(\bs{k}_1+\bs{g}) }.
\end{align}
We can next evaluate the momentum space occupation expectation value by using Eq.~\eqref{eq:si_qh_ahc_uniform_occupation}, which yields
\begin{align}
    \expval{ \psi^{\dagger}_{\bs{k},n} \psi_{\bs{k},n} } = \nu.
\end{align}
The kinetic energy is then 
\begin{align}
    \expval{(-\nabla^{2})^{\alpha}}
    &= \frac{1}{A_{\rm u.c.} N_{\mathrm{u.c.}}} \sum_{\bs{k}} 
    \frac{ \sum_{\bs{g}} \mathcal{P}_{n,\bs{0}}(\bs{k}+\bs{g}) |\bs{k}+\bs{g}|^{2\alpha}}{\mathcal{R}_{n,\bs{0}}(\bs{k}) }, \label{eq:si_full_kinetic_energy}
\end{align}
where we have used that $N_{\mathrm{u.c.}} = N_{e}/\nu$. This formula can be directly numerically evaluated. 

Although not entirely necessary, we find it convenient to use our Poisson-resummed form of $\mathcal{R}_{n,\bs{0}}(\bs{k})$ in Eq.~\eqref{eq:si_poisson_resummed_R} once again to gain further analytical insights. In particular, if we once again truncate the sum in the denominator to the zeroth harmonics as $\mathcal{R}_{n,\bs{0}}(\bs{k}) \longrightarrow \overline{\mathcal{R}_{n,\bs{0}}}$, the kinetic energy becomes 
\begin{align}
    \expval{(-\nabla^{2})^{\alpha}} \approx   \frac{\int d^2q \mathcal{P}_{n,\bs{0}}(\bs{q}) |\bs{q}|^{2\alpha}}{ \int d^2q \mathcal{P}_{n,\bs{0}}(\bs{q}) }.
\end{align}
This amounts to approximating the parent band momentum space occupation by $n(\bs{q}) \approx \mathcal{P}_{n,\bs{0}}(\bs{q}) \Big/ \int d^2q \mathcal{P}_{n,\bs{0}}(\bs{q}) $, in which case case the kinetic energy is $E_{\text{Kin}} = \int d^2q \mathcal{E}(\bs{q}) n(\bs{q})$ for a dispersion $\mathcal{E}(\bs{q})$. Directly evaluating the integral in the numerator, we find the approximate kinetic energy to be
\begin{align}
    \expval{(-\nabla^2)^{\alpha}} 
    &\approx l_B^{-2\alpha}2^{\alpha} \frac{ \sum_{\ell=0}^{N_{\ell}-1} (\gamma/l_B)^{2(\ell+n)} 2^{\ell + n} (\ell + n + \alpha)! }{ \sum_{\ell=0}^{N_{\ell}-1} (\gamma/l_B)^{2(\ell+n)} 2^{\ell+n} (\ell+n)! } 
    = (4 \pi n_e / \nu)^{\alpha} \frac{ \sum_{\ell=0}^{N_{\ell}-1} \gamma^{2\ell} (4 \pi n_e / \nu)^{\ell} (\ell + n + \alpha)! }{ \sum_{\ell=0}^{N_{\ell}-1} \gamma^{2\ell}  (4 \pi n_e / \nu)^{\ell} (\ell+n)!}. \label{eq:si_full_kinetic_energy_zeroth_order}
\end{align}

We can use this final form to study the dominant scaling in various limits. When $\gamma \ll l_B$, the $\ell = 0$ term dominates both sums. Expanding the ratio to first order in $\gamma^2/l_B^2$,
\begin{align}
    \expval{(-\nabla^2)^\alpha} \approx \frac{2^\alpha}{l_B^{2\alpha}} \frac{(n+\alpha)!}{n!} \cdot \frac{1 + \frac{2\gamma^2}{l_B^2}(n+1+\alpha) + \mathcal{O}(\gamma^4/l_B^4)}{1 + \frac{2\gamma^2}{l_B^2}(n+1) + \mathcal{O}(\gamma^4/l_B^4)} \approx \frac{2^{2\alpha} \pi^\alpha (1/\nu)^\alpha n_e^\alpha\, (n+\alpha)!}{n!} \left( 1 + \frac{2\alpha \gamma^2}{l_B^2} + \mathcal{O}\!\left(\frac{\gamma^4}{l_B^4}\right) \right). \label{eq:si_full_kinetic_energy_small_gamma}
\end{align}
In the opposite limit $\gamma \gg l_B$, the $\ell = N_{\ell} - 1$ term dominates and we obtain
\begin{align}
    \expval{(-\nabla^2)^\alpha} 
    \approx \frac{2^\alpha (N_{\ell} + n + \alpha - 1)!}{l_B^{2\alpha}\,(N_{\ell} + n - 1)!} 
    = \frac{2^{2\alpha} \pi^\alpha (1 / \nu)^\alpha n_e^\alpha\,(N_{\ell} + n + \alpha - 1)!}{(N_{\ell} + n - 1)!}. \label{eq:si_full_kinetic_energy_large_gamma}
\end{align}
In both limits, we find $\expval{(-\nabla^2)^\alpha} \propto (1/\nu)^\alpha$, as one would have \emph{a priori} expected. In the $\gamma \gg l_B$ limit, we also see that the $\gamma$ dependence is almost entirely suppressed, whereas a quadratic increase with $\gamma$ is expected in the $\gamma \ll l_B$ regime for all values of $\nu$, $\alpha$, $n$, and $N_\ell$.

\subsection{Interaction energy}
\label{si_subsec:fahc_interaction_energy}

We now evaluate the interaction energy of the IAHC and FAHC ansätze. We consider a density-density interaction
\begin{align}
    \hat{\mathcal H}^{\rm int}
    =
    \frac{1}{2A}
    \sum_{\bs q}
    V(\bs q):
    \hat{\tilde\rho}^{\rm AHC}_{\bs q}
    \hat{\tilde\rho}^{\rm AHC}_{-\bs q}: ,
\end{align}
where $A=N_{\rm u.c.}A_{\rm u.c.}$ and the projected density operator in the AHC basis is
\begin{align}
    \hat{\tilde\rho}^{\rm AHC}_{\bs q}
    =
    \sum_{\bs k\in{\rm BZ}}
    \mathcal F^{\rm AHC}_{n}(\bs k+\bs q,\bs k)\,
    \psi^\dagger_{\lceil\bs k+\bs q\rceil,n}
    \psi_{\bs k,n}.
    \label{eq:si_AHC_projected_density_general}
\end{align}
We will always assume that $V(\bs{q})$ describes an isotropic screened Coulomb interaction, such that we do not need to worry about proper background subtraction to obtain convergent results. Using the factorization of the AHC form factor~\eqref{eq:si_dressed_lll_orm_factor} into the LLL form factor and the AHC dressing factor,
\begin{align}
    \mathcal F^{\rm AHC}_{n}(\bs k+\bs q,\bs k)
    =
    \mathcal F^{\rm LLL}(\bs k+\bs q,\bs k)
    \mathcal W_{n,\bs q}(\bs k),
\end{align}
with
\begin{align}
    \mathcal W_{n,\bs q}(\bs k)
    =
    \frac{\mathcal R_{n,\bs q}(\bs k)}
    {
    \sqrt{
    \mathcal R_{n,\bs0}(\bs k)
    \mathcal R_{n,\bs0}(\lceil\bs k+\bs q\rceil)}
    },
    \label{eq:si_W_factor_interaction}
\end{align}
we can write
\begin{align}
    \hat{\tilde\rho}^{\rm AHC}_{\bs q}
    =
    \sum_{\bs k\in{\rm BZ}}
    \mathcal W_{n,\bs q}(\bs k)\,
    \hat{J}_{\mathbf{q}}^{(n)}(\mathbf{k}),
\end{align}
where
\begin{align}
    \hat{J}_{\mathbf{q}}^{(n)}(\mathbf{k})
    \equiv
    \mathcal F^{\rm LLL}(\bs k+\bs q,\bs k)
    \psi^\dagger_{\lceil\bs k+\bs q\rceil,n} \psi_{\bs k,n}, 
    \qquad  \hat{\varrho}_{\mathbf{q}}^{(n)} = \sum_{\bs{k}\in \mathrm{BZ}} \hat{J}_{\mathbf{q}}^{(n)}(\mathbf{k}).
\end{align}

Under the QH-AHC correspondence, the occupation-space correlation functions of $\psi_{\bs k,n}$ are identified with those of the corresponding LLL operators. Thus, the nontrivial difference between the AHC and the reference QH state enters through the single-particle dressing factor $\mathcal W_{n,\bs q}(\bs k)$. We then directly note that in the $0^{\mathrm{th}}$ harmonic approximation for $\mathcal{R}_{n,\bs{q}}$ (see Sec.~\ref{app:zeroth_harmonic_approx_R}) we get $\mathcal{W}_{n,\bs{q}}=1$ such that the interaction energy would be the exact same as for the corresponding QH state. Since we have argued that corrections to the $0^{th}$ harmonic approximation are typically small, we expect the AHC interaction energy to be very close to that of the associated QH state and, accordingly, almost independent of $\gamma$ or the underlying lattice geometry. 

Since $\mathcal W_{n,\bs q}(\bs k)$ is periodic in the crystal momentum, we expand it in direct-lattice harmonics
\begin{align}
    \mathcal W_{n,\bs q}(\bs k)
    =
    \sum_{\bs R}
    w_{n, \bs R}(\bs q)e^{i\bs k\cdot\bs R}, 
    \qquad
    w_{n, \bs R}(\bs q)
    =
    \frac{1}{\Omega_{\rm BZ}}
    \int_{\rm BZ}d^2k\,
    e^{-i\bs k\cdot\bs R}
    \mathcal W_{n,\bs q}(\bs k).   \label{eq:si_W_direct_lattice_expansion}
\end{align}
Each direct-lattice harmonic can be absorbed into a shift of the physical density momentum. For each direct-lattice vector $\bs R$, define the reciprocal vector $\bs G_{\bs R}$ by
\begin{align}
    l_B^2\,\bs G_{\bs R}\wedge\bs k
    =
    \bs k\cdot\bs R,
    \qquad
    \bs G_{\bs R}
    = \wedge \bs{R} / l_B^2 .
    \label{eq:si_GR_definition}
\end{align}
Since $A_{\rm u.c.}=2\pi l_B^2$, $\bs G_{\bs R}$ is a reciprocal lattice vector. The LLL magnetic-Bloch form factor then implies the identity
\begin{align}
    \sum_{\bs k\in{\rm BZ}}
    e^{i\bs k\cdot\bs R}
    \hat{J}_{\mathbf{q}}^{(n)}(\mathbf{k})
    =
    \alpha_{\bs R}(\bs q)\,
    \hat{\varrho}^{(n)}_{\bs q+\bs G_{\bs R}},
    \label{eq:si_LLL_link_harmonic_identity}
\end{align}
where the magnitude  of $\alpha_{\bs R}(\bs q)$ is
\begin{align}
    |\alpha_{\bs R}(\bs q)|^2
    =
    \exp\left[
    \frac{l_B^2}{2}
    \left(
        |\bs q+\bs G_{\bs R}|^2-|\bs q|^2
    \right)
    \right].
    \label{eq:si_alpha_R_abs2}
\end{align}
Combining Eqs.~\eqref{eq:si_W_direct_lattice_expansion} and \eqref{eq:si_LLL_link_harmonic_identity}, the AHC density operator becomes a sum of ordinary LLL projected densities at shifted physical momenta
\begin{align}
    \hat{\tilde\rho}^{\rm AHC}_{\bs q}
    =
    \sum_{\bs R}
    a_{n,\bs R}(\bs q)
    \hat{\varrho}^{(n)}_{\bs q+\bs G_{\bs R}},
    \qquad
    a_{n,\bs R}(\bs q)
    \equiv
    w_{n,\bs R}(\bs q)\alpha_{\bs R}(\bs q).
    \label{eq:si_AHC_density_shifted_LLL_densities}
\end{align}

We now split the AHC density into its expectation value and fluctuations,
\begin{align}
    \hat{\tilde\rho}^{\rm AHC}_{\bs q}
    =
    \bar\rho^{\rm AHC}_{\bs q}
    +
    \delta\hat{\tilde\rho}^{\rm AHC}_{\bs q},
    \qquad
    \bar\rho^{\rm AHC}_{\bs q}
    \equiv
    \left\langle
    \hat{\tilde\rho}^{\rm AHC}_{\bs q}
    \right\rangle .
\end{align}
The corresponding energy separates into a coherent Bragg contribution and a connected fluctuation contribution.

For the coherent term, we compute matrix expectation values using the AHC-QH correspondence and assuming the diagonal density matrix (see Sec.~\ref{si_subsubsec:one_body_density_matrix}), in which case $\langle\psi^\dagger_{\bs k,n}\psi_{\bs k',n}\rangle_{\mathrm{AHC}} = \langle c^\dagger_{\bs k,n} c_{\bs k',n}\rangle_{\mathrm{QH}} = \nu \delta_{\bs{k},\bs{k}'}$. Then the expectation value of the density is nonzero only at reciprocal lattice vectors $\bs q=\bs G$, in which case $\lceil\bs k+\bs G\rceil=\bs k$. For such reciprocal lattice vectors,
\begin{align}
    \bar\rho^{\rm AHC}_{\bs G}
    =
    N_e e^{i\omega_{\bs G}}
    e^{-l_B^2|\bs G|^2/4}
    \rho_n(\bs G),
\end{align}
where the dimensionless AHC density harmonic is
\begin{align}
    \rho_n(\bs G)
    =
    \frac{1}{N_{\rm u.c.}}
    \sum_{\bs k\in{\rm BZ}}
    e^{i l_B^2\bs G\wedge\bs k}
    \mathcal W_{n,\bs G}(\bs k)
    =
    \frac{1}{N_{\rm u.c.}}
    \sum_{\bs k\in{\rm BZ}}
    e^{i l_B^2\bs G\wedge\bs k}
    \frac{\mathcal R_{n,\bs G}(\bs k)}
    {\mathcal R_{n,\bs0}(\bs k)} .
    \label{eq:si_rho_n_density_harmonic}
\end{align}
The coherent Bragg contribution to the interaction energy is then
\begin{align}
    \frac{E_B}{N_e}
    =
    \frac{n_e}{2}
    \sum_{\bs G}
    V(\bs G)
    e^{-l_B^2|\bs G|^2/2}
    |\rho_n(\bs G)|^2 .
\end{align}

Now, for the connected part, we first recall that for a conventional QH state
\begin{align}
    \left\langle
    :
    \delta\hat{\tilde\rho}^{\rm LLL}_{\bs Q}
    \delta\hat{\tilde\rho}^{\rm LLL}_{\bs Q'}
    :
    \right\rangle_{\rm QH}
    =
    n_eN_e\,g_{\rm conn}^{\rm QH}(\bs Q)\,
    \delta_{\bs Q+\bs Q',\bs0},
    \label{eq:si_QH_connected_density_diagonal}
\end{align}
where $g_{\rm conn}^{\rm QH}$ is the background-subtracted QH pair correlation in momentum space. Such pair correlation functions for QH states are well known. Various results have been tabulated in the literature and can be used directly~\cite {girvin1984anomalous, girvin1986magneto, fulsebakke2023parametrization}. Using $\bs G_{-\bs R}=-\bs G_{\bs R}$, $a_{-\bs R}(-\bs q)=a_{\bs R}(\bs q)^*$ and the AHC-QH correspondence (see Sec.~\ref{si_sec:ahc_many_body_and_ahc_correspondence}), we obtain
\begin{align}
    g_{\rm conn}^{\rm AHC}(\bs q)
    &\equiv
    \frac{1}{n_eN_e}
    \left\langle
    :
    \delta\hat{\tilde\rho}^{\rm AHC}_{\bs q}
    \delta\hat{\tilde\rho}^{\rm AHC}_{-\bs q}
    :
    \right\rangle_{\mathrm{AHC}}
    =
    \sum_{\bs R}
    |a_{n,\bs R}(\bs q)|^2
    g_{\rm conn}^{\rm QH}(\bs q+\bs G_{\bs R}) .
    \label{eq:si_gconn_AHC_shifted_QH}
\end{align}

Combining the coherent Bragg contribution with the connected contribution gives the following interaction energy per particle for IAHC and FAHC
\begin{align}
    \frac{E_{\rm int}}{N_e}
    &=
    \frac{n_e}{2}
    \sum_{\bs G}
    V(\bs G)e^{-l_B^2|\bs G|^2/2}
    \left|\rho_n(\bs G)\right|^2
    +
    \frac{n_e}{2}
    \int\frac{d^2q}{(2\pi)^2}
    V(\bs q)
    \sum_{\bs R}
    \left|a_{n,\bs R}(\bs q)\right|^2
    g_{\rm conn}^{\rm QH}(\bs q+\bs G_{\bs R}) .
    \label{eq:si_Eint_FAHC_final}
\end{align}
Equivalently, the same result can be summarized by the momentum-space pair correlation
\begin{align}
    g^{\rm AHC}(\bs q)
    &=
    \sum_{\bs G}
    (2\pi)^2\delta^{(2)}(\bs q-\bs G)
    e^{-l_B^2|\bs G|^2/2}
    |\rho_n(\bs G)|^2
    +
    \sum_{\bs R}
    |a_{n,\bs R}(\bs q)|^2
    g_{\rm conn}^{\rm QH}(\bs q+\bs G_{\bs R}),
    \label{eq:si_gAHC_pair_correlation_summary}
\end{align}
with the interaction energy given by 
\begin{align}
    E_{\mathrm{int}}/N_e = \frac{n_e}{2} \int \frac{d^2q}{(2\pi)^2} V(\bs{q}) g^{\rm AHC}(\bs q).
\end{align}

We note once again that in the $0^{\mathrm{th}}$ harmonic approximation for $\mathcal{R}_{n,\bs{q}}$ (see Sec.~\ref{app:zeroth_harmonic_approx_R}) we instead simply have
\begin{align}
    g^{\rm AHC}(\bs q) \approx g^{\rm QH}(\bs q)
\end{align}
and the interaction is the same as in the associated QH state
\begin{align}
    E_{\mathrm{int}}/N_e 
    \approx \frac{n_e}{2} \int \frac{d^2q}{(2\pi)^2} V(\bs{q}) g^{\rm QH}(\bs q). \label{eq:si_eint_qh_zeroth_order_approx}
\end{align}

\subsection{Periodic potential}

We now consider a scalar periodic potential commensurate with the crystal lattice. In the first-harmonic approximation, we keep only the smallest reciprocal lattice vectors and write
\begin{align}
    U(\bs{r})
    &=
    \sum_{j=1}^{3}
    \left(
        U_j e^{i\bs{G}_j\cdot\bs{r}}
        +
        U_j^* e^{-i\bs{G}_j\cdot\bs{r}}
    \right)
    =
    \sum_{\bs{G}\in\{\pm \bs{G}_1,\pm \bs{G}_2,\pm \bs{G}_3\}}
    U_{\bs{G}} e^{i\bs{G}\cdot\bs{r}},
\end{align}
where $\{\pm \bs{G}_j\}$ with $j=1,2,3$ are the six smallest reciprocal lattice vectors and $U_{-\bs{G}}=U_{\bs{G}}^*$ for a real potential.

Using the projected density operator introduced above, the projected periodic-potential Hamiltonian is
\begin{align}
    \hat{\mathcal{H}}_{\mathrm{p.p.}}
    =
    \int d^2r \, U(\bs{r}) \hat{\rho}(\bs{r})
    =
    \sum_{\bs{G}\in\{\pm \bs{G}_1,\pm \bs{G}_2,\pm \bs{G}_3\}} U_{-\bs{G}} 
    \hat{\tilde\rho}^{\rm AHC}_{\bs{G}} .
\end{align}
The corresponding energy per particle is therefore
\begin{align}
    \frac{E_{\mathrm{p.p.}}}{N_e}
    =
    \frac{1}{N_e}
    \sum_{\bs{G}\in\{\pm \bs{G}_1,\pm \bs{G}_2,\pm \bs{G}_3\}} U_{\bs{G}} \,
    \expval{\hat{\tilde\rho}^{\rm AHC}_{\bs{G}}}.
\end{align}
This expression is the first-order energy shift for arbitrary phases of the Fourier components \(U_{\bs G}\). In the main text, we are interested in the maximal commensuration energy at fixed first-harmonic amplitude.  We therefore minimize over the relative registry between the imposed potential and the AHC.  Equivalently, if
\begin{align}
    \expval{\hat{\tilde\rho}^{\rm AHC}_{\bs G_j}}
    =
    \left|
    \expval{\hat{\tilde\rho}^{\rm AHC}_{\bs G_j}}
    \right|
    e^{i\theta_j},
    \qquad
    U_{\bs G_j}=|U_0|e^{i\alpha_j},
\end{align}
then the optimal independent phase choice is
\begin{align}
    \alpha_j=\theta_j+\pi
    \qquad
    (\mathrm{mod}\;2\pi),
\end{align}
or \(U_{\bs G_j}^{\rm opt}=-|U_0|e^{i\theta_j}\).  This gives
\begin{align}
    \frac{E_{\mathrm{p.p.}}^{\rm min}}{N_e}
    =
    -\frac{2|U_0|}{N_e}
    \sum_{j=1}^{3}
    \left|
    \expval{\hat{\tilde\rho}^{\rm AHC}_{\bs G_j}}
    \right|.
    \label{eq:si_periodic_potential_minimized_energy}
\end{align}
Since $\lceil \bs{k}+\bs{G}\rceil=\bs{k}$ for any reciprocal lattice vector $\bs{G}$, one has
\begin{align}
    \expval{ \hat{\tilde\rho}^{\rm AHC}_{\bs{G}} }
    =
    \sum_{\bs{k}\in \mathrm{BZ}}
    \mathcal{F}_n(\bs{k}+\bs{G},\bs{k})
    \expval{\psi^\dagger_{\bs{k}}\psi_{\bs{k}}} = \nu \sum_{\bs{k}\in \mathrm{BZ}}
    \mathcal{F}_n(\bs{k}+\bs{G},\bs{k}).
\end{align}
so that the first-order commensuration energy shift, within the AHC--QH correspondence, is
\begin{align}
    \frac{E_{\mathrm{p.p.}}}{N_e}
    &=
    \sum_{\bs{G}\in\{\pm \bs{G}_1,\pm \bs{G}_2,\pm \bs{G}_3\}}
    U_{-\bs{G}}
    e^{-l_B^2 |\bs{G}|^2/4}
    e^{i\omega_{\bs{G}}}
    \frac{1}{N_{\rm u.c.}} \sum_{\bs{k}\in \mathrm{BZ}}
    e^{i l_B^2 \bs{G}\wedge\bs{k}}
    \frac{\mathcal{R}_{n,\bs{G}}(\bs{k})}{\mathcal{R}_{n,\bs{0}}(\bs{k})}.
\end{align}

Finally, if the external modulation is incommensurate with the crystal, the first-harmonic contribution vanishes.

\section{Monte Carlo simulations}
\label{si_sec:monte_carlo}

We now evaluate the kinetic and interaction energies of the IAHC and Laughlin FAHC trial states directly by Monte Carlo (MC)~\cite{laughlin1983anomalous, morf1986microscopic, morf1986monte, levesque1984crystallization, jain1997composite, wang2019lattice, prange1990quantum, jain2007composite}. This provides a numerical check on the QH--AHC correspondence discussed in Sec.~\ref{si_sec:ahc_many_body_and_ahc_correspondence} and allows us to compare the energy of integer and fractional anomalous Hall crystals with the analytic expressions derived above. 

We consider trial wavefunctions of the form
\begin{align}
    \Psi^{(m)}_{\{\ell_i\}}(\{\bs r_i\})
    =
    \Phi^{\rm QH}_m(\{\bs r_i\})
    \prod_{i=1}^{N_e}
    \chi_{0,\ell_i}^{(\bs0)}(\bs r_i),
    \label{eq:si_mc_ahc_wavefunction}
\end{align}
where
\begin{align}
    \Phi^{\rm QH}_m(\{\bs r_i\})
    =
    \prod_{i<j}(z_i-z_j)^m
    \exp\left[
        -\sum_i\frac{|\bs r_i|^2}{4l_B^2}
    \right].
    \label{eq:si_mc_laughlin_factor}
\end{align}
Here $z_i=x_i+iy_i$, and $m$ is an odd integer. The case $m=1$ corresponds to the integer AHC, while $m=3,5,7,\ldots$ gives Laughlin-type FAHCs at filling $\nu=1/m$ electron per unit cell. 

\subsection{Algorithm}

Let
\begin{align}
    \mathcal C\equiv\{\bs r_1,\ldots,\bs r_{N_e}\},
    \qquad
    d\mathcal C\equiv\prod_{i=1}^{N_e}d^2r_i .
\end{align}
For a given particle configuration $\mathcal C$, the layer-resolved many-body wavefunction can be viewed as a vector
\begin{align}
    \vec\Psi(\mathcal C)
    =
    \left(
    \Psi_{\{0,0,\ldots,0\}}(\mathcal C),
    \Psi_{\{1,0,\ldots,0\}}(\mathcal C),
    \Psi_{\{0,1,\ldots,0\}}(\mathcal C),
    \ldots
    \right)^T .
\end{align}
The expectation value of an operator $\hat O$ can then be written as
\begin{align}
    \langle \hat O\rangle
    =
    \frac{
        \int d\mathcal C\,
        \vec\Psi^\dagger(\mathcal C)\hat O\vec\Psi(\mathcal C)
    }{
        \int d\mathcal C\,
        \vec\Psi^\dagger(\mathcal C)\vec\Psi(\mathcal C)
    }
    =
    \frac{
        \int d\mathcal C\,
        P_m(\mathcal C)\,
        O_{\rm loc}(\mathcal C)
    }{
        \int d\mathcal C\,
        P_m(\mathcal C)
    },
    \label{eq:si_mc_local_estimator_general}
\end{align}
where
\begin{align}
    P_m(\mathcal C)
    =
    \vec\Psi^\dagger(\mathcal C)\vec\Psi(\mathcal C),
    \qquad
    O_{\rm loc}(\mathcal C)
    =
    \frac{
        \vec\Psi^\dagger(\mathcal C)\hat O\vec\Psi(\mathcal C)
    }{
        \vec\Psi^\dagger(\mathcal C)\vec\Psi(\mathcal C)
    } .
\end{align}
Thus, we can estimate observables by generating random sample configurations according to the probability distribution given by the positive weight $P_m(\mathcal C)$. We generate such samples by constructing a Markov chain with single-particle updates that follow the Metropolis-Hastings algorithm~\cite{metropolis1953equation, ceperley1986quantum}.

For the wavefunctions of interest, the layer trace factorizes locally
\begin{align}
    P_m(\mathcal C)
    =
    \left|
        \Phi^{\rm QH}_m(\mathcal C)
    \right|^2
    \prod_{i=1}^{N_e} \left(\sum_{\ell=0}^{N_\ell-1}
    \left|
        \chi_{0,\ell}^{(\bs0)}(\bs r)
    \right|^2 \right)
    .
    \label{eq:si_mc_sampling_distribution}
\end{align}
This is the distribution sampled in the MC simulation. Importantly, we do not explicitly sample the layer indices. Instead, the sum over layers is performed analytically. The different layer components of the Abrikosov lattice wavefunctions are evaluated using Eq.~\eqref{eq:si_rewrite_chi_with_magnetic_bloch_states}. The magnetic Bloch functions entering this expression are evaluated using~\eqref{eq:si_magnetic_bloch_states}, with the double sum truncated over finite windows sufficiently large that increasing them does not change the measured observables within the MC error bars.

We work with a finite quantum Hall droplet geometry~\cite{morf1986monte}. For a Laughlin factor at filling $\nu=1/m$, the characteristic droplet radius is
\begin{align}
    R_{\rm drop}
    =
    l_B\sqrt{2mN_e}.
    \label{eq:si_mc_droplet_radius}
\end{align}
The initial configurations are generated in a larger nominal disk,
\begin{align}
    R_{\rm disk}
    =
    (1+\delta_{\rm disk})R_{\rm drop},
\end{align}
where $\delta_{\rm disk}$ is a buffer parameter. To initialize uniformly in area, we draw
\begin{align}
    r=R_{\rm disk}\sqrt{u},
    \qquad
    \theta=2\pi v,
\end{align}
with $u,v\in[0,1]$ independent uniform random variables. The resulting initial configuration is used only as the starting point for the Markov chain; all measurements are performed after thermalization.

Each Metropolis step proposes to move a single randomly chosen particle. The proposal is a symmetric displacement uniformly distributed in a disk of radius $\Delta$,
\begin{align}
    \bs r_i'
    =
    \bs r_i
    +
    \Delta\sqrt{u}\,
    (\cos\theta,\sin\theta),
    \qquad
    \theta=2\pi v,
\end{align}
again with $u,v$ uniform in $[0,1]$. Since the proposal distribution is symmetric, the Metropolis-Hastings acceptance probability reduces to
\begin{align}
    p_{\rm acc}
    =
    \min\left[
        1,
        \frac{P_m(\mathcal C')}{P_m(\mathcal C)}
    \right],
    \label{eq:si_mc_acceptance_probability}
\end{align}
with $\mathcal{C}'$ the proposed updated configuration and $\mathcal{C}$ the current electronic configuration. 

\subsection{Observables}
\label{si_subsec:mc_observables}

We now describe how observables are evaluated from the Markov chain. The MC samples are distributed according to the positive weight $P_m(\mathcal C)$ defined above, and any observable is estimated by averaging its local estimator over the production samples.

Because the simulations are performed in a finite disk geometry, we distinguish between the full droplet and the bulk region used for measurements. For an observable $X$, we define a circular measurement mask
\begin{align}
    \Theta_i^{(X)}
    =
    \Theta\!\left(R_X-|\bs r_i|\right),
    \qquad
    N_X(\mathcal C)
    =
    \sum_{i=1}^{N_e}\Theta_i^{(X)} ,
\end{align}
where $R_X$ is the mask radius and $\Theta\!\left(R_X-|\bs r_i|\right)$ is a Heaviside function. The mask suppresses edge effects by restricting the reference particles entering the estimator to the droplet's bulk. We use different mask radii as a diagnostic of residual edge dependence.

\subsubsection{Kinetic energy}
\label{si_subsubsec:mc_kinetic_energy}

The kinetic energy is evaluated by measuring local derivative estimators. For a dispersion expanded as a polynomial in momentum,
\begin{align}
    E(\bs p)
    =
    c_2 |\bs p|^2
    +
    c_4 |\bs p|^4
    +\cdots ,
\end{align}
the kinetic energy per particle is
\begin{align}
    \frac{E_{\rm kin}}{N_e}
    =
    c_2\left\langle -\nabla^2 \right\rangle
    +
    c_4\left\langle (-\nabla^2)^2 \right\rangle
    +\cdots .
    \label{eq:si_mc_kinetic_expansion}
\end{align}
Here
\begin{align}
    \left\langle(-\nabla^2)^\alpha\right\rangle
    =
    \left\langle
        K_\alpha^{\rm loc}(\mathcal C)
    \right\rangle_{\rho}
\end{align}
with a local estimator
\begin{align}
    K_\alpha^{\rm loc}(\mathcal C)
    =
    \frac{1}{N_K(\mathcal C)}
    \sum_{i=1}^{N_e}
    \Theta_i^{(K)}
    k_{\alpha,i}^{\rm loc}(\mathcal C),
    \label{eq:si_mc_masked_kinetic_estimator}
\end{align}
where
\begin{align}
    k_{\alpha,i}^{\rm loc}(\mathcal C)
    =
    \frac{
    \sum_{\{\ell_j\}}
    \Psi_{\{\ell_j\}}^*(\mathcal C)
    \left[(-\nabla_i^2)^\alpha
    \Psi_{\{\ell_j\}}(\mathcal C)\right]
    }{
    \sum_{\{\ell_j\}}
    |\Psi_{\{\ell_j\}}(\mathcal C)|^2
    } .
    \label{eq:si_mc_single_particle_kinetic_estimator}
\end{align}
The derivatives act only on the coordinate of particle $i$. In practice, the derivatives are evaluated analytically from the QH factor $\Phi_m^{\rm QH}$ and the Abrikosov factors $\chi_{0,\ell}^{(\bs0)}$. The layer sums are then reduced to local sums over $\chi_{0,\ell}^{(\bs0)}$ and its derivatives at the position $\bs r_i$.

For example, define
\begin{align}
    A_{i,\alpha}
    =
    \frac{\partial_{i,\alpha}\Phi_m^{\rm QH}}{\Phi_m^{\rm QH}},
    \qquad
    B_i
    =
    \frac{\nabla_i^2\Phi_m^{\rm QH}}{\Phi_m^{\rm QH}},
\end{align}
and the local Abrikosov sums
\begin{align}
    S_{00}(\bs r)
    &=
    \sum_{\ell}
    |\chi_{0,\ell}^{(\bs0)}(\bs r)|^2,\\
    S_{01,\alpha}(\bs r)
    &=
    \sum_{\ell}
    \chi_{0,\ell}^{(\bs0)*}(\bs r)
    \partial_{\alpha}\chi_{0,\ell}^{(\bs0)}(\bs r),\\
    S_{02}(\bs r)
    &=
    \sum_{\ell}
    \chi_{0,\ell}^{(\bs0)*}(\bs r)
    \nabla^2\chi_{0,\ell}^{(\bs0)}(\bs r).
\end{align}
Then the local estimator for the Laplacian of particle $i$ is
\begin{align}
    \frac{
    \sum_{\{\ell_j\}}
    \Psi_{\{\ell_j\}}^*
    \nabla_i^2
    \Psi_{\{\ell_j\}}
    }{
    \sum_{\{\ell_j\}}
    |\Psi_{\{\ell_j\}}|^2
    }
    =
    B_i
    +
    2A_{i,\alpha}
    \frac{S_{01,\alpha}(\bs r_i)}{S_{00}(\bs r_i)}
    +
    \frac{S_{02}(\bs r_i)}{S_{00}(\bs r_i)} ,
    \label{eq:si_mc_laplacian_local_estimator}
\end{align}
where repeated Cartesian indices are summed. The estimator for $-\nabla_i^2$ is minus the real part of this expression. The imaginary part has vanishing expectation value and is discarded in the numerical average.

For the quartic term, it is convenient to use integration by parts rather than explicitly applying four derivatives. Since the droplet wavefunction is exponentially suppressed at large radius, boundary terms vanish, and
\begin{align}
    \left\langle(-\nabla^2)^2\right\rangle
    =
    \frac{
    \int d\mathcal C
    \sum_{\{\ell_j\}}
    \frac{1}{N_e}\sum_i
    \left|
        \nabla_i^2\Psi_{\{\ell_j\}}(\mathcal C)
    \right|^2
    }{
    \int d\mathcal C
    \sum_{\{\ell_j\}}
    |\Psi_{\{\ell_j\}}(\mathcal C)|^2
    }.
    \label{eq:si_mc_bilaplacian_by_parts}
\end{align} 
This yields a positive local estimator that involves only the Laplacian already computed in Eq.~\eqref{eq:si_mc_laplacian_local_estimator}, together with the corresponding derivatives of the Abrikosov factors. Higher polynomial terms in the dispersion can be treated in the same way.

\subsubsection{Interaction energy}
\label{si_subsubsec:mc_interaction_energy}

The interaction energy is evaluated directly from the sampled particle coordinates. For a two-body interaction $V(r)$, the full finite-droplet local estimator is
\begin{align}
    V_{\rm loc}(\mathcal C)
    =
    \frac{1}{N_e}
    \sum_{i<j}
    V(|\bs r_i-\bs r_j|).
    \label{eq:si_mc_full_interaction_estimator}
\end{align}
This estimator gives the interaction energy per particle of the finite droplet. However, because the droplet has an edge, the full estimator contains edge contributions. For bulk estimates, we instead use an interaction window. We define an
inner reference region and an outer partner region,
\begin{align}
    \mathcal I(\mathcal C)
    =
    \{i: |\bs r_i|<R_{\rm in}\},
    \qquad
    \mathcal O(\mathcal C)
    =
    \{j: |\bs r_j|<R_{\rm out}\},
\end{align}
with $R_{\rm out}\ge R_{\rm in}$. The local interaction-window
estimator is
\begin{align}
    V_{\rm loc}^{\rm win}(\mathcal C)
    =
    \frac{1}{N_{\rm in}(\mathcal C)}
    \sum_{i\in\mathcal I(\mathcal C)}
    \frac{1}{2}
    \sum_{\substack{j\in\mathcal O(\mathcal C)\\ j\ne i}}
    V(|\bs r_i-\bs r_j|),
    \label{eq:si_mc_window_interaction_estimator}
\end{align}
where $N_{\rm in}(\mathcal C)=|\mathcal I(\mathcal C)|$. Thus, the reference particle must lie within the inner bulk window, while its interaction partners are included up to the larger radius $R_{\rm out}$. This separates the reference-particle bulk cut from the finite-droplet outer cutoff. Because $N_{\rm in}(\mathcal C)$ fluctuates between MC samples, we do not quote the simple mean of $V_{\rm loc}^{\rm win}$. Instead, for recorded samples $\mathcal C_s$, we accumulate the numerator and denominator separately and use the particle-weighted estimator
\begin{align}
    \varepsilon_{\rm int}^{\rm MC}
    = \frac{E_{\rm int}}{N_e} =
    \frac{
        \sum_s
        N_{\rm in}(\mathcal C_s)
        V_{\rm loc}^{\rm win}(\mathcal C_s)
    }{
        \sum_s N_{\rm in}(\mathcal C_s)
    } .
    \label{eq:si_mc_window_ratio_estimator}
\end{align}
This choice keeps the local environment of the reference particle close to that of the thermodynamic bulk while still including its interactions with the full droplet. 

Since the MC estimator averages the local interaction energy of particles inside a bulk mask, $|{\bs r}|<R_{\rm in}$, while not considering particles outside $R_{\rm out}$, it should slightly underestimate the positive, non-background-subtracted bulk interaction energy. A simple upper-scale estimate is obtained by adding the continuum tail that a masked particle would feel from uniform density outside the finite droplet,
\begin{align}
    (\Delta E_{\rm int})_{\rm tail}/N_e
    \simeq
    \frac{n_e}{2}\frac{1}{\pi R_{\rm in}^2}
    \int_{|{\bf r}_0|<R_{\rm in}} d^2 r_0
    \int_{|{\bf r}|>R_{\rm out}} d^2 r\,
    V(|{\bf r}-{\bf r}_0|). \label{eq:si_estimate_error_interaction_mc}
\end{align}
For the parameters used below, such contributions from beyond the droplet radius are negligible and always smaller than the statistical uncertainty. 

The same framework applies to different interaction kernels. Here, we report results for a dual-gated-screened interaction with gate distance $d_s$ and dielectric constant $\epsilon$ of the form 
\begin{align}
    V(r)
    = V_0
    \sum_{n=-\infty}^{\infty}
    \frac{(-1)^n}{\sqrt{r^2 + (2 n d_s)^2}}, 
    \qquad V(q) = \frac{2\pi V_0}{q}\tanh(q d_s), \label{eq:si_dual_gated_screened_interaction}
\end{align}
with $V_0 = e^2/\epsilon$. The code evaluates the corresponding radially screened kernel from a precomputed interpolation table.

\subsubsection{Thermalization and uncertainties}
\label{si_subsubsec:mc_uncertainties}

Before measuring any observable, we discard an initial thermalization segment of the Markov chain. In the simulations reported below, the burn-in consists of $N_{\rm therm}\sim 10^3$--$10^4$ sweeps, with one sweep defined as $N_e$ attempted single-particle updates. During thermalization, the proposal step size is adjusted to achieve a reasonable acceptance rate of approximately $50\%$. Once production sampling begins, the proposal parameters are kept fixed. We check equilibration by monitoring the acceptance rate, the radial density profile, and the running averages of the measured kinetic and interaction estimators. After equilibration, the sampled configurations are distributed according to $P_m(\mathcal C)$, and observables are accumulated every fixed number of sweeps.

Successive MC samples are correlated, so the uncertainty should not be estimated from the naive standard error of the raw time series. For each measured observable $O_s$, where $s$ labels the recorded samples, we compute the normalized autocorrelation function~\cite{becca2017quantum, sandvik2010computational}
\begin{align}
    C_O(t)
    =
    \frac{
        \left\langle
        (O_s-\bar O)(O_{s+t}-\bar O)
        \right\rangle_s
    }{
        \left\langle
        (O_s-\bar O)^2
        \right\rangle_s
    } .
    \label{eq:si_mc_autocorrelation_function}
\end{align}
The integrated autocorrelation time is estimated as
\begin{align}
    \tau_{\rm int}
    =
    \frac12+\sum_{t>0}C_O(t),
    \label{eq:si_mc_integrated_autocorrelation_time}
\end{align}
with the sum truncated using an automatic windowing criterion. This gives an effective number of statistically independent samples:
\begin{align}
    N_{\rm eff}
    \simeq
    \frac{N_{\rm samp}}{2\tau_{\rm int}}.
\end{align}
As an independent check, we perform a binning analysis. The production time series is divided into blocks of increasing size, the mean is computed within each block, and the uncertainty is extracted from the variance of the block means. The quoted MC error bars are taken from the plateau reached when the block size is large compared with $\tau_{\rm int}$. Consistency between the autocorrelation estimate and the blocking plateau is used as a final check that the statistical errors are reliable. For the interaction-window estimator in Eq.~\eqref{eq:si_mc_window_ratio_estimator}, the same numerator-over-denominator ratio is computed within each bin. The reported interaction-energy error bars are obtained from the variance of these block-ratio estimates.

\subsection{Numerical results}

\begin{figure}
    \centering
    \includegraphics[width=1.00\linewidth]{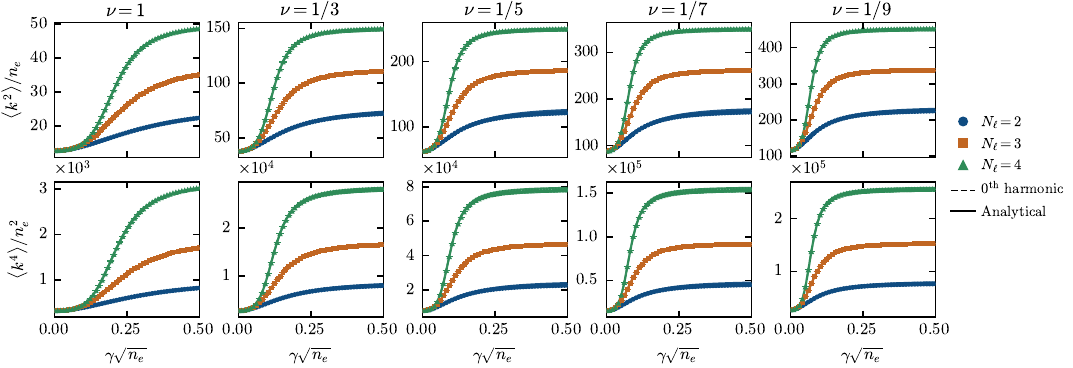}\vspace{-3mm}
    \caption{Dependence of $\expval{k^2}$ (first row) and $\expval{k^4}$ (second row) on $\gamma$ with $N_\ell=2,3,4$ for IAHC (first column) and FAHC at $\nu=1/3$, 1/5, 1/7, and 1/9 (second, third, fourth, and fifth columns). We compare results obtained with MC (data point), the full analytical expression~\eqref{eq:si_full_kinetic_energy} (solid line), and the $0^{\mathrm{th}}$ harmonic approximation~\eqref{eq:si_full_kinetic_energy_zeroth_order} (dashed line).}
    \label{si_fig:mc_kinetic_energy_1}
\end{figure}

\begin{figure}
    \centering
    \includegraphics[width=1.00\linewidth]{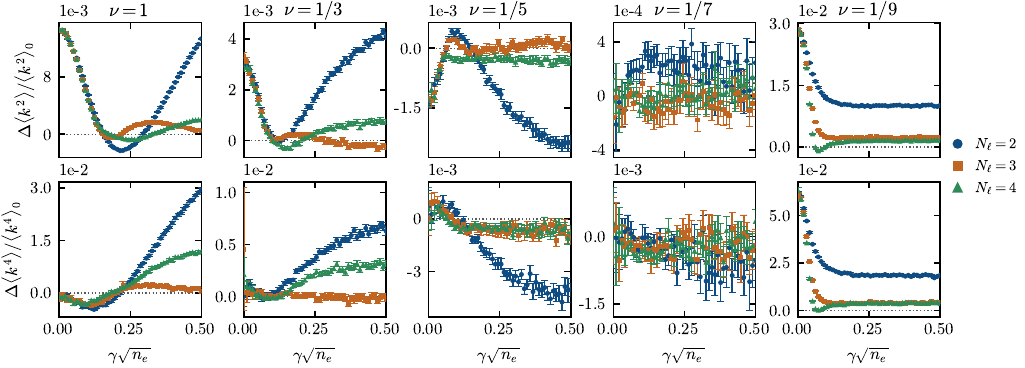}\vspace{-3mm}
    \caption{Relative difference given in~\eqref{eq:si_relative_difference_kinetic_energy} between the MC data and the $0^{\mathrm{th}}$ order approximation \eqref{eq:si_full_kinetic_energy_zeroth_order} for $\expval{k^2}$ (first row) and $\expval{k^4}$ (second row) as a function of $\gamma$ with $N_\ell=2,3,4$. The comparison is done for IAHC (first column) and FAHC at $\nu=1/3$, 1/5, 1/7, and 1/9 (second, third, fourth, and fifth columns).}
    \label{si_fig:mc_kinetic_energy_2}
\end{figure}

We now discuss some of the numerical results obtained using the MC algorithm described above. We first compute the kinetic energy for a system of 300 electrons as a function of $\gamma \sqrt{n_e}$ for IAHC and FAHC at $\nu=1/3$, 1/5, 1/7, and $1/9$. The results presented in Fig.~\ref{si_fig:mc_kinetic_energy_1} are obtained by sampling the system over $1.5 \cdot 10^6$ sweeps after performing $10^{3}$ initial thermalization sweeps. We use a mask with radius $R_K=R_{\mathrm{drop}}/2$ to reduce boundary effects. We see excellent agreement between the MC results, the full analytical expression~\eqref{eq:si_full_kinetic_energy}, and the $0^{\mathrm{th}}$ harmonic approximation~\eqref{eq:si_full_kinetic_energy_zeroth_order} over the whole range of $\gamma\sqrt{n_e}$ and all values of $\nu$ and $N_\ell$ studied. In particular, for all values of $\nu$ and $N_\ell$, we observe the quadratic rise with $\gamma$ of $\expval{k^{2\alpha}}$ in the small $\gamma$ regime as well as the $\gamma$-independent behavior at large values of $\gamma$ we had anticipated theoretically (i.e., Eqs.~\eqref{eq:si_full_kinetic_energy_small_gamma} and \eqref{eq:si_full_kinetic_energy_large_gamma}). Since the three curves do not appear to deviate at all on the scale of Fig.~\ref{si_fig:mc_kinetic_energy_1}, we present in Fig.~\ref{si_fig:mc_kinetic_energy_2} the relative difference
\begin{align}
    \frac{\Delta\expval{k^{2\alpha}}}{\expval{k^{2\alpha}}_0} = \frac{\expval{k^{2\alpha}}_{\mathrm{MC}} - \expval{k^{2\alpha}}_{0}  }{\expval{k^{2\alpha}}_0} , \label{eq:si_relative_difference_kinetic_energy}
\end{align}
between the kinetic energy calculated with Monte-Carlo $\expval{k^{2\alpha}}_{\mathrm{MC}}$ and the $0^{\mathrm{th}}$-order approximation~\eqref{eq:si_full_kinetic_energy_zeroth_order}. We see that the relative difference remains small for all values of $\nu$, $\gamma$, and $N_\ell$. This confirms the validity of our approximation~\eqref{eq:si_full_kinetic_energy_zeroth_order}. 

\begin{figure}
    \centering
    \includegraphics[width=1.00\linewidth]{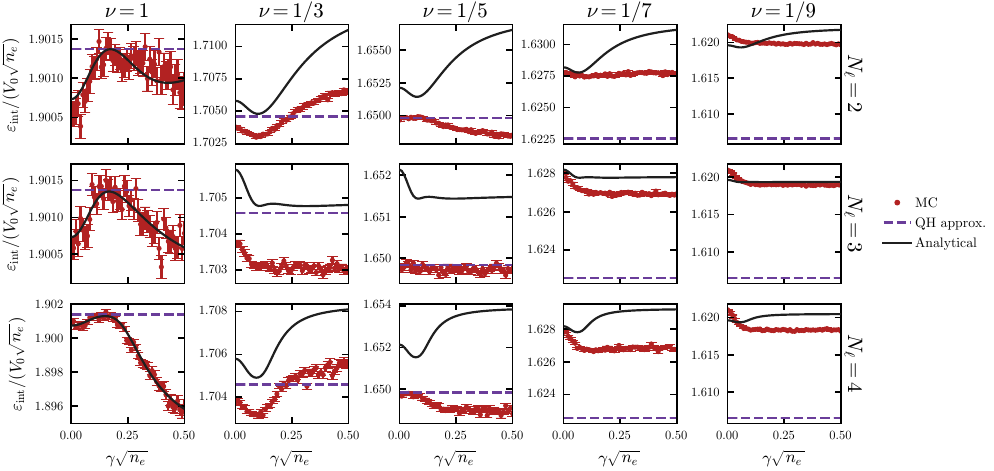}\vspace{-3mm}
    \caption{Comparison between the interaction energy evaluated with MC (red markers), the full thermodynamic analytical estimate based on the AHC--QH correspondence in Eq.~\eqref{eq:si_Eint_FAHC_final} (black solid line), and the zeroth-order QH approximation in Eq.~\eqref{eq:si_eint_qh_zeroth_order_approx} (purple dashed line). The MC data use the native particle-weighted window estimator Eq.~\eqref{eq:si_mc_window_ratio_estimator} for $N_e=500$, with $R_{\rm in}\sqrt{n_e}=3.154$ and $R_{\rm out}\sqrt{n_e}=12.616$. Results are shown for $N_\ell=2,3,4$ (first, second, and third row) for IAHC (first column) and FAHC at $\nu=1/3$, 1/5, 1/7, and 1/9 (second, third, fourth, and fifth columns). Simulations use a dual-gated screened interaction~\eqref{eq:si_dual_gated_screened_interaction} with a gate distance of $d_s=1/\sqrt{n_e}$.}
    \label{si_fig:mc_interaction_energy_1}
\end{figure}

For the interaction energy, we present in Fig.~\ref{si_fig:mc_interaction_energy_1} a comparison between interaction energy per particle $\varepsilon_{\rm int} = E_{\rm int}/N_e$ obtained with MC, the analytical formula~\eqref{eq:si_Eint_FAHC_final}, and the zeroth-order approximation~\eqref{eq:si_eint_qh_zeroth_order_approx} for dual gated screened interaction with $d_s\sqrt{n_e}=1$. The MC simulations use $N_e=500$ electrons and six independent replicas, each with $2.5\times 10^5$ production sweeps after $10^3$ thermalization sweeps. Both theoretical estimates are evaluated using the quantum Hall pair correlation functions tabulated in Ref.~\cite{fulsebakke2023parametrization}. Overall, from Fig.~\ref{si_fig:mc_interaction_energy_1}, we see an extremely weak dependence of the interaction energy on both $\gamma$ and $N_\ell$. We also note that the interaction energy decreases with $\nu$. The agreement between the Monte Carlo theoretical estimate from both~\eqref{eq:si_Eint_FAHC_final} and its zeroth-order approximation ~\eqref{eq:si_eint_qh_zeroth_order_approx} is also particularly good. In particular, the relative difference between the MC results and our theoretical estimate always remains below $0.4\%$ over the entire parameter range shown in Fig.~\ref{si_fig:mc_interaction_energy_1}. Notably, the theoretical estimate appears to capture the weak $\gamma$ dependence of the data for the $\nu=1$ and $\nu=1/3$ states quite accurately. This broad agreement confirms once again the validity of the QH-AHC correspondence (see Sec.~\ref{si_sec:ahc_many_body_and_ahc_correspondence}) used to derive our analytical results.

In the figure~\ref{si_fig:mc_interaction_energy_1}, we take $R_{\rm out}=R_{\rm drop}$ for $N_e=500$, so the outer window coincides with the simulated droplet radius. We also evaluated additional exact-radius windows with smaller $R_{\rm out}$ and larger $R_{\rm in}$ to diagnose the sensitivity to the reference window and the outer cutoff. These checks show that the qualitative comparison to the thermodynamic analytical estimate is stable under the window choices used here.

\section{Hartree-Fock energy of the Fermi liquid in rhombohedral multilayer graphene} \label{si_sec:schf_on_fl_in_rng}

In the main text, we evaluate the energy of Fermi liquids with disk and annulus Fermi surfaces for relevant models of rhombohedral N-layer graphene. Here, we briefly comment on how this Hartree-Fock energy is evaluated. 

We assume that the FL is described by a Slater determinant that is rotationally symmetric in momentum space. The one-flavor Fermi wavevector is
\begin{equation}
  k_F = \sqrt{4\pi n_e}.
\end{equation}
The variational shape is an annulus controlled by $\eta \geq 1$
\begin{equation}
  k_{\max} = \eta k_F,
  \qquad
  k_{\min} = \sqrt{k_{\max}^2 - k_F^2}.
\end{equation}
Thus the occupied area satisfies
\begin{equation}
  \pi(k_{\max}^2-k_{\min}^2) = \pi k_F^2 = 4\pi^2 n_e,
\end{equation}
so the particle density is fixed while $\eta$ changes the shape. The case $\eta=1$ is the ordinary disk Fermi sea.

\begin{figure}
    \centering
    \includegraphics[width=1.00\linewidth]{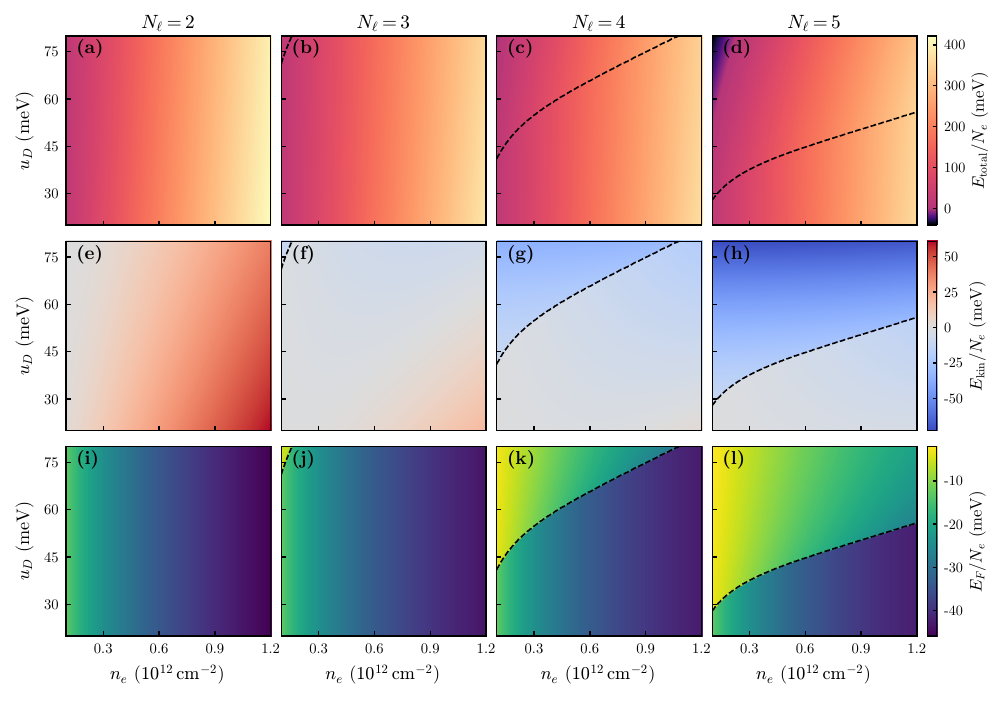}\vspace{-3mm}
    \caption{Hartree-Fock energy decomposition of the optimized spin- and valley-polarized Fermi liquid in rhombohedral multilayer graphene for a dual-gated screened interaction with $\epsilon=4$ and $d_s=30$~nm. Columns show layer number $N_\ell=2,3,4,5$, and rows show the total energy $E_{\mathrm{total}}$, kinetic energy $E_{\mathrm{kin}}$, and Fock exchange energy $E_F$, all per electron. The horizontal axis is electron density $n_e$, and the vertical axis is displacement field $u_D$. Dashed black curves mark the disk-to-annulus transition of the optimized Fermi surface, defined by $\eta_{\mathrm{opt}}=1$. For $N_\ell=2$, the optimized Fermi surface remains disk-like over the plotted range, so no transition curve appears. Color scales are shared across layer number within each energy row.}
    \label{si_fig:fl_energy}
\end{figure}

For a displacement field $u_D$, the rhombohedral graphene model supplies the dispersion $\mathcal{E}(\bs{k})$. For convenience, we always measure the energy of the dispersion relative to $\mathcal{E}(\bs{0})$ (i.e., we fix $\mathcal{E}(\bs{0})=0$). The kinetic energy per particle for a given annulus is then
\begin{equation}
  E_{\mathrm{kin}}(\eta)/N_e
  =
  \frac{1}{n_e}
  \int_{k_{\min} \leq |\mathbf{k}| \leq k_{\max}}
  \frac{d^2k}{(2\pi)^2}\,
  \mathcal{E}(\bs{k})
  =
  \frac{1}{2 \pi n_e}
  \int_{k_{\min}}^{k_{\max}} dk\, k\, \mathcal{E}(k).
\end{equation}

Since we consider a Slater determinant in momentum space here, the interaction energy splits into Hartree and Fock contributions. Given a two-dimensional momentum-space interaction $V(q)$, the Hartree term for a uniform FL is the $\bs{q}=0$ direct contribution
\begin{equation}
  E_{\mathrm{H}}/N_e = \frac{n_e}{2} V(0).
\end{equation}
The Fock exchange contribution is
\begin{equation}
  E_{\mathrm{F}}(\eta)/N_e
  =
  -\frac{1}{2 n_e}
  \int_{\mathrm{occ}}
  \frac{d^2k}{(2\pi)^2}
  \int_{\mathrm{occ}}
  \frac{d^2k'}{(2\pi)^2}
  V(|\mathbf{k}-\mathbf{k}'|)
  |F_P(\mathbf{k},\mathbf{k}')|^2 .
\end{equation}
The total interaction energy is then
\begin{equation}
  E_{\mathrm{int}}(\eta)/N_e = E_{\mathrm{H}}/N_e + E_{\mathrm{F}}(\eta)/N_e.
\end{equation}

We present in  Fig.~\ref{si_fig:fl_energy} a summary of the total, kinetic, and exchange energy of the optimized FL for a minimal model of Rhombohedral N-layer graphene with a fixed geometry set by $\gamma=v_D/t_1$ and the microscopic dispersion of the first conduction band that depends on $(u_D,t_1,v_D)$. The dashed line indicates the disk-to-annulus transition of the Fermi surface. This transition gets pushed down to smaller displacement fields as the number of layers increases.

%